\documentclass[iop]{emulateapj}

\usepackage{amsmath,natbib,graphicx}
\usepackage{epsf}
\usepackage{epstopdf}
\usepackage{epsfig}
\usepackage{color}
\DeclareGraphicsExtensions{.jpg,.pdf,.png,.eps,.ps}
\usepackage{scrextend}
\usepackage{hyperref}

\newcommand{\III}{~{\sc iii}}

\shorttitle{}

\begin{document}

\title{Is extended hard X-ray emission ubiquitous in Compton-thick AGN?}

\author{Jingzhe Ma$^{1}$, Martin Elvis$^{1}$, G. Fabbiano$^{1}$, Mislav Balokovi{\'c}$^{1,2}$, W. Peter Maksym$^{1}$, Mackenzie L. Jones$^{1}$, Guido Risaliti$^{3}$  }

\altaffiltext{1}{Center for Astrophysics $|$ Harvard \& Smithsonian, 60 Garden St, Cambridge, MA 02138, USA; \href{jingzhe.ma@cfa.harvard.edu}{jingzhe.ma@cfa.harvard.edu}}
\altaffiltext{2}{Black Hole Initiative at Harvard University, 20 Garden Street, Cambridge, MA 02138, USA}
\altaffiltext{3}{INAF- Osservatorio Astrofisico di Arcetri, Largo E. Fermi 5 50125 Firenze, Italy}


\begin{abstract}

The recent {\it Chandra} discovery of extended $\sim$kpc-scale hard ($>$ 3 keV) X-ray emission in nearby Compton-thick (CT) active galactic nuclei (AGN) opens a new window to improving AGN torus modeling and investigating how the central super massive black hole interacts with and impacts the host galaxy. Since there are only a handful of detections so far, we need to establish a statistical sample to determine the ubiquity of the extended hard X-ray emission in CT AGN, and quantify the amount and extent of this component. In this paper, we present the spatial analysis results of a pilot {\it Chandra} imaging survey of 7 nearby ($0.006 < z < 0.013$) CT AGN selected from the {\it Swift}-BAT spectroscopic AGN survey. We find that five out of the seven CT AGN show extended emission in the 3-7 keV band detected at $>$ 3$\sigma$ above the {\it Chandra} PSF with $\sim$12\% to 22\% of the total emission in the extended components. ESO 137-G034 and NGC 3281 display biconical ionization structures with extended hard X-ray emission reaching kpc-scales ($\sim$ 1.9 kpc and 3.5 kpc in diameter). The other three show extended hard X-ray emission above the PSF out to at least $\sim$360 pc in radius. We find a trend that a minimum 3-7 keV count rate of 0.01 cts/s and total excess fraction $>$20\% is required to detect a prominent extended hard X-ray component. Given that this extended hard X-ray component appears to be relatively common in this uniformly selected CT AGN sample, we further discuss the implications for torus modeling and AGN feedback.

\end{abstract}


\section{Introduction}
\label{sec:intro}

\begin{table*}
\centering
\caption{Observation Log}
\scriptsize
\begin{tabular}{lccccclccc}
\hline\hline
Sourcename & $z$   &ObsID    & Instrument    &  $T_{\rm exp}$ (ks)&  PI  & Date                    & $d$x (pix)   & $d$y (pix)     & Net Counts \\  
 \hline
NGC 424 (TOLOLO 0109-383)         &0.01176   &  21417 & ACIS-S          & 15.28                & Elvis &   2019  Feb 7    &---  &---& 748 $\pm$ 28\\    
                       &                &  3146   & ACIS-S          & 9.18                  & Matt   &  2002  Feb 4    & -0.308 & 0.048                  & 1238 $\pm$ 35 \\
NGC 1125     & 0.01093    &  21418   & ACIS-S         & 53.15              & Elvis & 2019 Oct 25      & ---&---                                & 696 $\pm$ 27 \\  
NGC 3281    &  0.01067  & 21419   & ACIS-S              & 9.13               & Elvis    & 2019 Jan 24   & ---&---                                &431 $\pm$ 21 \\		
NGC 4500    & 0.01038   & 21420  & ACIS-S             & 18.13                 & Elvis  & 2019 Aug 11    &--- &---				& 274 $\pm$ 17 \\		
ESO 005-G004 & 0.00623 & 21421  & ACIS-S            & 21.94               & Elvis  & 2019 Feb 18    &--- &--- 				& 189 $\pm$ 15\\			
ESO 137-G034 & 0.0090 & 21422  & ACIS-S           & 44.76                 & Elvis  & 2019 Oct 16     &---&---				& 1119 $\pm$ 34\\                 
2MASXJ00253292+6821442 & 0.0120 & 21423 & ACIS-S & 29.54   & Elvis   & 2019 Oct 10       &--- &---				& 230 $\pm$ 16\\					
                                              &           & 12862  & ACIS-S     & 4.9 & Mushotzky & 2010 Dec 3  & 1.369  & -1.327               & 53 $\pm$ 8 \\
\hline
\end{tabular}
\tablecomments{$d$x and $d$y are the positional shifts relative to the reference image in the ACIS instrumental pixels (0.492$\arcsec$). The total counts at 0.3-7 keV above the background are listed in the last column. }
\label{table1}
\end{table*}

The recent {\it Chandra} discovery of hard ($>$ 3 keV) X-ray emission in nearby Compton-thick (CT) active galactic nuclei (AGN) extending out from the central supermassive black hole (SMBH) to $\sim$kpc-scales (e.g., \citealt{Arevalo2014,Bauer2015,Maksym2017,Fabbiano2017,Fabbiano2018a,Fabbiano2018b,Fabbiano2019,Jones2020}) challenges the established picture that the characteristic hard X-ray continuum and fluorescent Fe K$\alpha$ lines are confined to the nuclear surroundings, as would be expected in the standard model of an AGN enshrouded by a CT obscuring torus (e.g., \citealt{Chou2007,Marinucci2012, Marinucci2017}). The extended diffuse emission is not only detected in the direction of the ionization cones but is also present in the region perpendicular to the ionization cones, i.e., cross-cones (e.g., \citealt{Fabbiano2018a,Fabbiano2018b,Fabbiano2019,Jones2020}), where the obscuring torus in the standard AGN unified model should completely obscure the nucleus (e.g., \citealt{Antonucci1993, Urry1995,Netzer2015}).

The spatial extent of the hard X-rays and the presence in both the ionization cones and cross-cones have several implications, linking the structure of the obscuring torus with the impact on the host galaxy, i.e., AGN feedback. First, measuring the amount of extended hard X-ray emission will help quantify potential bias in spectral modeling of the reprocessed emission from the torus \citep{Balokovic2014,Bauer2015,Farrah2016}. Ignoring its presence will bias the spectral modeling of the AGN torus as this emission will be incorrectly attributed to the torus, leading to biased estimates of the structural parameters of the torus, such as its covering factor. The presence of extended hard X-ray emission in the cross-cones indicates that the torus is likely porous. Second, the extent of hard X-ray emission suggests interaction of nuclear photons with the interstellar medium (ISM) of the host galaxy. The hard continuum and Fe K$\alpha$ extended emission observed in ESO 428-G014, a well-studied nearby CT AGN, is likely caused by scattering off dense molecular clouds in the host galaxy of photons escaping the nuclear region (e.g., \citealt{Fabbiano2017,Fabbiano2018a,Fabbiano2018b,Fabbiano2019}). The kpc-scale spatial extent is larger at the lower energies, suggesting that the optically thick molecular clouds responsible for the scattering of the higher energy photons are more concentrated in the inner radii. Finally, the fraction of CT AGN with extended hard emission will constrain the duty cycle for AGN feedback onto the host galaxy ISM.

Since there are only a handful of such detections so far, we need to build a statistical sample to answer the following questions: (1) Is the extended hard X-ray emission ubiquitous in CT AGN? (2) What is the typical extent and amount of extended hard X-ray emission relative to the total emission? And is there a limit on the physical scales that the extended hard X-ray can reach?  (3) What is the origin of the extended hard X-ray emission and its implications on AGN models? 

In an effort to address these questions, we have been conducting a {\it Chandra} survey of nearby ($z < 0.02$) CT AGN with joint {\it NuSTAR} observations. CT AGN have the advantage that the nuclear continuum emission is suppressed by CT obscuring materials such that the diffuse extended emission can be revealed. {\it Chandra}'s combined sub-arcsecond resolution and sensitivity are essential for this study, as we can spatially resolve the extended component on $< 100$ pc scales, quantitatively measure the extent and fraction of this component, and compare with optical, submillimeter, or radio observations to investigate the origin. These measurements from the {\it Chandra} observations will provide constraints on torus parameters, e.g., opening angle and covering factor, and test the validity of torus modeling with the {\it NuSTAR} spectra.

In this paper, we present the results from our pilot {\it Chandra} ACIS survey of 7 nearby CT AGN selected from the {\it Swift}-BAT spectroscopic AGN survey \citep{Koss2017}. We describe the sample selection, {\it Chandra} observations, and data reduction in Section \ref{sec:obs}. Section \ref{sec:methods} introduces the spatial analysis methods we used in this work. We discuss our results in Section \ref{sec:results} and compare with other extended hard X-ray detected CT AGN in the literature in Section \ref{sec:discussion}. We also discuss implications for torus modeling and AGN feedback. We summarize the conclusions and implications for future observations in Section \ref{sec:conclusions}.

Throughout this paper, we adopt a concordance $\Lambda$CDM cosmological model with $H_0$ = 70 km s$^{-1}$Mpc$^{-1}$, $\Omega$$_{\Lambda}$ = 0.7, and $\Omega$$_{\rm m}$ = 0.3.

\section{Sample selection, observations, and data reduction}
\label{sec:obs}

The seven CT AGN are the targets of our pilot joint Cycle 20 {\it Chandra} and {\it NuSTAR} program (P.I. M. Elvis). They are drawn from the {\it Swift}-BAT spectroscopic AGN survey 70-month catalog \citep{Koss2017}, based on the following criteria: (1) we selected those with $z < 0.013$, $D$ $<$ 50 Mpc, which gives a plate scale of 1$\arcsec$ $\sim$ 250 pc, such that the extended emission can be spatially resolved with sub-arcsecond {\it Chandra} ACIS-S imaging. (2) we used a 2-10 keV BASS flux cut of $>$ 4 $\times$ 10$^{-13}$ erg cm$^{-2}$ s$^{-1}$ to ensure an adequate count rate. (3) we restricted the source list to those with X-ray spectra that indicate log $N_{\rm H}$ $>$ 10$^{23.9}$ cm$^{-2}$ \citep{Ricci2017}.

Table \ref{table1} summarizes the {\it Chandra} observations used in this work. We first reprocessed the data using CIAO\footnote{CIAO; http://cxc.harvard.edu/ciao/} (v4.12) and CALDB\footnote{CALDB; http://cxc.harvard.edu/caldb/} (v4.9.0) \citep{Fruscione2006}, provided by the {\it Chandra} X-ray Center (CXC). In addition to the {\it Chandra} Cycle 20 observations (ObsIDs 21417-21423), NGC 424 and  2MASXJ00253292+6821442 (J0025+6821 hereafter) also have archival observations that can be merged and included in the analysis. We inspected each observation and made centroid shifts (Table \ref{table1}), with the longer observation as the astrometric reference to better align the images. We merged the observations following the CIAO merge threads\footnote{http://cxc.harvard.edu/ciao/threads/combine/}$^{,}$\footnote{http://cxc.harvard.edu/ciao/threads/merge all/}. High background flares (3 $\sigma$) were examined and the entire data set was acceptable. Pile-up is not a concern given the count rates and the (1/4) sub-array configurations of our observations. PIMMS\footnote{PIMMS v4.10; https://cxc.harvard.edu/toolkit/pimms.jsp} estimates that the pile-up fraction in ACIS-S is at most 1\% for all except NGC 424, which has an acceptable pile-up fraction of $\sim$3\%.

\section{Methods used in spatial analysis}
\label{sec:methods}

\subsection{Sub-pixel imaging and adaptive smoothing}

We used the CIAO image analysis tools installed in SAOImage DS9\footnote{ds9; http://ds9.si.edu} to investigate the X-ray morphological properties of the CT AGN. We created images in the 0.3-7.0 keV band (full band), the 0.3-3.0 keV band (soft band), the 3.0-7.0 keV band (hard band), and the 6.0-7.0 keV band (where the Fe K$\alpha$ line dominates) if possible for the following analysis. We employed the sub-pixel binning technique to push for the highest spatial resolution, which has been tested and frequently applied to imaging studies of X-ray jets and extended emission (e.g., \citealt{Wang2011a,Wang2011b,Wang2011c,Paggi2012}). A fine pixel size of 0.062$\arcsec$ (1/8 of the ACIS native pixel size) was used when producing the images. We further generated adaptively smoothed images using {\it dmimgadapt}\footnote{https://cxc.harvard.edu/ciao/gallery/smooth.html} to demonstrate morphological structures on different scales. The adaptively smoothed images were built from the 1/8 sub-pixel data using a 0.5-15 pixel scale with 5 counts under the kernel for 30 iterations. The smoothing parameters were chosen to optimize the details of the extended diffuse emission. Almost all the smoothed images reveal high surface brightness cores and large-scale, low surface brightness in the outer regions. The X-ray emission is more extended in the soft band; the hard X-ray emission is more concentrated in the center (Section \ref{sec:results}).

\subsection{Radial profiles}

To quantitatively measure the extent and amount of extended emission, we generated radial surface brightness profiles in different energy bands and different azimuthal sectors (when possible), following the procedure described in \cite{Fabbiano2017} (see also \citealt{Fabbiano2018a,Jones2020}). Concentric annuli were used to extract radial profiles out to a radius of 8$\arcsec$, reaching the background level. Off-nuclear point sources within the 8$\arcsec$ radius circle were all removed before generating the radial profiles. We started with an annular bin size of 0.5$\arcsec$ and increased the bin size at larger radii to maintain a minimum of 10 counts in each bin. To gauge the magnitude and significance of the extended emission, we compared the radial profiles to the {\it Chandra} Point Spread Functions (PSFs) for the corresponding energy bands. We modeled the PSF for each given centroid position and energy band using ChaRT\footnote{https://cxc.harvard.edu/ciao/PSFs/chart2/} and MARX 5.5.0\footnote{https://space.mit.edu/cxc/marx} following the CIAO PSF simulation thread\footnote{https://cxc.cfa.harvard.edu/ciao/threads/psf.html}. The radial profile plots show the background-subtracted (measured from large off-source areas) surface brightness distribution in units of counts per arcsecond$^2$ in each energy band. The PSF radial profiles were generated in the same energy bands and were normalized to the counts in the central 0.5$\arcsec$ radius bin. To avoid potential contamination from a nuclear component, we focus on analyzing the region outside the central 1.5$\arcsec$ radius circle. Table \ref{table2} lists the excess counts over the {\it Chandra} PSF with associated Poisson statistical errors (including the background error), and the fraction of the extended emission in the 1.5$\arcsec$-8$\arcsec$ annular region in each energy band. The extended fraction is defined as the ratio of the excess counts above the {\it Chandra} PSF in the 1.5$\arcsec$-8$\arcsec$ annular region to the background-subtracted, total counts within the 8$\arcsec$ radius circle at the given energy band. We also list in Table \ref{table2} the excess counts within the central 1.5$\arcsec$ region as some sources do not show excess emission beyond 1.5$\arcsec$. We define the total excess fraction to be the ratio of the total excess counts above the PSF (including the 0.5$\arcsec$-1.5$\arcsec$ region) to the total net counts within the 8$\arcsec$ radius circle at the given energy band. This includes all the extended emission that does not belong to the central point source. 

In cases where we were able to identify azimuthal sectors, e.g., ionization cones and cross-cones, we also produced radial profiles in each cone region out to 8$\arcsec$, following the same procedure described above. 

We discuss the morphology, extended component, and radial profiles of individual sources in the following section.

\section{Results}
\label{sec:results}

\begin{table*}
\centering
\caption{The excess counts over the {\it Chandra} PSF, extended fractions, and total excess fractions. }
\begin{tabular}{lccccc}
\hline\hline
Sourcename &  \bf{ 0.3-3.0 keV}                                        &\bf{ 0.3-3.0 keV}                          &\bf{ 0.3-3.0 keV}                      & \bf{ 0.3-3.0 keV}                                                   &\bf{ 0.3-3.0 keV}    \\
            &     excess counts 1.5$\arcsec$-8$\arcsec$         & excess counts $\leq$ 1.5$\arcsec$ & total excess counts            & extended fraction 1.5$\arcsec$-8$\arcsec$    & total excess fraction    \\
\hline
NGC 424          &  85.1 $\pm$ 11.3 (7.5$\sigma$) & 300.8 $\pm$ 22.5 (13.4$\sigma$)     & 385.9 $\pm$ 25.2 (15.3$\sigma$)     &  5.9\% $\pm$ 0.8\%              &     25.4\% $\pm$ 1.9\%       \\
NGC 1125$^a$& 84.2 $\pm$ 10.9 (7.7$\sigma$) &131.3 $\pm$ 13.2 (10.0$\sigma$)        & 215.5 $\pm$ 17.1 (12.6$\sigma$)   & 22.8\% $\pm$ 3.2\%               &58.4\% $\pm$ 5.6\%      \\
NGC 1125$^b$ &  155.8 $\pm$ 13.8 (11.3$\sigma$) &128.6 $\pm$ 13.1 (9.8$\sigma$)    & 284.4 $\pm$ 19.0 (14.9$\sigma$)   &35.3\% $\pm$ 3.6\%             &64.5\% $\pm$ 5.4\%     \\
NGC 3281       &   41.4 $\pm$ 7.1 (5.9$\sigma$) &23.1 $\pm$ 5.2 (4.4$\sigma$)              & 64.5 $\pm$ 8.8 (7.3$\sigma$)         &55.5\% $\pm$ 11.6\%           &80.5\% $\pm$ 15.0\%             \\
NGC 4500       &  33.2 $\pm$ 7.1 (4.7$\sigma$) &56.8 $\pm$ 8.1 (7.0$\sigma$)              & 90.0 $\pm$ 10.8 (8.4$\sigma$)       &26.1\% $\pm$ 6.1\%           &70.8\% $\pm$ 10.8\%         \\
ESO 005-G004 &  8.4 $\pm$ 5.2 (1.6$\sigma$) &13.3 $\pm$ 4.7 (2.9$\sigma$)               & 14.4 $\pm$ 5.9 (2.4$\sigma$)   &49.3\% $\pm$ 35.6\%          &84.8\% $\pm$ 46.1\%          \\
ESO 137-G034 & 341.1 $\pm$ 19.3 (17.7$\sigma$) &222.0 $\pm$ 15.8 (14.0$\sigma$)  & 563.2 $\pm$ 25.0 (22.6$\sigma$)  &49.4\% $\pm$ 3.3\%           &81.6\% $\pm$ 4.8\%  \\
J0025+6821     & 8.6 $\pm$ 5.4 (1.6$\sigma$) &12.4 $\pm$ 5.4 (2.3$\sigma$)                & 18.0 $\pm$ 6.9 (2.6$\sigma$) &15.9\% $\pm$ 10.4\%         &33.3\% $\pm$ 13.9\%             \\
\hline

             &     \bf{3.0-7.0 keV}   & \bf{3.0-7.0 keV}            & \bf{3.0-7.0 keV}         &     \bf{3.0-7.0 keV}   & \bf{3.0-7.0 keV}  \\
           &     excess counts 1.5$\arcsec$-8$\arcsec$         & excess counts $\leq$ 1.5$\arcsec$   &total excess counts          & extended fraction 1.5$\arcsec$-8$\arcsec$    & total excess fraction    \\
\hline 
NGC 424          &      $<$ 15.9                          &88.5 $\pm$ 14.3 (6.2$\sigma$)          &65.9 $\pm$ 15.3 (4.3$\sigma$)        & $<$ 2.9\%                     & 12.1\% $\pm$ 2.9\% \\
NGC 1125$^a$   &  $<$ 18.3                                  &32.7 $\pm$ 9.7 (3.4$\sigma$)      &38.2 $\pm$ 11.5 (3.3$\sigma$)                & $<$ 7.7\%                     &  16.1\% $\pm$ 5.0\%    \\
NGC 1125$^b$   &  23.5 $\pm$ 7.5 (3.1$\sigma$) &23.8 $\pm$ 9.4 (2.5$\sigma$)      &47.3 $\pm$ 12.0 (3.9$\sigma$)             & 9.2\% $\pm$ 3.0\%     &18.5\% $\pm$ 4.9\%     \\
NGC 3281       &  12.0 $\pm$ 6.2 (1.9$\sigma$)    &36.1 $\pm$ 11.0 (3.3$\sigma$)     & 48.1 $\pm$ 12.6 (3.8$\sigma$)         &  3.4\% $\pm$ 1.8\%     &13.5\% $\pm$ 3.6\%  \\
NGC 4500       &   $<$ 12.2                                     &22.4 $\pm$ 7.7 (2.9$\sigma$)      & 21.5 $\pm$ 8.4 (2.6$\sigma$)         &  $<$ 8.3\%                  & 14.6\% $\pm$ 5.9\%   \\
ESO 005-G004 & 7.3 $\pm$ 4.9 (1.5$\sigma$)       &22.1 $\pm$ 7.9 (2.8$\sigma$)      & 29.5 $\pm$ 9.3 (3.2$\sigma$)       &  4.3\% $\pm$ 2.8\%      &17.1\% $\pm$ 5.5\%  \\
ESO 137-G034 &  33.1 $\pm$ 8.4 (3.9$\sigma$)    &62.1 $\pm$ 12.4 (5.0$\sigma$)    & 95.2 $\pm$ 15.0 (6.4$\sigma$)      & 7.7\% $\pm$ 2.0\%       & 22.2\% $\pm$ 3.7\%  \\
J0025+6821     & 5.4 $\pm$ 5.3 (1.0$\sigma$)        &11.8 $\pm$ 8.2 (1.4$\sigma$)      & 17.1 $\pm$ 9.8 (1.7$\sigma$)      &    2.4\% $\pm$ 2.3\%   & 7.5\% $\pm$ 4.3\%    \\
\hline

             &     \bf{6.0-7.0 keV}     & \bf{6.0-7.0 keV}              & \bf{6.0-7.0 keV}                                                                         &     \bf{6.0-7.0 keV}     & \bf{6.0-7.0 keV}     \\
            &     excess counts 1.5$\arcsec$-8$\arcsec$         & excess counts $\leq$ 1.5$\arcsec$  &total excess counts           & extended fraction 1.5$\arcsec$-8$\arcsec$    & total excess fraction    \\
\hline 
ESO 137-G034 &   8.7 $\pm$ 4.7 (1.9$\sigma$)             &25.3 $\pm$ 8.0 (3.2$\sigma$)   &   34.0 $\pm$ 9.3 (3.7$\sigma$)                &   5.1\% $\pm$ 2.8\%                                        &20.1\% $\pm$ 5.7\% \\
\hline
            &    \bf{0.3-7.0 keV}     &  \bf{0.3-7.0 keV}                 &  \bf{0.3-7.0 keV}                                                                                     &     \bf{0.3-7.0 keV}     &  \bf{0.3-7.0 keV}   \\
           &     excess counts 1.5$\arcsec$-8$\arcsec$         & excess counts $\leq$ 1.5$\arcsec$  &total excess counts           & extended fraction 1.5$\arcsec$-8$\arcsec$    & total excess fraction    \\
\hline
NGC 424          &  92.6 $\pm$ 12.5 (7.4$\sigma$)  & 423.4 $\pm$ 26.7 (15.9$\sigma$) &516.0 $\pm$ 27.7 (18.6$\sigma$) & 4.7\% $\pm$ 0.6\%    & 24.6\% $\pm$ 1.6\% \\
NGC 1125$^a$& 94.8 $\pm$ 12.5 (7.6$\sigma$)    & 171.9 $\pm$ 16.3 (10.5$\sigma$)& 266.7 $\pm$ 20.6 (13.0$\sigma$)  & 15.6\% $\pm$ 2.2\%   & 44.0\% $\pm$ 3.9\%\\
NGC 1125$^b$ & 184.7 $\pm$ 15.7 (11.7$\sigma$) &160.6 $\pm$ 16.1 (10.0$\sigma$)& 345.3 $\pm$ 22.5 (15.3$\sigma$) & 26.5\% $\pm$ 2.5\% & 49.6\% $\pm$ 3.8\% \\
NGC 3281       &  59.5 $\pm$ 9.4 (6.3$\sigma$)      & 69.9 $\pm$ 12.0 (5.8$\sigma$)    & 129.4 $\pm$ 15.2 (8.5$\sigma$) &13.8\% $\pm$ 2.3\%   & 30.0\% $\pm$ 3.8\%\\
NGC 4500       & 37.7 $\pm$ 8.2 (4.6$\sigma$)    & 83.9 $\pm$ 10.9 (7.7$\sigma$)      & 121.6 $\pm$ 13.7 (8.9$\sigma$)&13.8\% $\pm$ 3.1\%   & 44.4\% $\pm$ 5.7\%\\
ESO 005-G004 & 16.8 $\pm$ 7.1 (2.3$\sigma$)     &31.3 $\pm$ 8.3 (3.8$\sigma$)      & 48.1 $\pm$ 11.0 (4.4$\sigma$) &8.9\% $\pm$ 3.8\%   & 25.4\% $\pm$ 6.1\%\\
ESO 137-G034 & 384.2 $\pm$ 21.1 (18.1$\sigma$)  & 302.9 $\pm$ 20.0 (15.1$\sigma$)& 687.1 $\pm$ 29.1 (23.6$\sigma$)&34.3\% $\pm$ 2.2\%&61.4\% $\pm$ 3.2\% \\
J0025+6821     &15.2 $\pm$ 7.6 (2.0$\sigma$)      & 21.8 $\pm$ 9.3 (2.4$\sigma$)     & 37.1 $\pm$ 12.0 (3.1$\sigma$)  &5.4\% $\pm$ 2.7\%  & 13.1\% $\pm$ 4.3\%\\
\hline

\end{tabular}
\label{table2}
\tablecomments{Extended fraction $\equiv$ (Source 1.5$\arcsec$-8$\arcsec$ counts $-$ PSF 1.5$\arcsec$-8$\arcsec$ counts) / (Total 8$\arcsec$ source counts). Total excess fraction $\equiv$ (Source 0.5$\arcsec$-8$\arcsec$ counts $-$ PSF 0.5$\arcsec$-8$\arcsec$ counts) / (Total 8$\arcsec$ source counts). We place a 3$\sigma$ upper limit if there are no excess counts above the PSF (i.e. $<$ 1$\sigma$). (a). The two nearby X-ray sources are removed from the measurements. (b). The two nearby X-ray sources are included in the measurements. }
\end{table*}

\begin{table*}
\centering
\caption{ESO 137-G034: The excess counts over the {\it Chandra} PSF, extended fractions, and total excess fractions in the NW-SE ionization and NE-SW cross-cones.}
\begin{tabular}{lcccc}
\hline\hline
Energy &  NW-SE \bf{ionization cone} & NW-SE \bf{ionization cone} &   NW-SE \bf{ionization cone} & NW-SE  \bf{ionization cone} \\
(keV)            &    excess counts 1.5$\arcsec$-8$\arcsec$ &excess counts $\leq$1.5$\arcsec$ & extended fraction 1.5$\arcsec$-8$\arcsec$& total excess fraction \\
\hline 
0.3-3.0         & 271.6 $\pm$ 16.9 (16.1$\sigma$)   & 149.7 $\pm$ 12.8 (11.7$\sigma$)   &39.4\% $\pm$ 2.9\%$^a$, 55.0\% $\pm$ 4.2\%$^b$  &61.1\% $\pm$ 3.9\%$^a$, 85.3\% $\pm$ 5.8\%$^b$     \\
\bf{3.0-7.0}  & 23.9 $\pm$ 6.5 (3.7$\sigma$)          & 42.7 $\pm$ 9.4 (4.6$\sigma$)         &5.6\% $\pm$ 1.5\%, 10.2\% $\pm$ 2.9\%                  &15.6\% $\pm$ 2.8\%, 28.4\% $\pm$ 5.2\%                         \\
\bf{6.0-7.0}  & 5.9 $\pm$ 3.6 (1.6$\sigma$)            & 20.4 $\pm$ 6.3 (3.2$\sigma$)         &3.5\% $\pm$ 2.2\%, 5.9\% $\pm$ 3.7\%                   & 15.5\% $\pm$ 4.4\%, 26.2\% $\pm$ 7.7\%                        \\
0.3-7.0        & 299.8 $\pm$ 18.1 (16.5$\sigma$)    & 205.5 $\pm$ 15.9 (12.9$\sigma$)    &26.8\% $\pm$ 1.8\%, 41.0\% $\pm$ 2.9\%              & 45.2\% $\pm$ 2.6\%, 69.0\% $\pm$ 4.2\%                    \\
\hline
Energy &  NE-SW \bf{cross-cone}  & NE-SW \bf{cross-cone} &   NE-SW \bf{cross-cone}  & NE-SW \bf{cross-cone} \\
(keV)            &     excess counts 1.5$\arcsec$-8$\arcsec$ & excess counts $\leq$1.5$\arcsec$  &extended fraction 1.5$\arcsec$-8$\arcsec$& total excess fraction \\
\hline 
0.3-3.0      &     65.9 $\pm$ 8.9 (7.4$\sigma$)        &   60.0 $\pm$ 8.9 (6.7$\sigma$)           & 9.6\% $\pm$ 1.3\%$^c$, 33.8\% $\pm$ 5.2\%$^d$  & 18.3\% $\pm$ 2.0\%$^c$, 64.5\% $\pm$ 8.0\%$^d$   \\
\bf{3.0-7.0}      & 7.8 $\pm$ 5.0 (1.5$\sigma$)       &  $<$ 21.8                                              & 1.8\% $\pm$ 1.2\%, 4.0\% $\pm$ 2.7\%                  & $<$ 6.2\%, $<$ 13.8\%       \\
\bf{6.0-7.0}      & $<$ 8.6                                        &  $<$ 13.1                                             & $<$ 5.0\%, $<$12.4\%                                              & $<$ 9.3\%, $<$ 22.9\%  \\
0.3-7.0       &   78.5 $\pm$ 10.3 (7.6$\sigma$)      & 62.4 $\pm$ 11.4 (5.5$\sigma$)            & 7.0\% $\pm$ 0.9\%, 20.4\% $\pm$ 2.9\%                 & 12.6\% $\pm$ 1.4\%, 36.5\% $\pm$ 4.4\%     \\
\hline

\end{tabular}
\label{table3}
\tablecomments{We place a 3$\sigma$ upper limit if there are no excess counts above the PSF (i.e. $<$ 1$\sigma$). (a). The first fraction is the excess counts in the extended region of the NW-SE ionization bicone relative to the total counts within the 8$\arcsec$ radius circle at the given energy band. (b). The second fraction is the excess counts in the extended region of the NW-SE ionization bicone relative to the total counts in the bicone itself at the given energy band. (c). The first fraction is the excess counts in the extended region of the NE-SW cross-cones relative to the total counts within the 8$\arcsec$ radius circle at the given energy band. (d). The second fraction is the excess counts in the extended region of the NE-SW cross-cones relative to the total counts in the cross-cones at the given energy band.}
\end{table*}

\begin{table*}
\centering
\caption{NGC 3281: The excess counts over the {\it Chandra} PSF, extended fractions, and total excess fractions in the N-S ionization and E-W cross-cones.}
\begin{tabular}{lcccc}
\hline\hline
Energy &  N-S  \bf{ionization cone}  & N-S  \bf{ionization cone} &   N-S  \bf{ionization cone}  & N-S  \bf{ionization cone}    \\
(keV)            &   excess counts 1.5$\arcsec$-8$\arcsec$&excess counts $\leq$1.5$\arcsec$&  extended fraction 1.5$\arcsec$-8$\arcsec$& total excess fraction   \\
\hline 
0.3-3.0          &   31.2 $\pm$ 5.8 (5.3$\sigma$)   & 13.7 $\pm$ 3.9 (3.5$\sigma$)    &  41.7\% $\pm$ 9.3\%$^a$, 62.0\% $\pm$ 14.7\%$^b$  &60.0\% $\pm$ 11.9\%$^a$, 89.2\% $\pm$ 19.0\%$^b$       \\
\bf{3.0-7.0}   & 16.4 $\pm$ 5.2 (3.2$\sigma$)     & 37.5 $\pm$ 8.4 (4.5$\sigma$)     &4.6\% $\pm$ 1.5\%, 9.7\% $\pm$ 3.2\%                          &15.2\% $\pm$ 2.9\%, 32.0\% $\pm$ 6.4\%                              \\
0.3-7.0       &   50.1 $\pm$ 7.8 (6.4$\sigma$)      & 57.7 $\pm$ 9.3 (6.2$\sigma$)      &11.6\% $\pm$ 1.9\%, 22.8\% $\pm$ 3.9\%                     &25.0\% $\pm$ 3.1\%, 49.2\% $\pm$ 6.5\%                       \\
\hline
Energy &  E-W \bf{cross-cone} & E-W \bf{cross-cone}&   E-W \bf{cross-cone} & E-W \bf{cross-cone}  \\
(keV)            &   excess counts  1.5$\arcsec$-8$\arcsec$& excess counts  $\leq$1.5$\arcsec$  & extended fraction 1.5$\arcsec$-8$\arcsec$ &total excess fraction    \\
\hline 
0.3-3.0          &   10.3 $\pm$ 3.8 (2.7$\sigma$)    & 5.0 $\pm$ 2.6 (1.9$\sigma$)   & 13.8\% $\pm$ 5.3\%$^c$, 42.1\% $\pm$ 18.0\%$^d$    & 20.4\% $\pm$ 6.7\%$^c$, 62.4\% $\pm$ 23.3\%$^d$    \\
\bf{3.0-7.0}   & $<$ 9.6                                         & $<$ 21.8                                    &  $<$ 2.7\%, $<$ 5.1\%                                                &  $<$ 6.7\%, $<$ 12.7\%     \\
0.3-7.0       &  9.7 $\pm$ 5.0 (2.0$\sigma$)        & 16.0 $\pm$ 7.7 (2.1$\sigma$)   & 2.3\% $\pm$ 1.2\%, 4.6\% $\pm$ 2.4\%                        & 6.0\% $\pm$ 2.2\%, 12.1\% $\pm$ 4.4\%      \\
\hline
\end{tabular}
\label{table4}
\tablecomments{We place a 3$\sigma$ upper limit if there are no excess counts above the PSF (i.e. $<$ 1$\sigma$). (a). The first fraction is the excess counts in the extended region of the N-S ionization bicone relative to the total counts within the 8$\arcsec$ radius circle at the given energy band. (b). The second fraction is the excess counts in the extended region of the N-S ionization bicone relative to the total counts in the bicone itself at the given energy band. (c). The first fraction is the excess counts in the extended region of the E-W cross-cones relative to the total counts within the 8$\arcsec$ radius circle at the given energy band. (d). The second fraction is the excess counts in the extended region of the E-W cross-cones relative to the total counts in the cross-cones at the given energy band.}
\end{table*}

\begin{table*}
\centering
\caption{Literature comparison sample of extended hard X-ray emission: the excess counts over the {\it Chandra} PSF, extended fractions, and total excess fractions.}
\scriptsize
\begin{tabular}{lcccccccc}
\hline\hline
Sourcename &  $z$  & \bf{3.0-7.0 keV}                                                                  &  \bf{3.0-7.0 keV}             &  \bf{3.0-7.0 keV}                                                                  &  \bf{3.0-7.0 keV}          & \bf{3.0-7.0 keV}        \\
            &                   &excess counts 1.5$\arcsec$-8$\arcsec$& excess counts $\leq$1.5$\arcsec$ &total excess counts   &extended fraction 1.5$\arcsec$-8$\arcsec$ & total excess fraction\\
\hline 
ESO 428-G014      &  0.00566&    318.4 $\pm$ 19.9 (16.0$\sigma$)  & 144.6 $\pm$ 24.3 (6.0$\sigma$)&  462.9 $\pm$ 31.4 (14.8$\sigma$)  &   20.8\% $\pm$ 1.4\% &30.2\% $\pm$ 2.2\%       \\
NGC 7212             & 0.0266   &    226.0 $\pm$ 15.0 (15.1$\sigma$)  &  323.5 $\pm$ 18.0 (18.0$\sigma$)& 549.5 $\pm$ 23.4 (23.5$\sigma$)  &14.7\% $\pm$ 1.0\% &  35.7\% $\pm$ 1.8\%   \\
\hline
 &  &  \bf{6.0-7.0 keV} & \bf{6.0-7.0 keV}   &\bf{6.0-7.0 keV} &\bf{6.0-7.0 keV} & \bf{6.0-7.0 keV}\\
            &                   & excess counts 1.5$\arcsec$-8$\arcsec$ & excess counts $\leq$1.5$\arcsec$  & total excess counts  &extended fraction 1.5$\arcsec$-8$\arcsec$& total excess fraction \\
\hline 
ESO 428-G014      &  & 52.4 $\pm$ 9.1 (5.7$\sigma$) & 20.5 $\pm$ 13.3 (1.5$\sigma$) & 72.8 $\pm$ 16.2 (4.5$\sigma$)  &11.8\% $\pm$ 2.1\%   & 16.4\% $\pm$ 3.7\% \\
NGC 7212             &   & 27.3 $\pm$ 5.2 (5.3$\sigma$)  & 68.4 $\pm$ 8.3 (8.2$\sigma$) &95.7 $\pm$ 9.8 (9.8$\sigma$)  &7.3\% $\pm$ 1.4\%     &25.5\% $\pm$ 2.9\%   \\
\hline
\end{tabular}
\label{table5}
\end{table*}

\subsection{ESO 137-G034}

\begin{figure} 
\centering
\includegraphics[width=7cm]{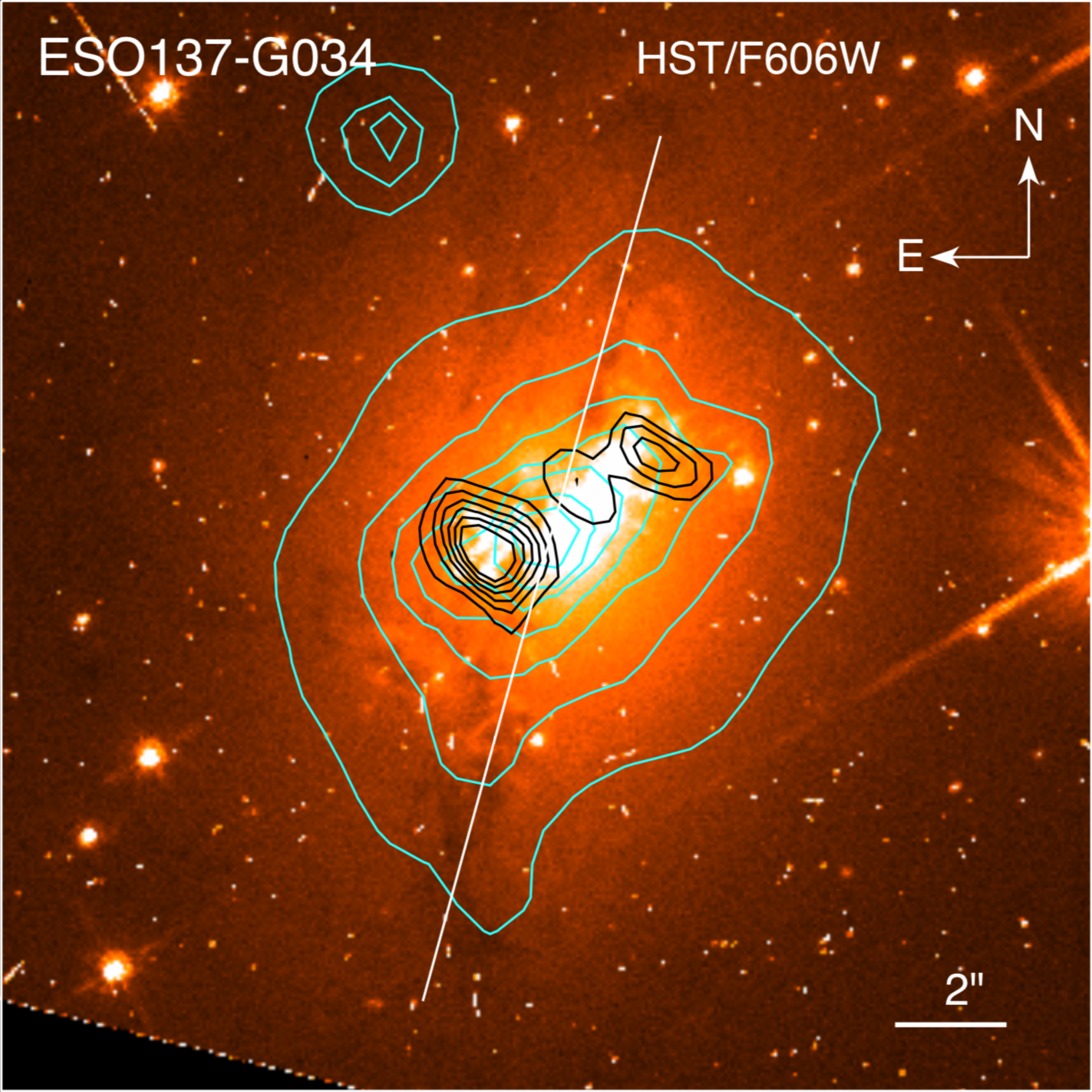}
\caption{20$\arcsec$ $\times$ 20$\arcsec$ {\it HST}/F606W image of ESO 137-G034 \citep{Malkan1998}, an S0/a galaxy hosting a Seyfert 2 nucleus. The {\it Chandra} 0.3-7.0 keV X-ray contours are overlaid in cyan, extending to the host galaxy out to 8$\arcsec$ in radius. The black contours are extracted from the ATCA radio map at 8.6 GHz in \cite{Morganti1999}. There is a long dust lane passing close to the nucleus, parallel to the photometric major axis (P.A. $\sim$165$^{\circ}$; white line) of the galaxy.   }
\label{HST_ESO137-G034}
\end{figure}

ESO 137-G034 is an S0/a galaxy (Figure \ref{HST_ESO137-G034}; \citealt{Malkan1998,Ferruit2000}) hosting a Seyfert 2 nucleus at $z$ = 0.009 ($D$ $\sim$ 38.8 Mpc; 1" $\sim$ 185 pc) with a CT column density of log $N_{\rm H}$ = 24.3-24.5 cm$^{-2}$ \citep{Ricci2015,Ricci2017,Georgantopoulos2019}. There is a long dust lane passing close to the nucleus, parallel to the photometric major axis of the galaxy.

\subsubsection{Extended soft X-ray and optical line emission}
\begin{figure*} 
\centering
\includegraphics[width=15cm]{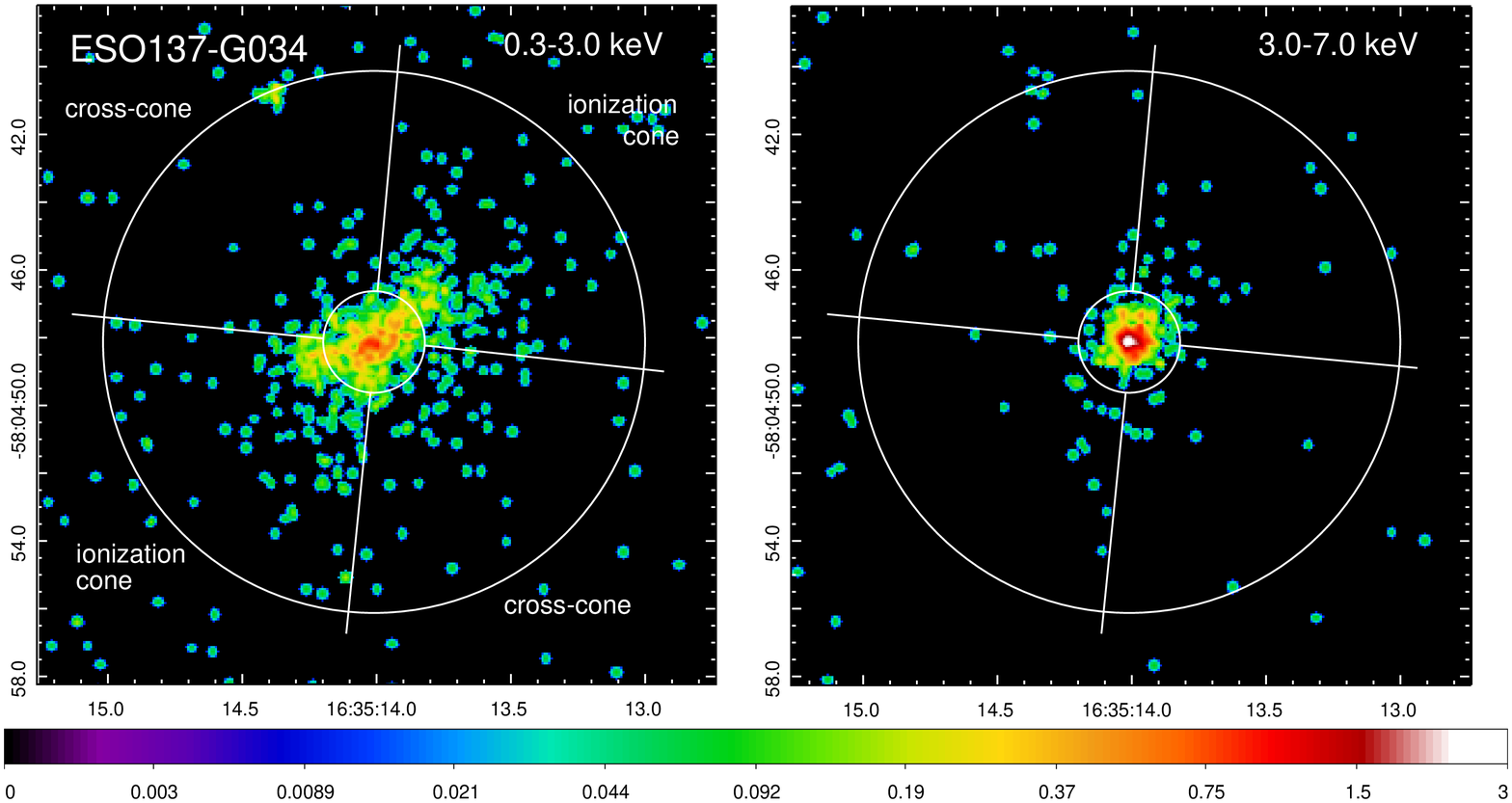}
\includegraphics[width=15cm]{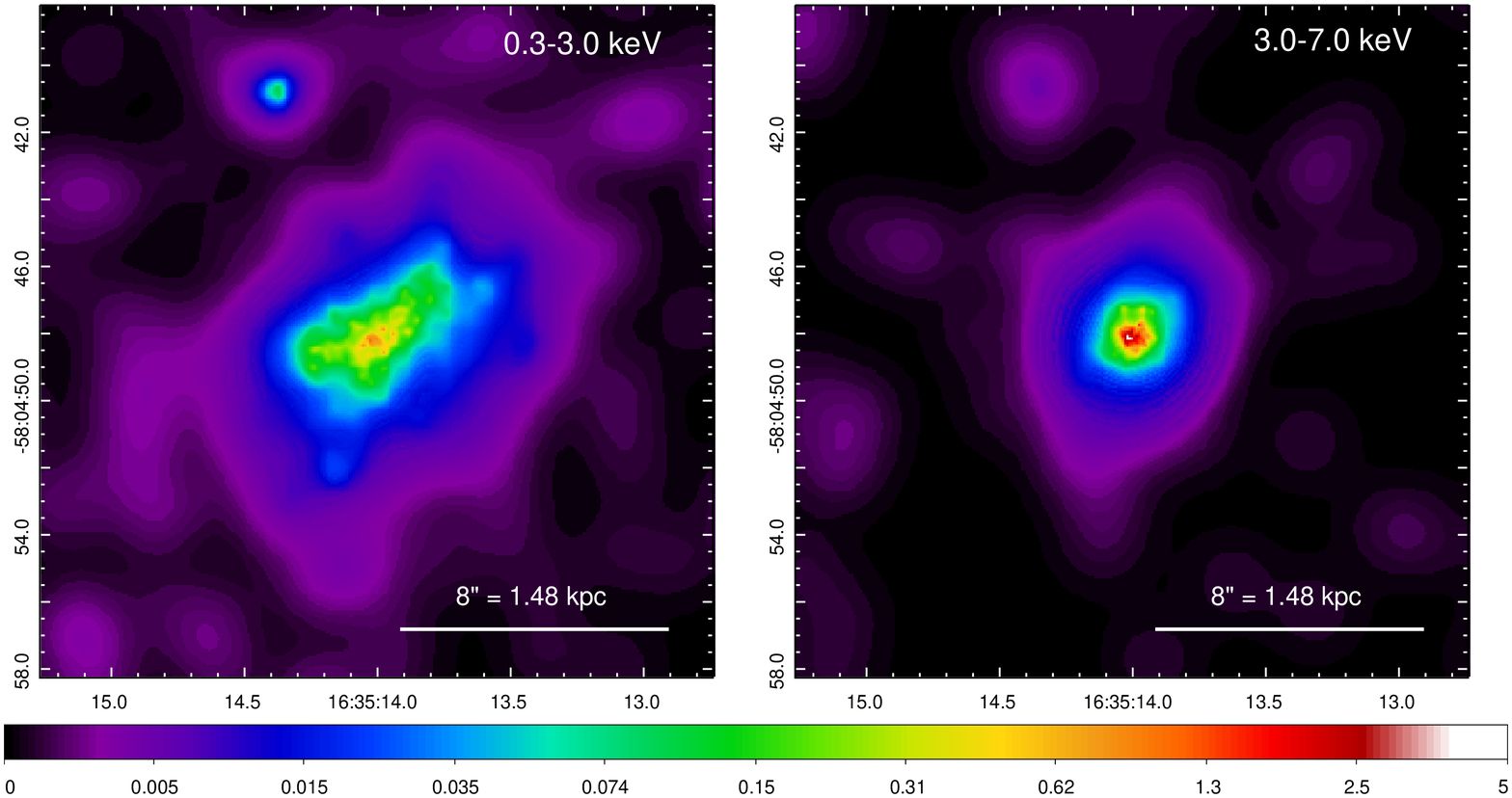}
\caption{{\bf Top:} 20$\arcsec$ $\times$ 20$\arcsec$ {\it Chandra} ACIS-S 0.3-3.0 keV (left) and 3.0-7.0 keV (right) band images of ESO 137-G034 at 1/8 subpixel binning. Based on the azimuthal dependence of the soft X-ray emission, the 8$\arcsec$ radius circular region is divided into the NW-SE ionization cones with an opening angle of 90$^\circ$ and the NE-SW cross-cones. The ionization bicone axis is orientated $\sim$135$^{\circ}$ measured from North through East. The inner 1.5$\arcsec$ radius circle and the outer 8$\arcsec$ circle define the region in between for extracting excess counts in the extended emission. {\bf Bottom:} Adaptively smoothed images ({\it dmimgadapt}; 0.5-15 pixel scales, 5 counts under kernel, 30 iterations) on 1/8 binned pixel. All the images are displayed in logarithmic scale with colors corresponding to the counts per image pixel. }
\label{ESO137-G034_chandra}
\end{figure*}

Figure \ref{ESO137-G034_chandra} (top) shows the {\it Chandra} 0.3-3 keV soft band and 3-7 keV hard band images of ESO 137-G034 at 1/8 subpixel binning, and the adaptively smoothed versions are shown at the bottom row. ESO 137-G034 has the most prominent extended X-ray emission among the 7 CT AGN. The soft band emission is clearly seen as extended and elongated along the northwest-southeast (NW-SE) direction, while less extended along the northeast-southwest (NE-SW) direction. Based on the azimuthal dependence, we divided the data into two biconical regions opening $\sim$ 90$^\circ$ outward from the central nucleus in the NW-SE direction (i.e., ionization cones), and in the NE-SW direction (i.e., cross-cones). The ionization bicone axis is orientated $\sim$130$^{\circ}$ measured from North through East. We generated radial profiles over all azimuthal angles (Figure \ref{ESO137-G034_radial_profiles}) and in the ionization cones and cross-cones defined above (Figure \ref{ESO137-G034_cone_profiles}). To assess the significance of the extended emission, we measured the excess counts over the PSF in all the cone regions and calculated the extended fractions both in the bicone regions and in the 8$\arcsec$ radius circular region (Table \ref{table2}). Figure \ref{ESO137-G034_radial_profiles} and Figure \ref{ESO137-G034_cone_profiles} show that the soft X-ray emission has excess counts above the PSF at all radii in all the cone and cross-cone regions. About half of the total, background-subtracted counts at 0.3-3.0 keV are in the extended component (i.e., 1.5$\arcsec$-8$\arcsec$ annular region), with $\sim$40\% in the ionization cones and $\sim$10\% in the cross-cones. The diffuse soft X-ray component extends to at least 8$\arcsec$ ($\sim$1.5 kpc) in radius in the ionization cone direction.

Previous {\it HST} optical [O\III]$\lambda$5007 and $H\alpha$ line maps show a similar morphology to the soft X-ray emission in the inner 4$\arcsec$ radius region, although more detailed structures are revealed in the optical line maps \citep{Ferruit2000}. Outside the 4$\arcsec$ radius region, instead of further extending along the NW-SE direction, the optical line emission displays extended tails at the extreme north and south ends of the emission line region, forming a Z-shaped morphology (see Figure 21 in \citealt{Ferruit2000}). Our {\it Chandra} data also show this pattern in the soft X-ray emission of further extended tails to the north and south out to $\sim$8$\arcsec$ measured from the center. This would not be surprising as previous studies already show a strong morphological correlation between the extended soft X-ray associated with AGN and extended [O\III] emission, likely due to a single photoionized medium giving rise to both the [O\III] and soft X-rays (e.g., \citealt{Bianchi2006,Levenson2006,Koss2015}). More recent studies suggest that the ionized medium is more complex, and this correlation is due to some combination of photoionization and collisional excitation in a multiphase medium (e.g., \citealt{Paggi2012,Maksym2017,Maksym2019}). \cite{Ferruit2000} reported an integrated [O\III]$\lambda$5007 line flux of 6.5 $\times$ 10$^{-13}$ erg cm$^{-2}$ s$^{-1}$ in a 14$\arcsec$$\times$22$\arcsec$ extraction box around the entire line emitting region. We measured the integrated flux of the soft X-ray emission (using {\it srcflux} in CIAO) in the same extraction box to be $f_{\rm 0.5-2 keV}$ = (1.4 $\pm$ 0.1) $\times$ 10$^{-13}$ erg cm$^{-2}$ s$^{-1}$. The [O\III]/soft X-ray flux ratio of the integrated emission is $\sim$4.8, which is very similar to the ratios of other Seyfert CT AGN that also have {\it Chandra} and {\it HST} observations, e.g., NGC 3393, Mrk 3 \citep{Bianchi2006}. Our {\it Chandra} data are not deep enough though to allow us to examine the variations of the [O\III]/soft X-ray ratio cloud by cloud or radially as a function of distance from the nucleus as in e.g., NGC 4151 \citep{Wang2011a,Wang2011b,Wang2011c} and Mrk 573 \citep{Paggi2012}. 

The Z-shaped structure of high-excitation gas surrounds a three-component radio structure as observed with the Australia Telescope Compact Array (ATCA) at 8.6 GHz (Figure \ref{HST_ESO137-G034}; \citealt{Morganti1999}), coincident with the inner part of the line emitting gas. Deeper {\it Chandra} data is necessary to reveal whether there are tails and enable a comparison of the detailed structures of the X-rays, optical line emission and radio emission.

\subsubsection{Extended hard X-ray emission}

ESO 137-G034 displays the most prominent extended hard X-ray emission among the 7 CT AGN, with an excess of 33.1 $\pm$ 8.4 counts over the PSF at 3-7 keV and an extended fraction of 7.7\% in the 1.5$\arcsec$-8$\arcsec$ region (Table \ref{table2}). Almost all of the excess counts come from the ionization bicone with a 3.7$\sigma$ detection and a 5.6\% extended fraction (Table \ref{table3}). The inner 1.5$\arcsec$ region also has significant excess emission detected at $\sim$5$\sigma$, contributing to a total of 95.2 $\pm$ 15.0 excess counts above the PSF and a total excess fraction of 22.2\%, the highest fraction among the sample. The total extent of the diffuse hard X-ray emission is about 10$\arcsec$ in diameter, or 1.85 kpc, which is roughly half the size of the extended soft X-ray emission. 

\begin{figure*} 
\centering
\includegraphics[width=7.5cm]{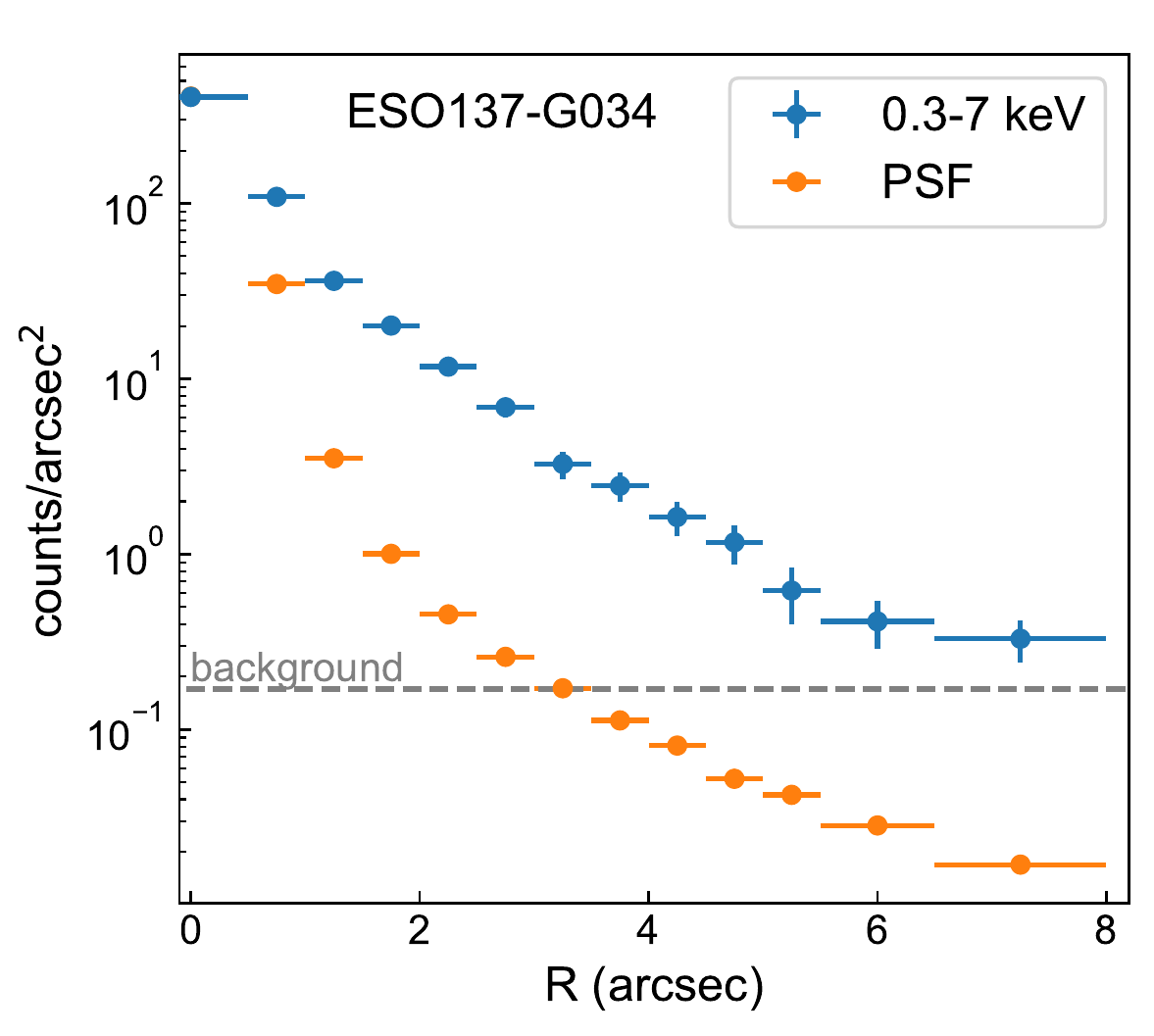} 
\includegraphics[width=7.5cm]{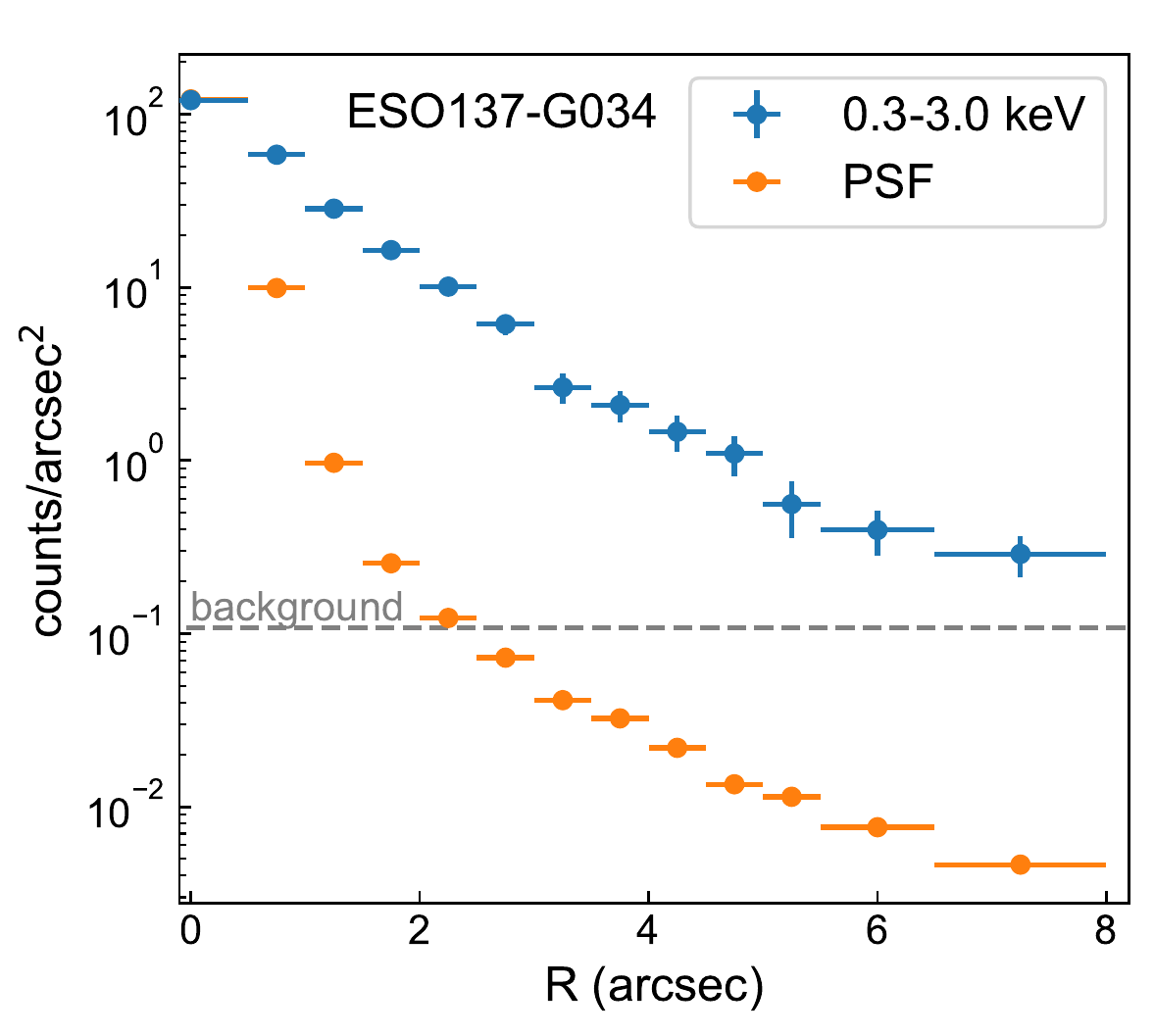} 
\includegraphics[width=7.5cm]{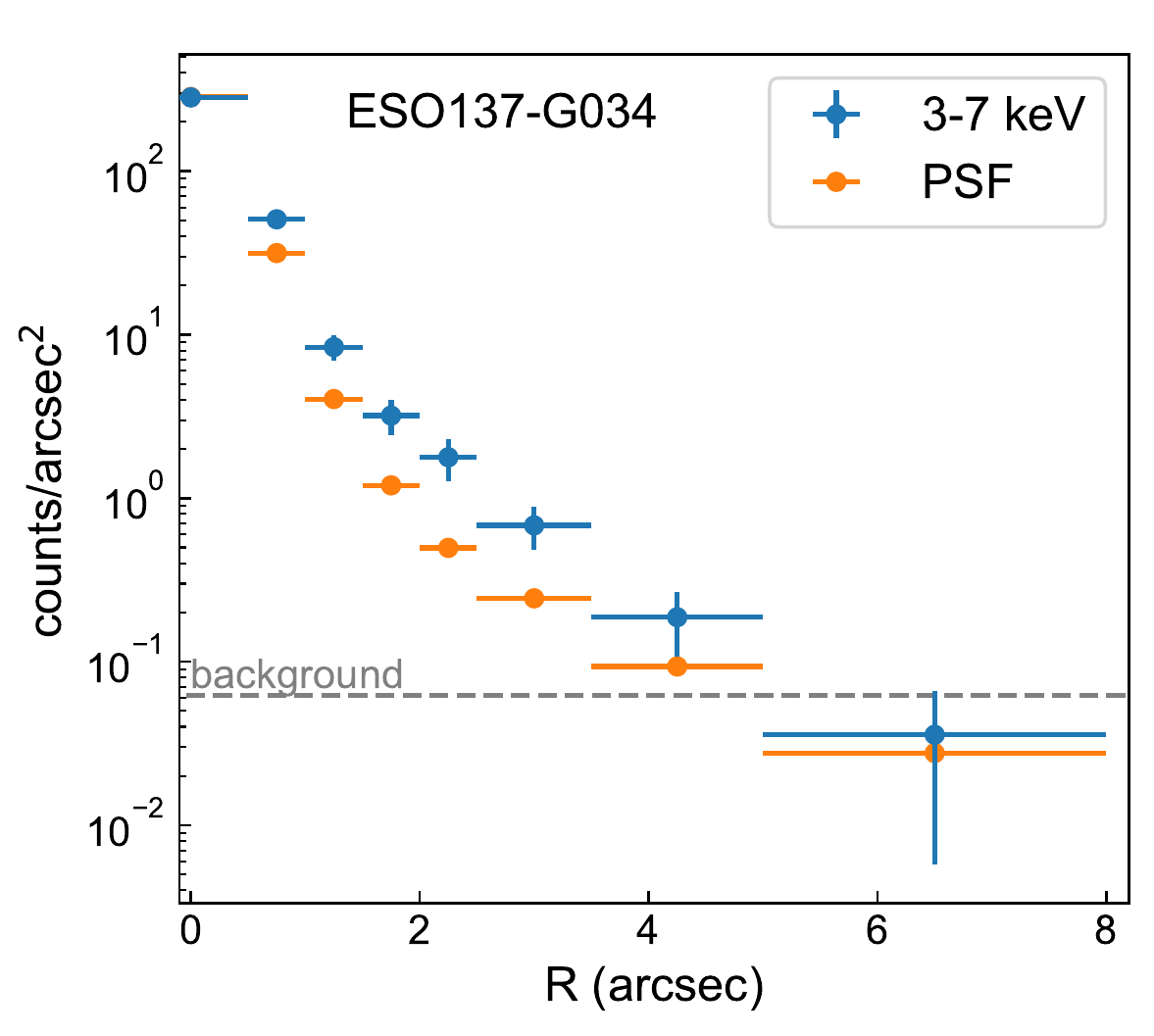}
\includegraphics[width=7.5cm]{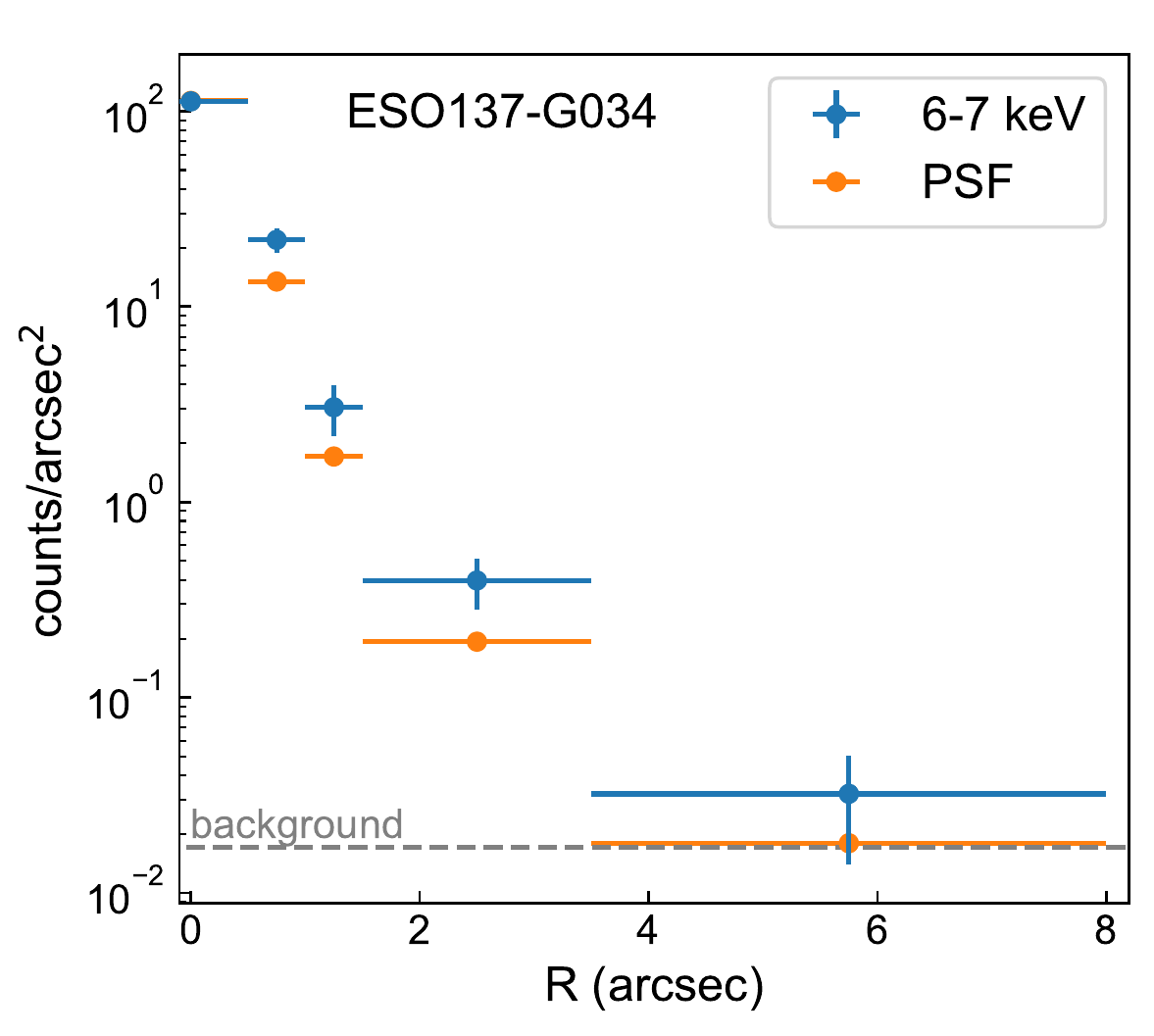}
\caption{Radial profiles of ESO 137-G034 for the full energy band (0.3-7.0 keV), soft band (0.3-3.0 keV), hard band (3.0-7.0 keV), and the 6-7 keV band where Fe K$\alpha$ line dominates. The background has been subtracted off from the radial profiles, and the level of which is indicated as the grey dashed horizontal line. The PSF is normalized to the counts in the central 0.5$\arcsec$ radius bin. We find excess emission above the PSF outside the nuclear region (0.5$\arcsec$, 93 pc) in all the energy bands.   }
\label{ESO137-G034_radial_profiles}
\end{figure*}

\begin{figure*} 
\centering
\includegraphics[width=7.5cm]{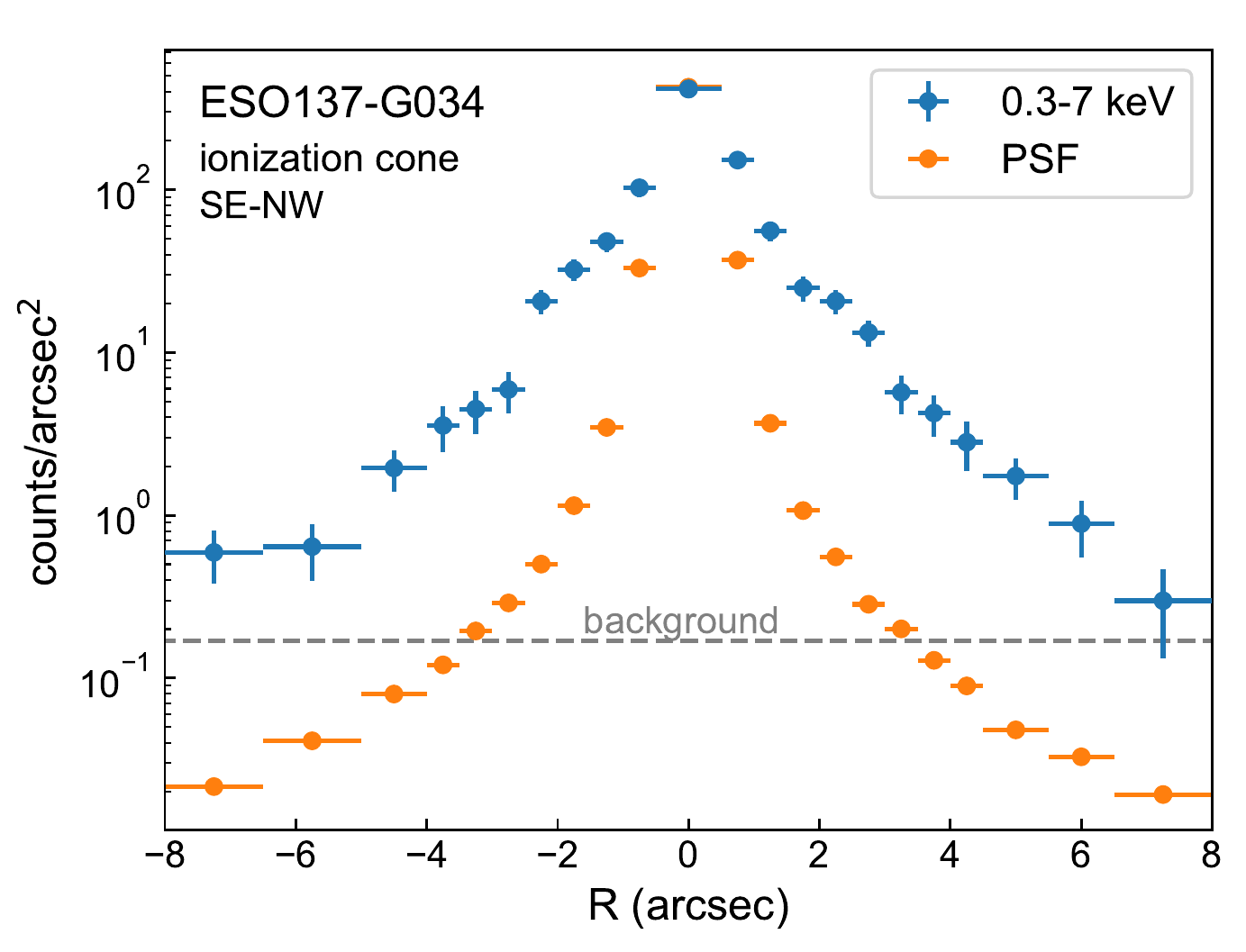}
\includegraphics[width=7.5cm]{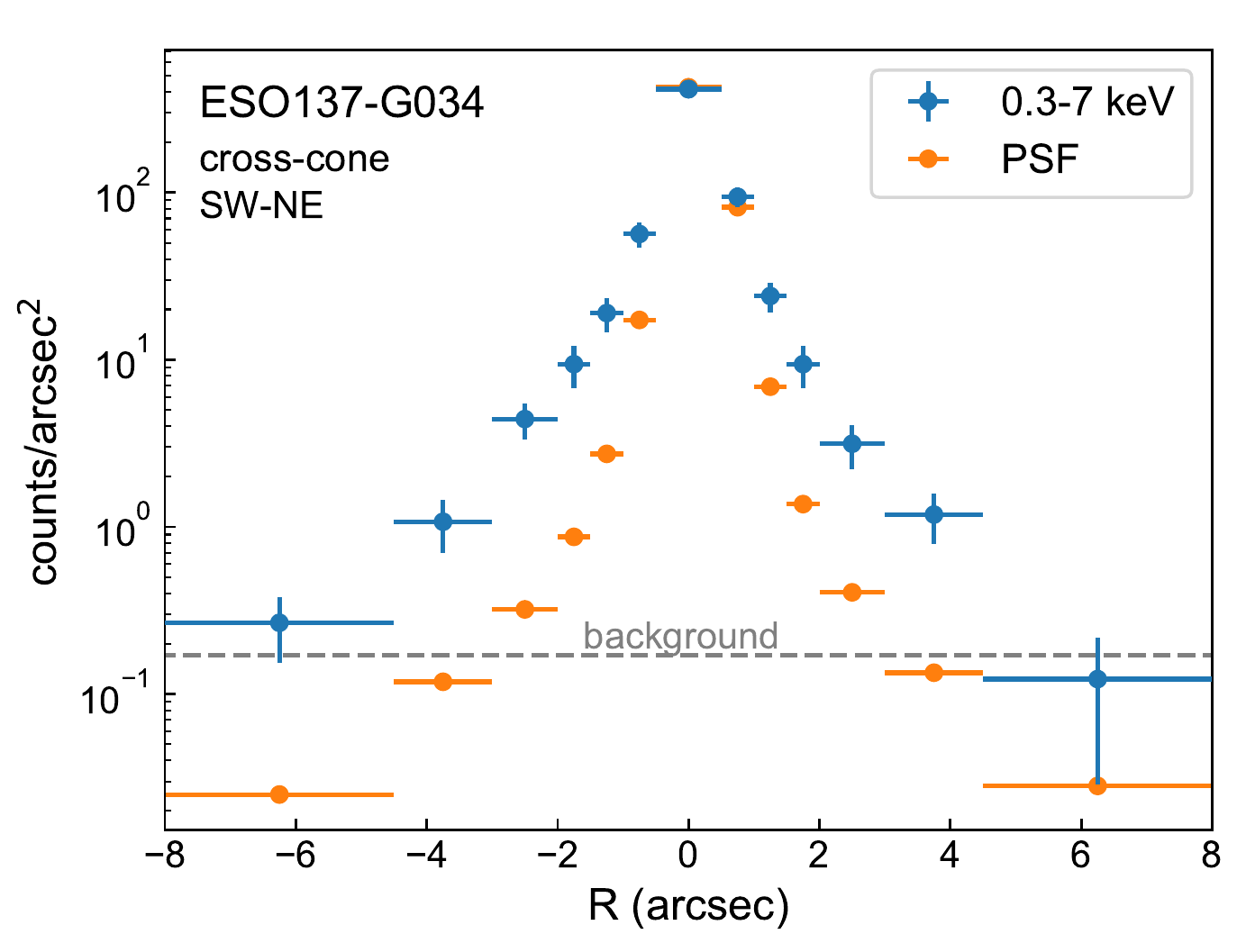}
\includegraphics[width=7.5cm]{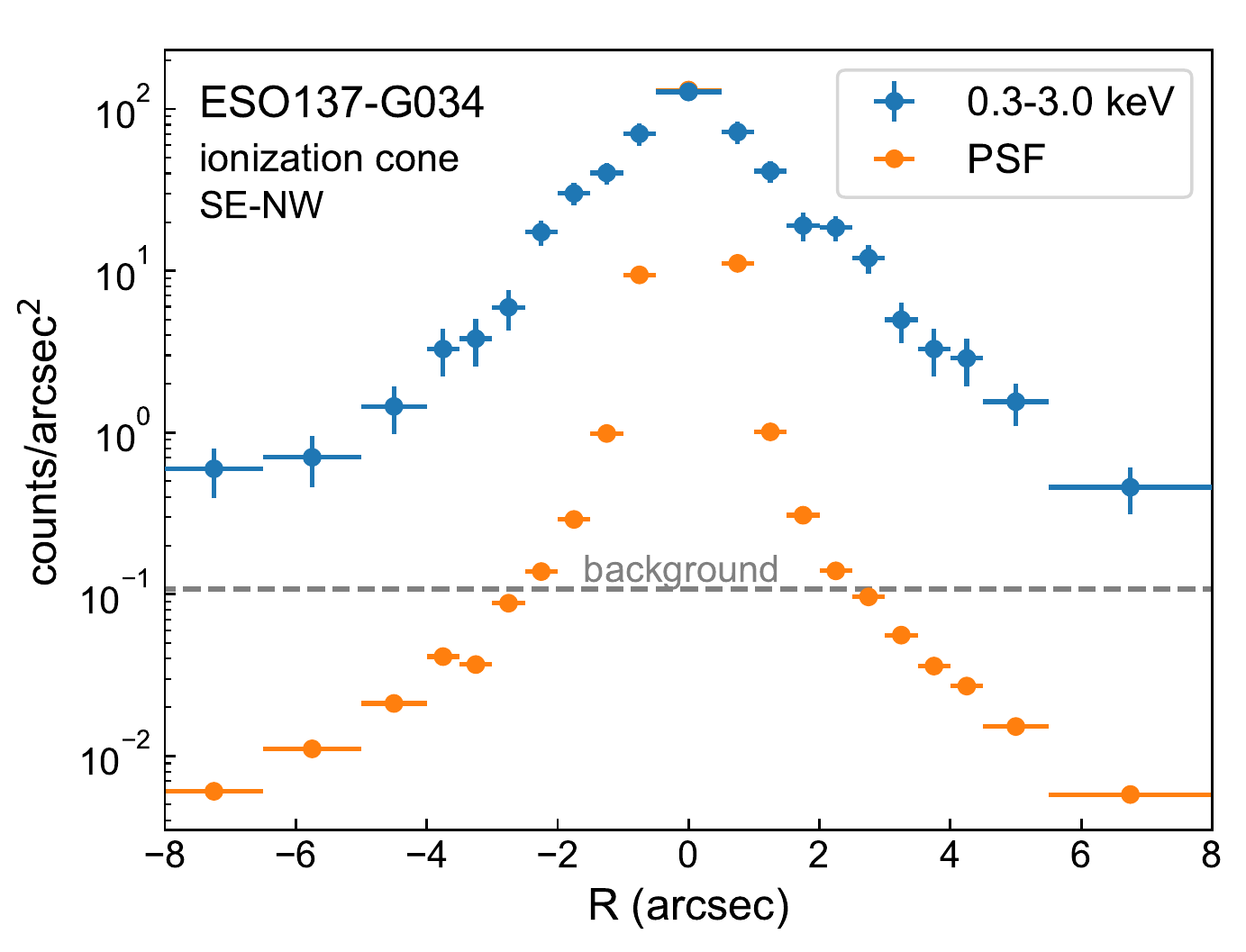}
\includegraphics[width=7.5cm]{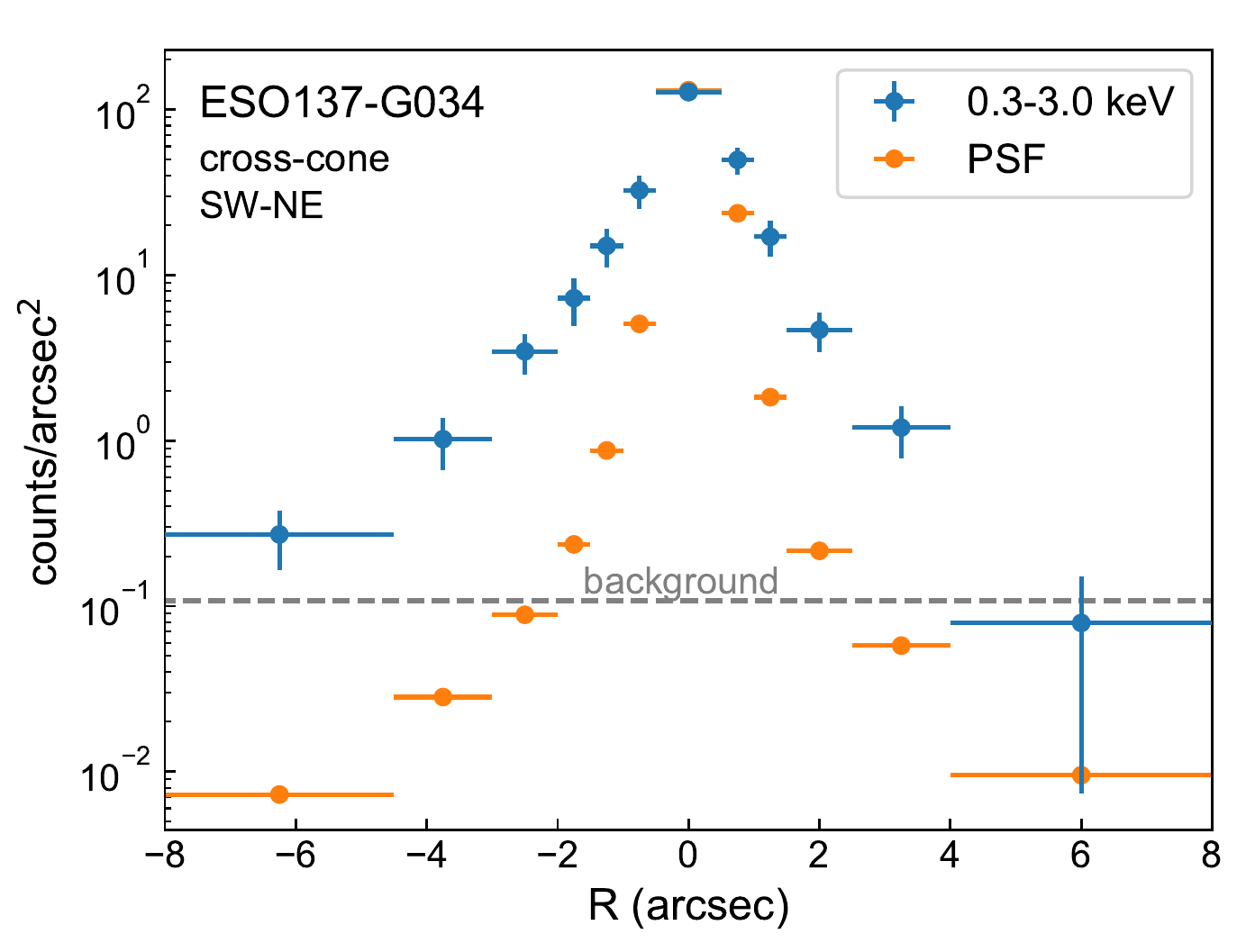}
\includegraphics[width=7.5cm]{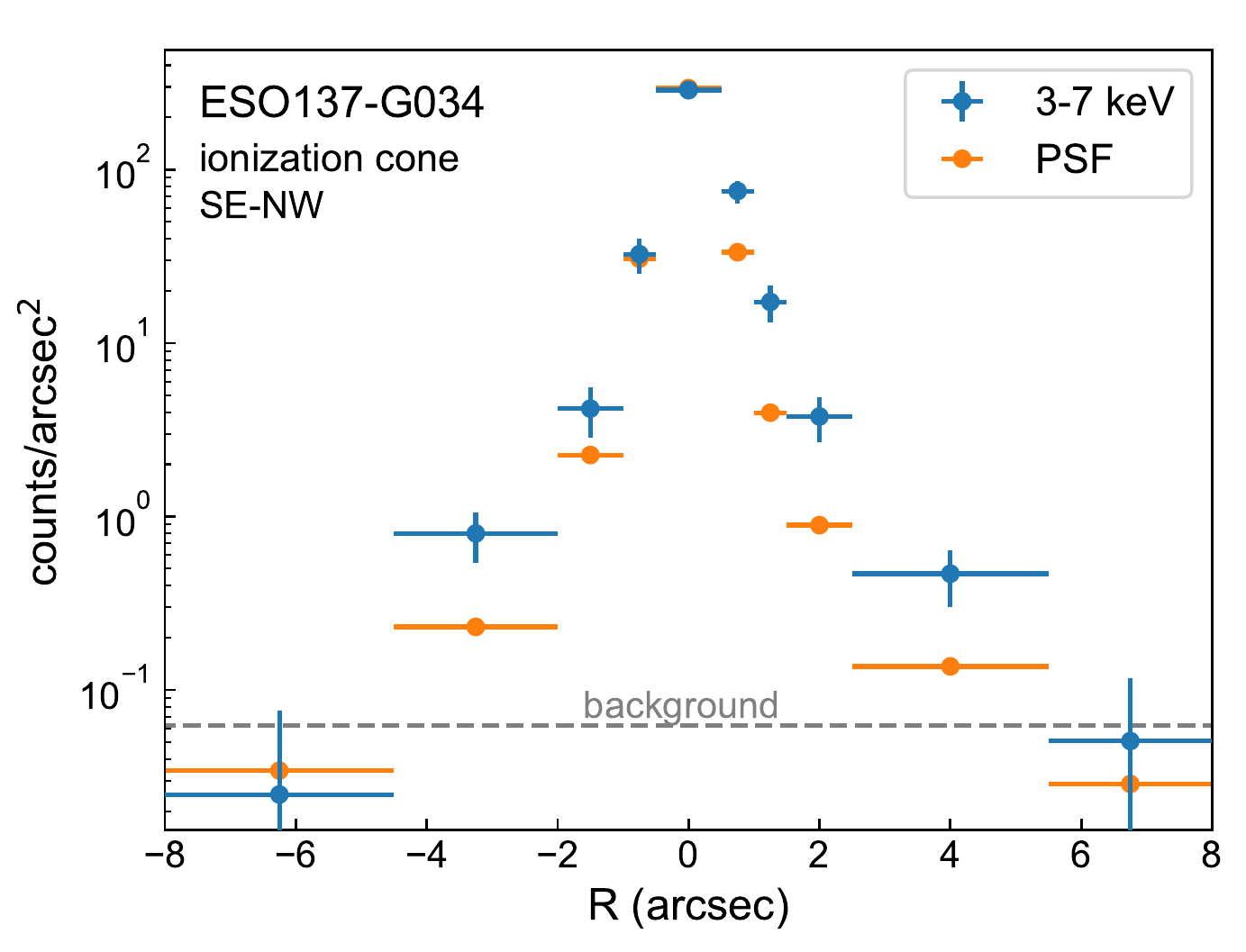}
\includegraphics[width=7.5cm]{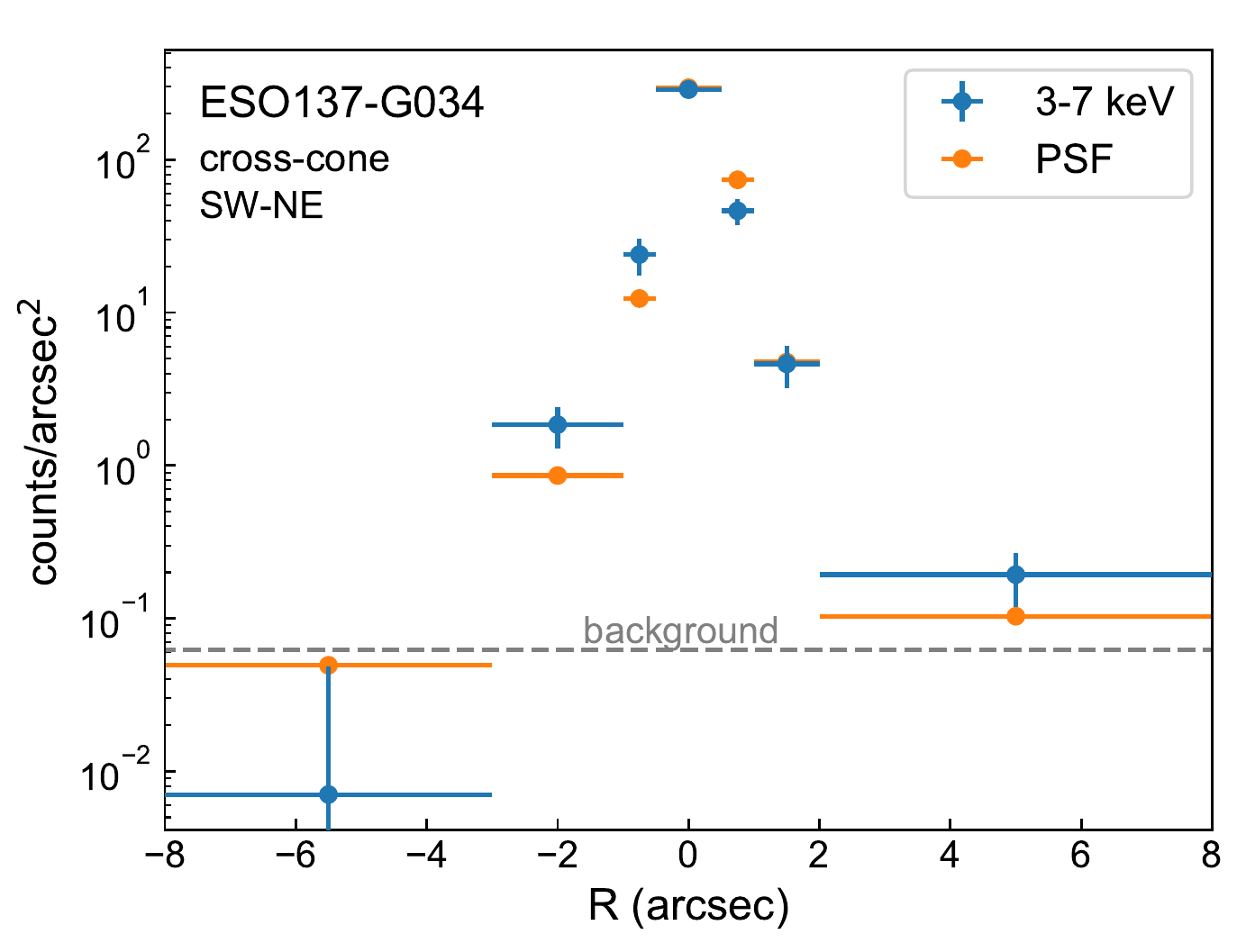}
\caption{Radial profiles of ESO 137-G034 in the ionization and cross-cones for the full band, soft band, and hard band. The background has been subtracted off from the radial profiles, and the level of which is indicated as the grey dashed horizontal line. The PSF is normalized to the counts in the central 0.5$\arcsec$ radius bin. The extended soft X-ray emission is well detected in both the ionization cones and the cross-cones, while the hard X-ray emission is only detected in the ionization cones. }
\label{ESO137-G034_cone_profiles}
\end{figure*}

The 6.0-7.0 keV band, where the Fe K$\alpha$ line dominates, has a tentative $\sim$2$\sigma$ detection of 5.1\% extended emission with an excess of 8.7 $\pm$ 4.7 counts over the PSF in the 1.5$\arcsec$-8$\arcsec$ region. The total excess counts, 34.0 $\pm$ 9.3, are better detected at above 3$\sigma$, accounting for $\sim$20\% of the total emission due to the contribution from the inner 1.5$\arcsec$ region. The surface brightness (Figure \ref{ESO137-G034_radial_profiles}) lies above the PSF out to 8$\arcsec$ in radius, but we need deeper {\it Chandra} data to get a higher signal-to-noise to confirm the total extent of this component.

\subsection{NGC 3281}

\begin{figure} 
\centering
\includegraphics[width=7cm]{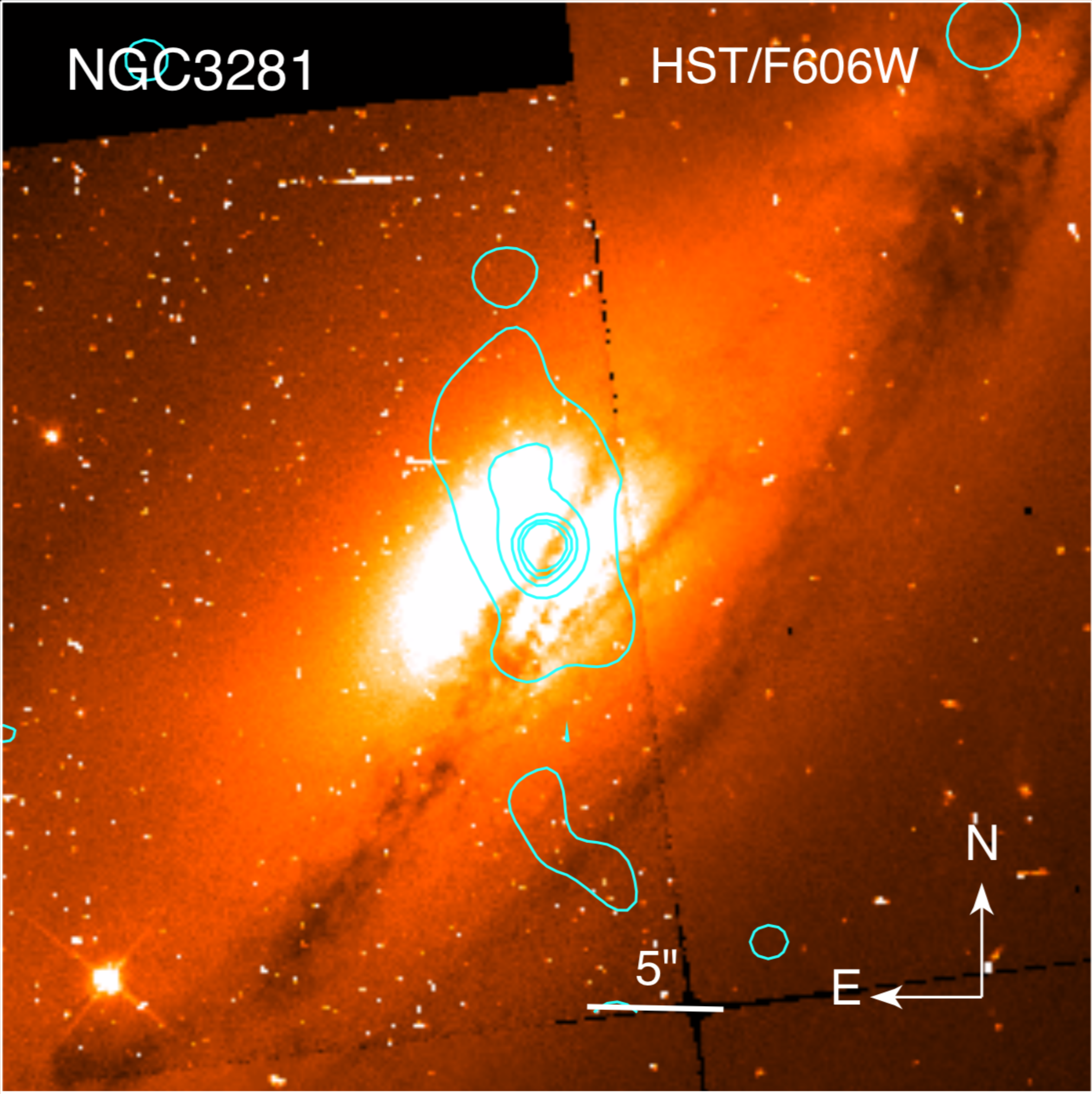}
\caption{40$\arcsec$ $\times$ 40$\arcsec$ {\it HST}/F606W image of NGC 3281 \citep{Malkan1998}, an SAab galaxy hosting a Seyfert 2 nucleus. The {\it Chandra} 0.3-3.0 keV X-ray contours are overlaid in cyan. }
\label{HST_NGC3281}
\end{figure}

\begin{figure*} 
\centering
\includegraphics[width=15cm]{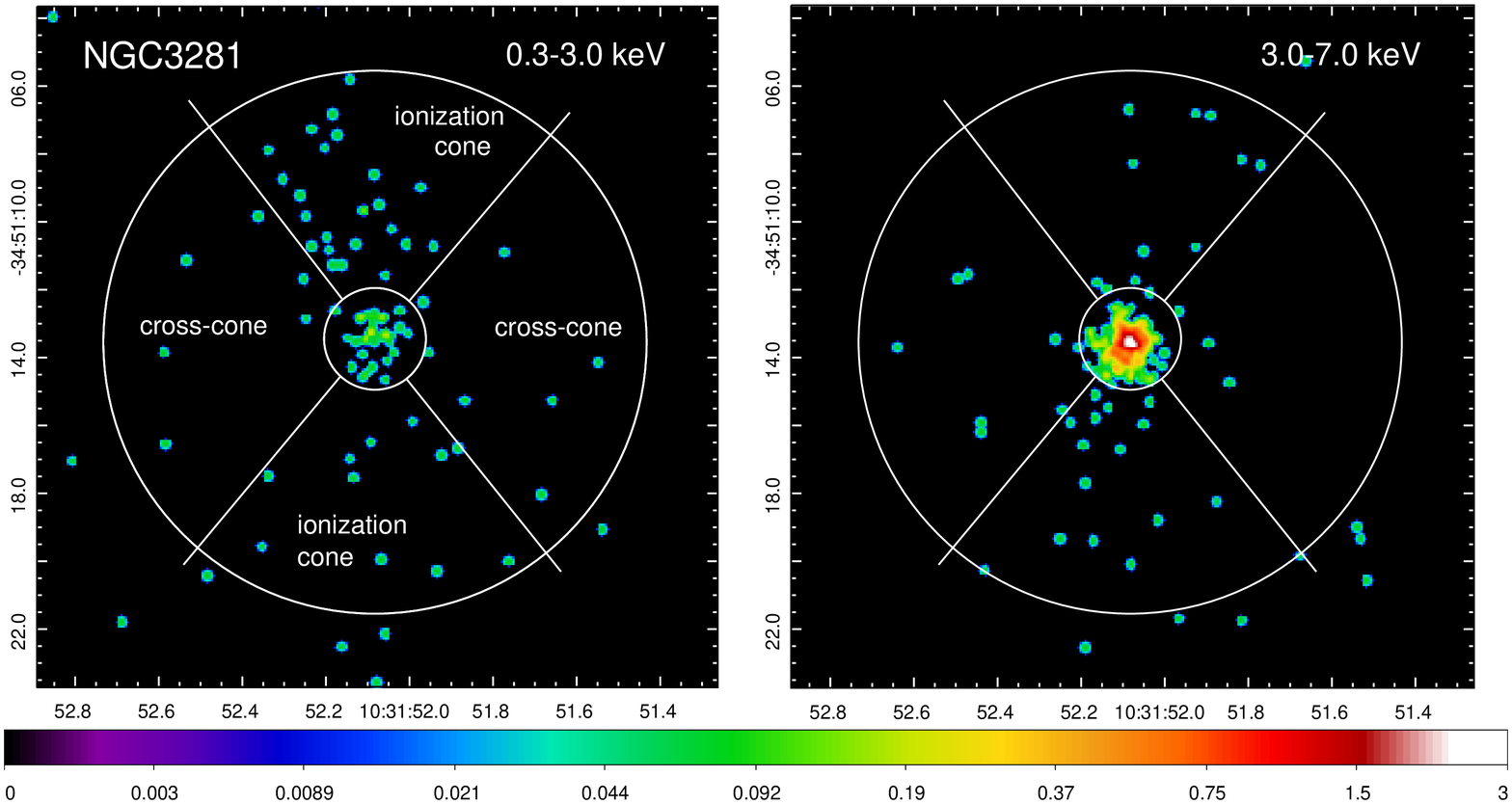}
\includegraphics[width=15cm]{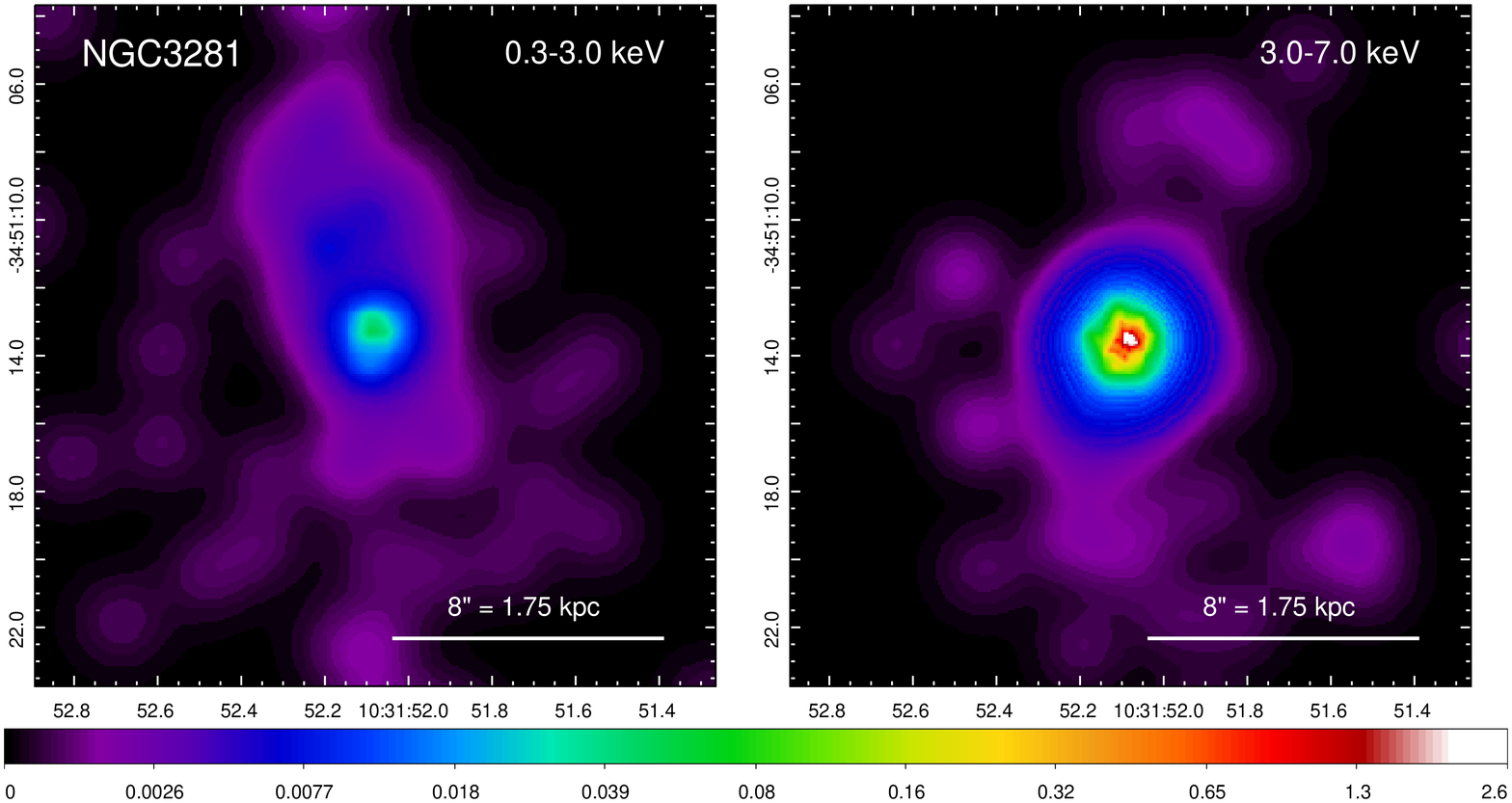}
\caption{{\bf Top:} 20$\arcsec$ $\times$ 20$\arcsec$ {\it Chandra} ACIS-S 0.3-3.0 keV (left) and 3.0-7.0 keV (right) band images of NGC 3281 at 1/8 subpixel binning. The inner 1.5$\arcsec$ radius circle and the outer 8$\arcsec$ circle define the region in between for extracting excess counts in the extended emission. Given the azimuthal dependence of the soft and hard X-ray emission, the 8$\arcsec$ radius circular region is split into the N-S ionization cones with an half-opening angle of 39$^\circ$ and the E-W cross-cones. The X-ray bicone axis is at an position angle of $\sim$178$^{\circ}$ East of North. {\bf Bottom:} Adaptively smoothed images ({\it dmimgadapt}; 0.5-15 pixel scales, 5 counts under kernel, 30 iterations) on 1/8 binned pixel. All the images are displayed in logarithmic scale with colors corresponding to the counts per image pixel.}
\label{NGC3281_chandra}
\end{figure*}

\begin{figure*} 
\centering
\includegraphics[width=5.952cm]{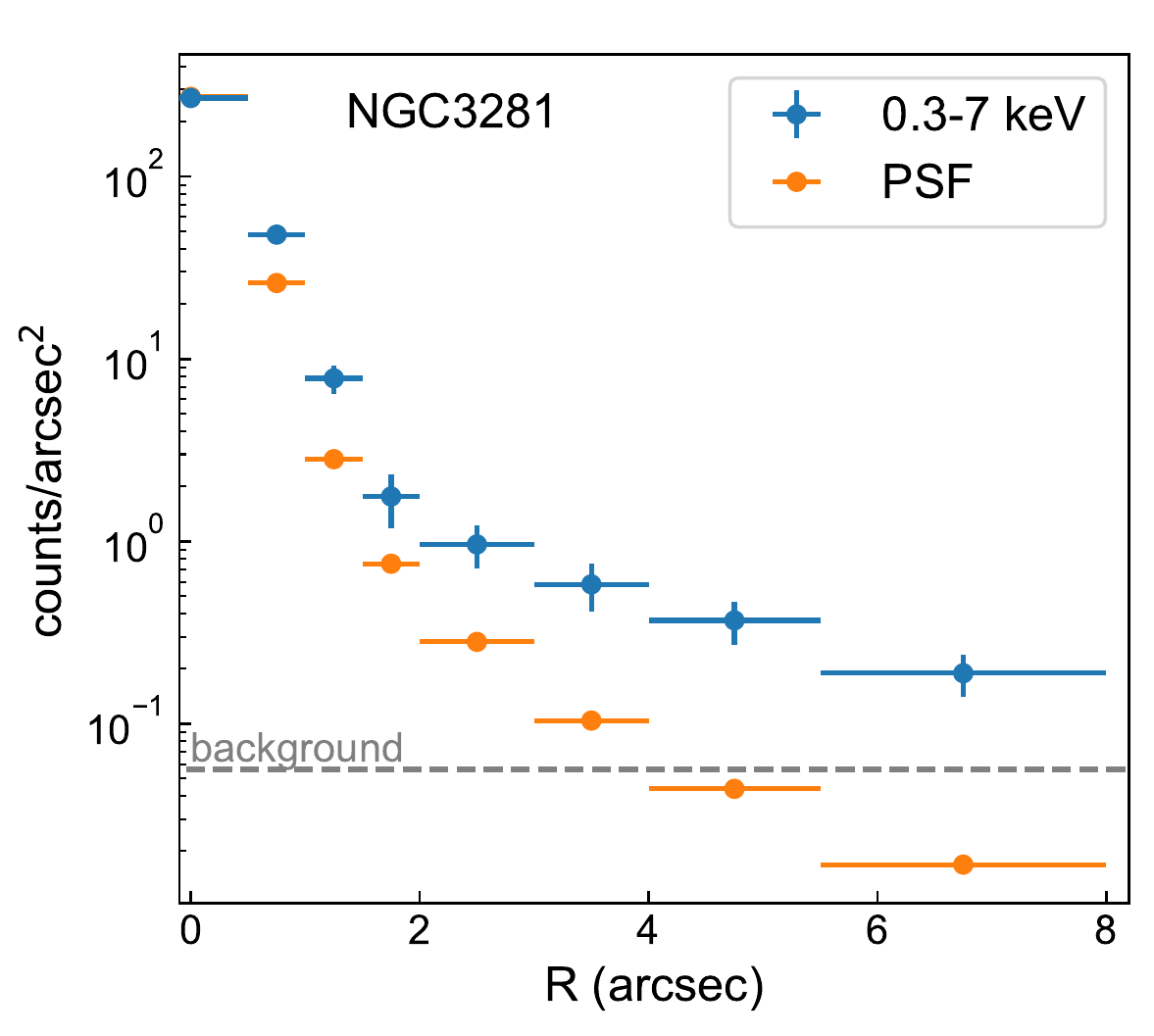}
\includegraphics[width=5.952cm]{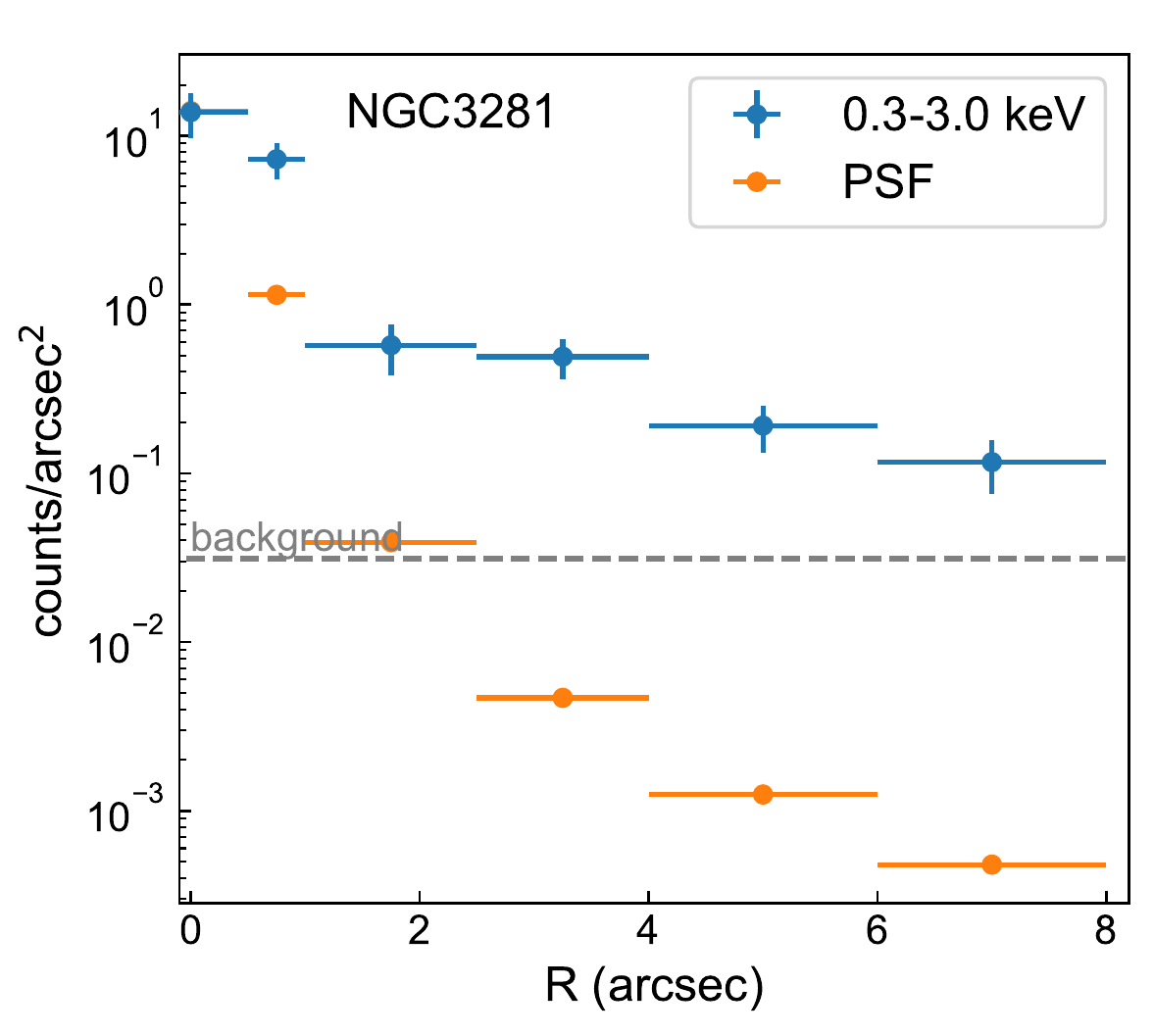}
\includegraphics[width=5.952cm]{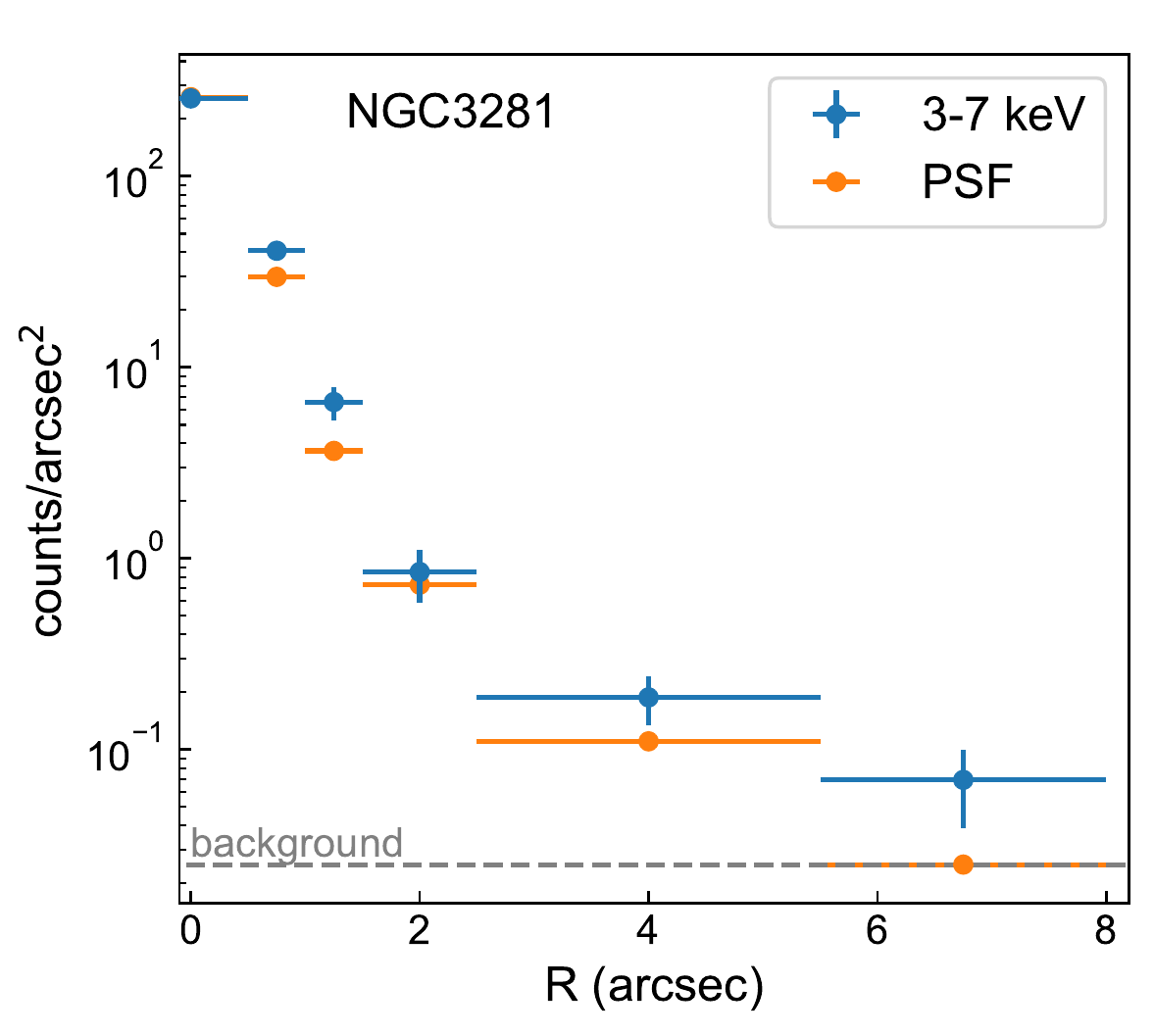}
\caption{Radial profiles of NGC 3281 for the full band, soft band, and hard band. The background has been subtracted off from the radial profiles, and the level of which is indicated as the grey dashed horizontal line. The PSF is normalized to the counts in the central 0.5$\arcsec$ radius bin.}
\label{NGC3281_radial_profiles}
\end{figure*}

\begin{figure*} 
\centering
\includegraphics[width=7.5cm]{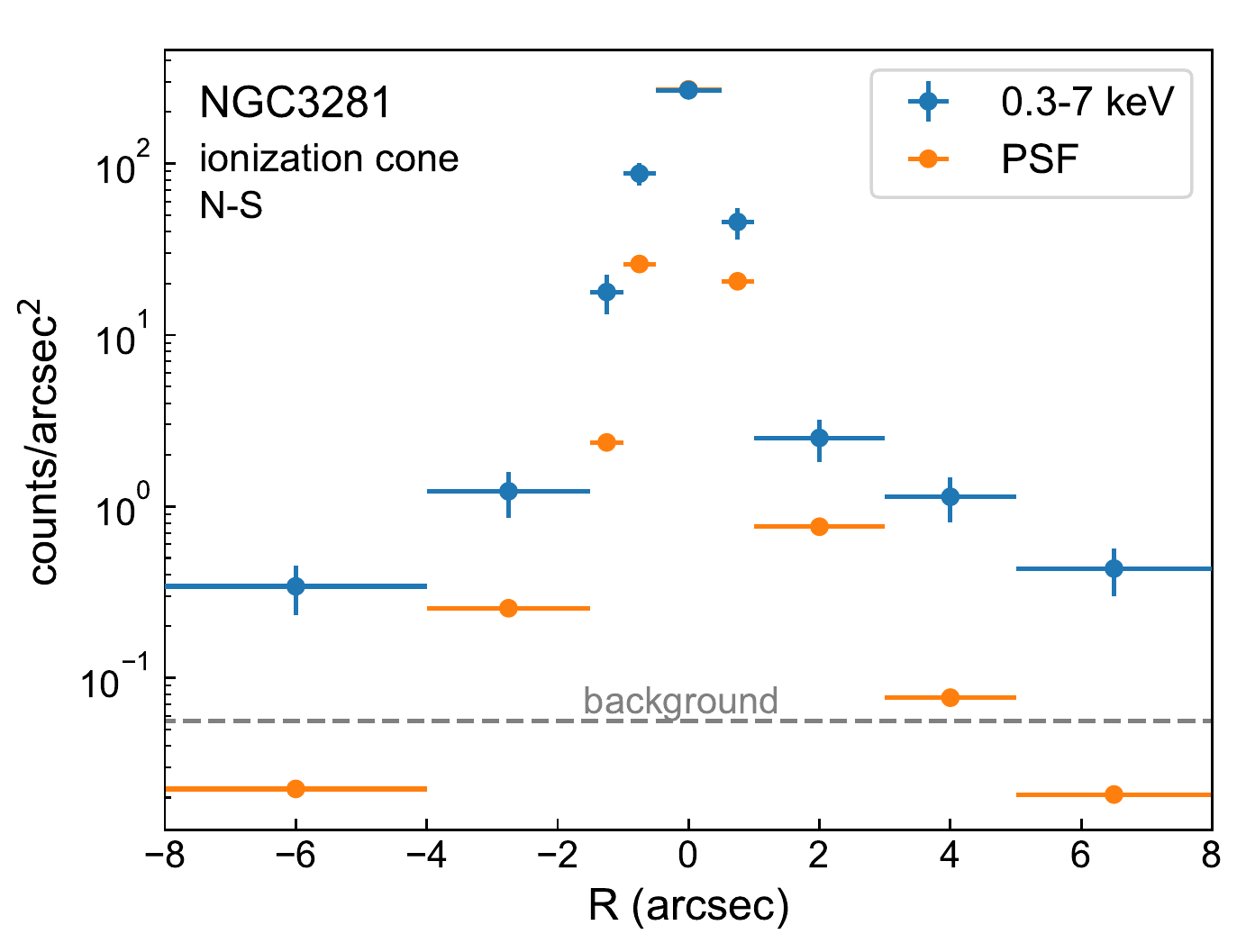}
\includegraphics[width=7.5cm]{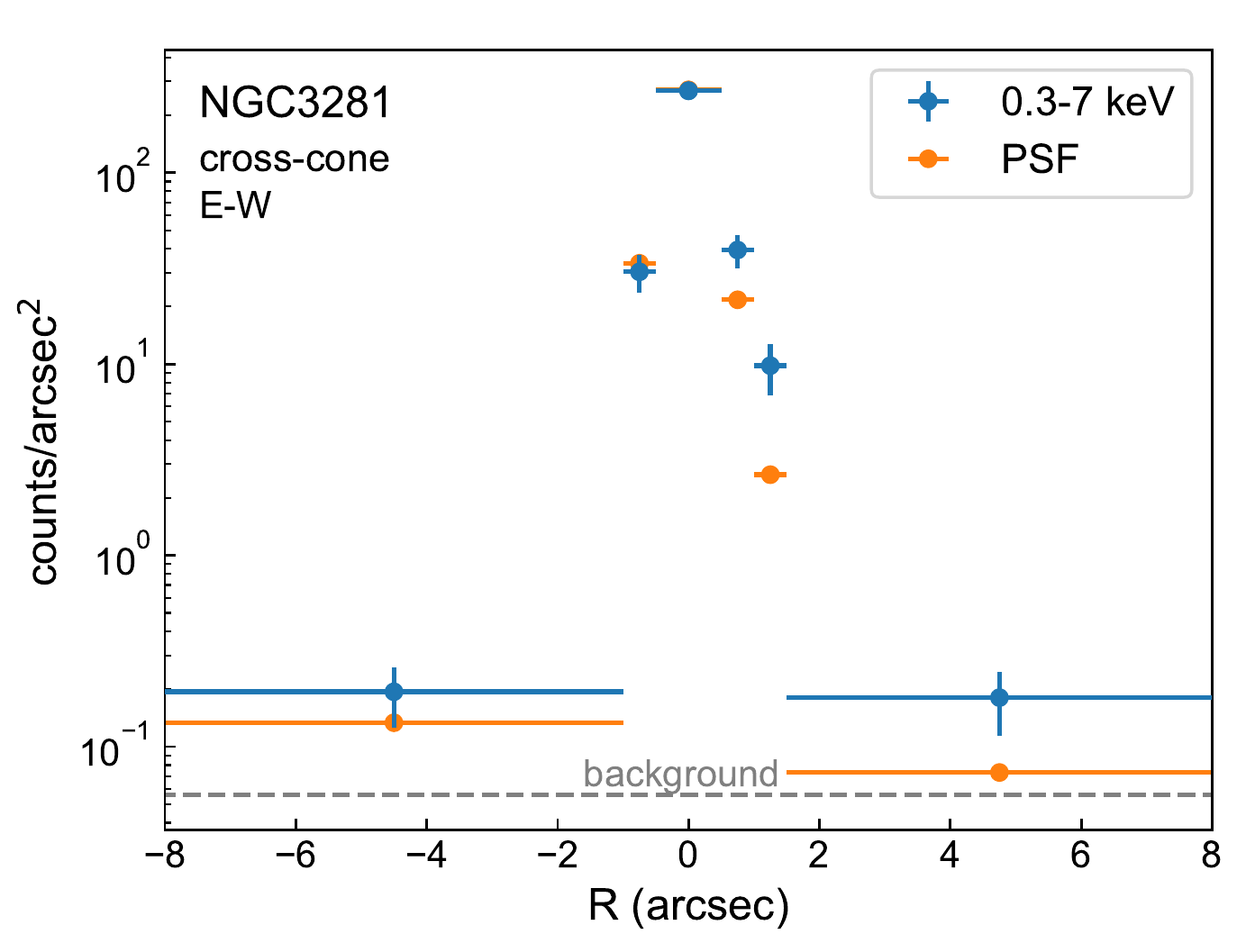}
\includegraphics[width=7.5cm]{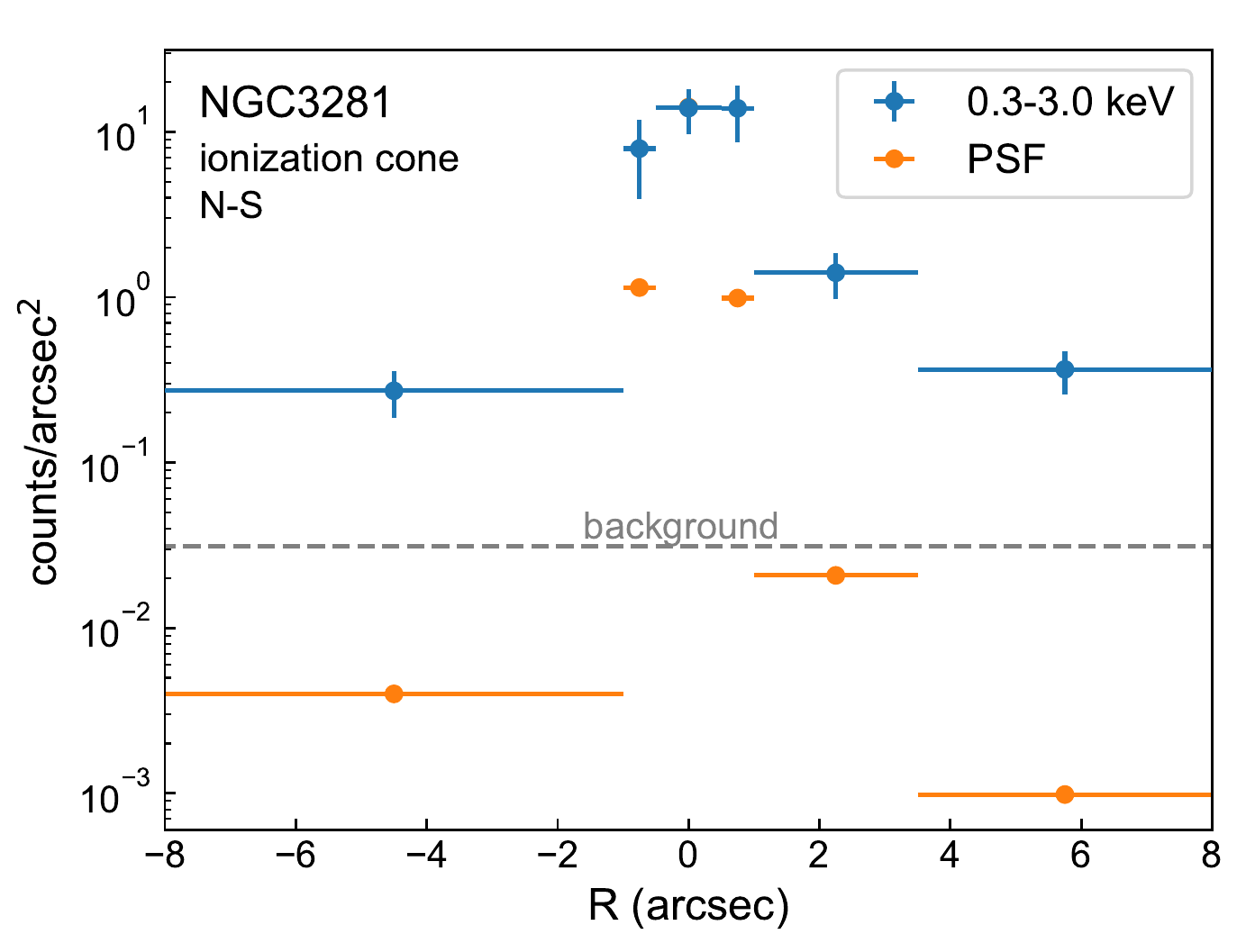}
\includegraphics[width=7.5cm]{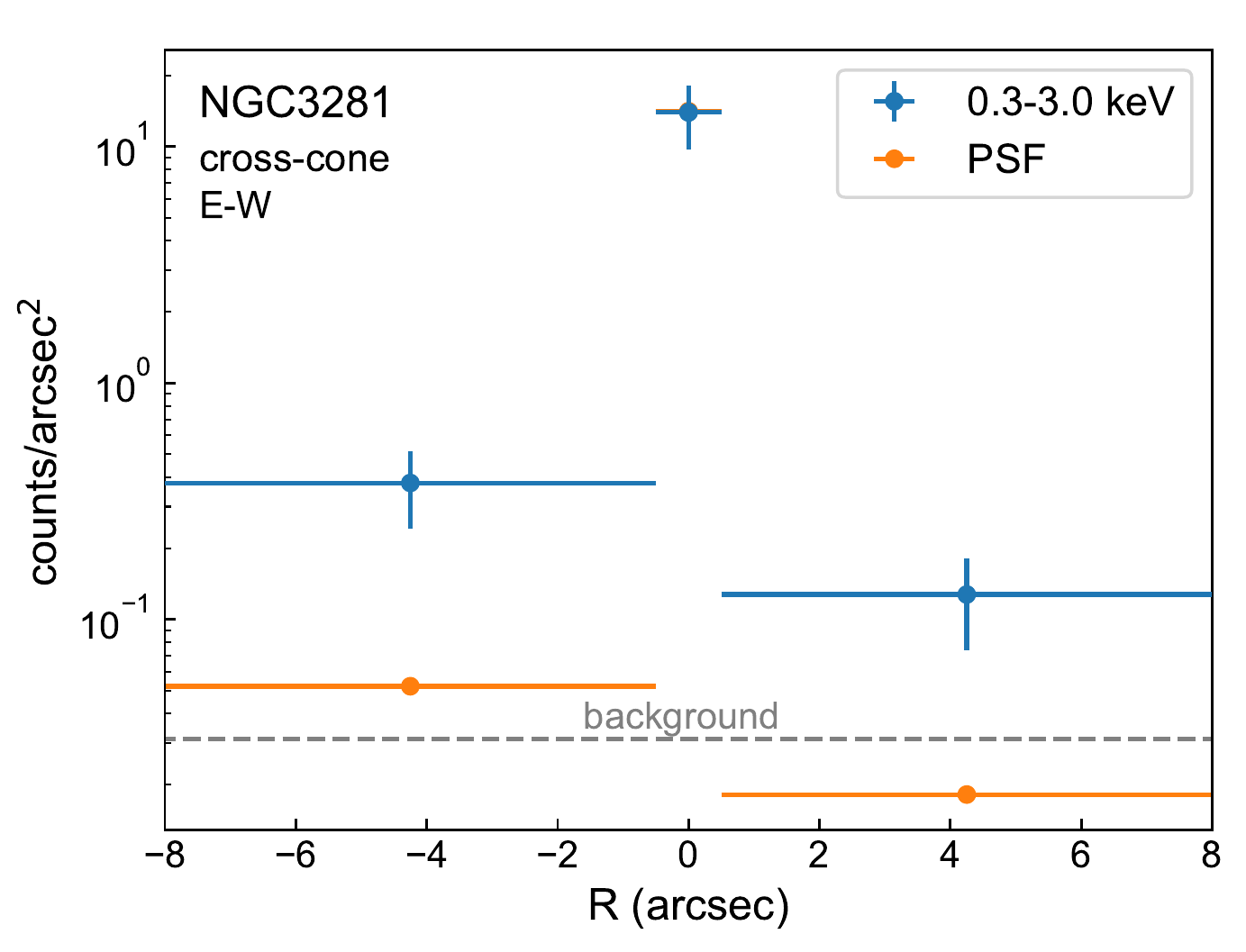}
\includegraphics[width=7.5cm]{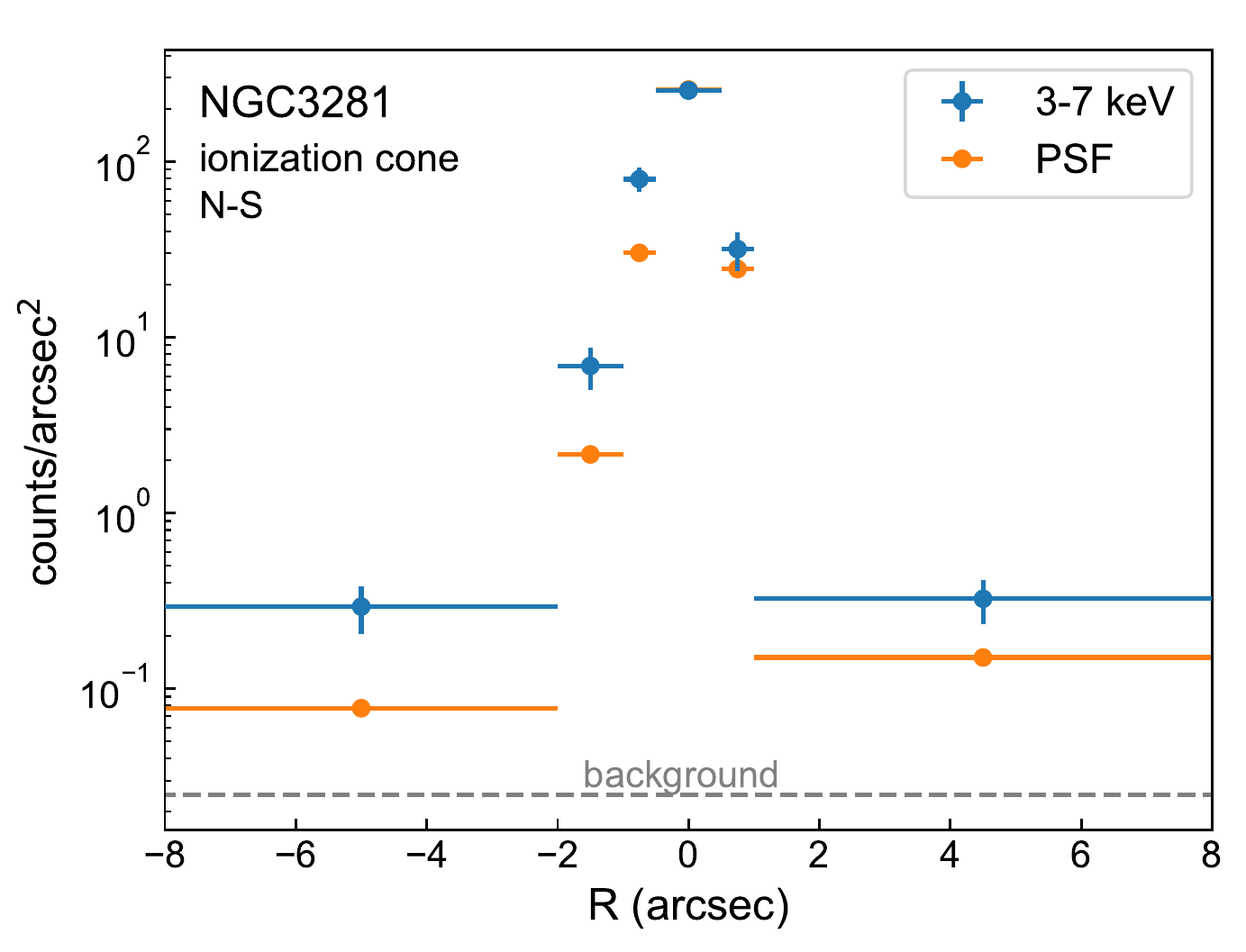}
\includegraphics[width=7.5cm]{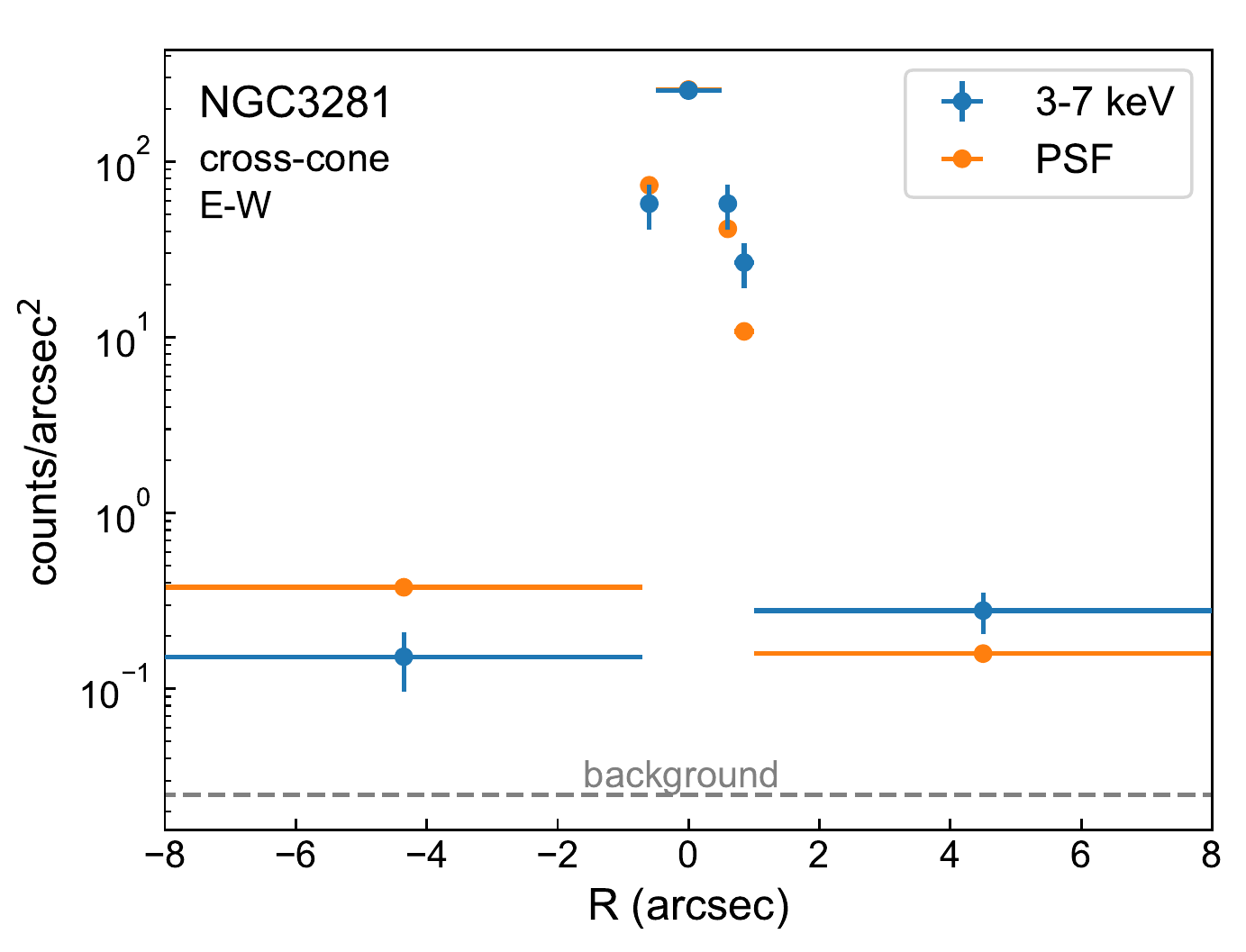}
\caption{Radial profiles of NGC 3281 in the ionization and cross-cones for the full band, soft band, and hard band. The background has been subtracted off from the radial profiles, and the level of which is indicated as the grey dashed horizontal line. The PSF is normalized to the counts in the central 0.5$\arcsec$ radius bin.}
\label{NGC3281_cone_profiles}
\end{figure*}

NGC 3281 is morphologically classified as an SAab galaxy (Figure \ref{HST_NGC3281}) at $z = 0.01067$ ($D$ $\sim$ 46.1 Mpc; 1" $\sim$ 219 pc). It hosts a Seyfert 2 nucleus with a CT column density of log $N_{\rm H}$ = 23.9-24.3 cm$^{-2}$ (\citealt{Vasylenko2013,Ricci2017,Balokovic2017}).

Figure \ref{NGC3281_chandra} shows that NGC 3281 has a relatively low surface brightness core in the soft band, and the soft X-ray emission is elongated and distributed asymmetrically along the N-S direction, with a larger Northern extent ($\sim$8$\arcsec$, 1.8 kpc) than Southern ($\sim$4$\arcsec$, 0.9 kpc). There is barely any emission in the perpendicular E-W direction. The hard X-ray band has a concentrated surface brightness distribution in the center and also has a vertically distributed, extended diffuse component, although less prominent than the soft X-rays. Since the X-ray emission has a strong azimuthal dependence in both the soft and hard bands, we split the data into two biconical regions with the bicone axis at a position angle of $\sim$178$^{\circ}$ East of North: one in the N-S direction (i.e., ionization cones) with a half-opening angle of $\sim$ 39$^\circ$ and one in the E-W direction (i.e., cross-cones). We generated radial profiles in both the 8$\arcsec$ circular region and the cone regions defined above (Figure \ref{NGC3281_radial_profiles} and Figure \ref{NGC3281_cone_profiles}). The extended hard X-ray emission is detected at $3.2\sigma$ over the PSF with an extended fraction of 9.7\% in the ionization bicone and a physical extent of $\sim$ 3.5 kpc in diameter (Table \ref{table4}). No extended emission is detected along the direction of the cross-cones. Overall there are 48.1 $\pm$ 12.6 excess counts above the PSF at 3-7 keV, taking 13.5\% of the total counts in this band.

NGC 3281 is one of the Seyfert galaxies selected in an {\it HST} survey of extended [O\III]$\lambda$5007 emission by \cite{Schmitt2003a,Schmitt2003b}. The [O\III] line map displays a conically shaped narrow line region preferentially distributed highly asymmetrically towards the N-E direction extending up to $\sim$1.3 kpc (see Figure 9 in \citealt{Schmitt2003a}), overlapping with the {\it Chandra} X-ray emission. \cite{Schmitt2003a} also noted that the ground-based observations by \cite{Bergmann1992} showed more extended emission in the NE direction (up to 2 kpc from the disk), and the less extended emission from {\it HST} could be due to the relatively small field of view ($\sim$ 13$\arcsec$ in diameter) of the ramp filter. The lack of [O\III] emission and also X-ray emission in the south is most likely caused by obscuration of the dust lanes in the host galaxy disk as seen in Figure \ref{HST_NGC3281}. We measured the integrated soft X-ray flux, $f_{\rm 0.5-2 keV}$ = (1.1 $\pm$ 0.3) $\times$ 10$^{-13}$ erg cm$^{-2}$ s$^{-1}$, and used the [O\III] line flux of 2.5 $\times$ 10$^{-13}$ erg cm$^{-2}$ s$^{-1}$ in \cite{Schmitt2003a} to calculate the [O\III]/soft X-ray ratio. The flux ratio of $\sim$2.3 is comparable to typical ratios of Seyfert galaxies \citep{Bianchi2006}.

\subsection{NGC 424}

NGC 424 is an SB0/a galaxy at $z = 0.01176$ ($D$ $\sim$ 50.8 Mpc; 1" $\sim$ 241 pc). The nuclear activity is classified as Seyfert 1/Seyfert 2 with a CT column density of log $N_{\rm H}$ = 24.1-24.3 cm$^{-2}$ \citep{Ricci2015,Paltani2017,Ricci2017}. \cite{Balokovic2014} suggest that $N_{\rm H}$ could be even higher. 

The {\it Chandra} 0.3-3.0 keV band image (Figure \ref{NGC424_chandra}) shows a very high surface brightness in the central region with the extended emission tending to distribute more along the E-W direction. NGC 424 is the only CT AGN in the sample that contains a higher central surface brightness core in the soft X-ray band than in the hard band. While the soft band has excess emission over the PSF at all radii (up to 8$\arcsec$, 1.9 kpc), the hard band surface brightness drops quickly to the level of the PSF beyond 1.0$\arcsec$ (Figure \ref{NGC424_radial_profiles}). Although there is no extended hard X-ray emission detected beyond 1.5$\arcsec$, the inner 1.5$\arcsec$ region (up to $\sim$362 pc in radius) does show significant excess counts detected at 6.2$\sigma$ above the PSF (Table \ref{table2}). Overall the excess counts take up 13.5\% of the total counts in the 3-7 keV band.

\subsection{NGC 1125}

\begin{figure} 
\centering
\includegraphics[width=7cm]{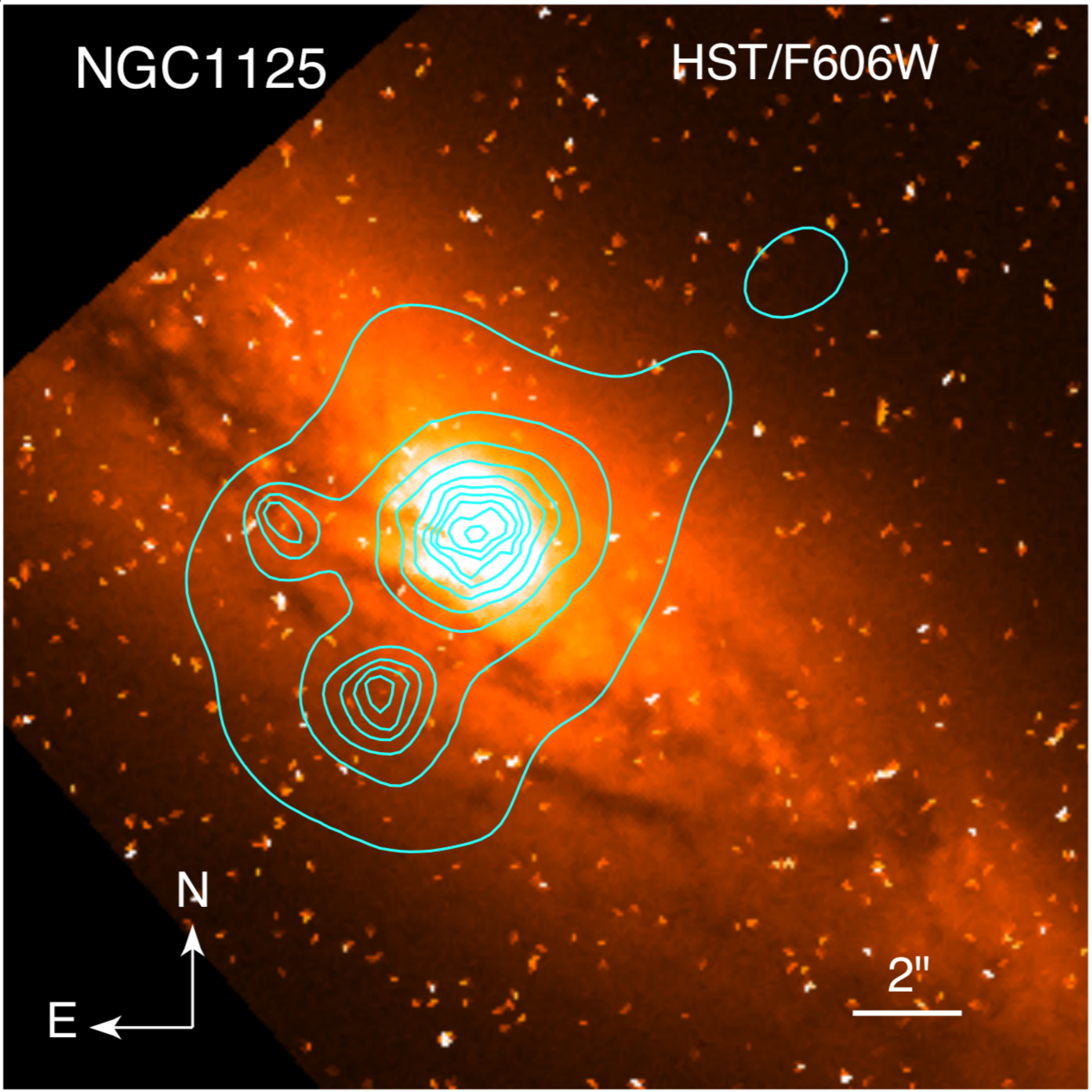}
\caption{20$\arcsec$ $\times$ 20$\arcsec$ {\it HST}/F606W image of NGC 1125 \citep{Malkan1998}, an SB0/a galaxy hosting a Seyfert 2 nucleus. The {\it Chandra} 0.3-7.0 keV X-ray contours are overlaid in cyan, extending perpendicular to the major axis of the host galaxy out to 8$\arcsec$ in radius. The nearby two X-ray detected sources have no optical counterparts.  }
\label{NGC1125_HST}
\end{figure}

NGC 1125 is an SB0/a galaxy (Figure \ref{NGC1125_HST}) at $z = 0.01093$ ($D$ $\sim$ 47.2 Mpc; 1" $\sim$ 224 pc), containing a Seyfert 2 nucleus \citep{Veron2006} with a CT column density of log $N_{\rm H}$ = 24.2-24.3 cm$^{-2}$ \citep{Ricci2015,Ricci2017}. 

The {\it Chandra} data (Figure \ref{NGC1125_chandra}) exhibit an elongated morphology along the NW-SE direction in both the soft and hard bands, extending perpendicular to the major axis of the host galaxy shown in Figure \ref{NGC1125_HST}. There are two X-ray sources detected off-center, but they have no optical counterparts in the {\it HST}/F606W image (Figure \ref{NGC1125_HST}). We estimated their 0.3-7 keV X-ray luminosities to be $\sim$3 $\times$ 10$^{39}$erg s$^{-1}$ and $\sim$3 $\times$ 10$^{38}$erg s$^{-1}$, which suggest that they could be high-mass X-ray binaries within the NGC 1125 host galaxy. However, we cannot completely rule out the possibility that these two blobs are part of an ionization cone structure. 

Figure \ref{NGC1125_radial_profiles} shows the radial profiles extracted after subtracting the two sources. The extended emission is well detected in the soft X-ray band, with a physical scale of $\sim$1.8 kpc in radius. While the surface brightness of the hard X-ray emission lies above the PSF out to $\sim$ 5$\arcsec$ ($\sim$ 1.1 kpc) in radius, the measured excess counts in the 1.5$\arcsec$-8$\arcsec$ region are negligible compared to the statistical error. The inner 1.5$\arcsec$ region does show excess counts above the PSF at $>$3$\sigma$, making a total excess fraction of 16.1\%. We also extracted excess counts including the two sources (Table \ref{table1}), which would make the 3-7 keV band a 3$\sigma$ detection in the extended emission.

\subsection{NGC 4500}

NGC 4500 is an SBa galaxy at $z = 0.01038$ ($D$ $\sim$ 44.8 Mpc; 1" $\sim$ 213 pc). The optical morphology is similar to that shown by Seyfert galaxies with a dominant bright, star-like nucleus. But the nuclear activity can be explained by a starburst, and NGC 4500 is classified as having a starburst nucleus or a composite AGN \citep{Balzano1983,Koss2017} with log $N_{\rm H}$ = 23.9 cm$^{-2}$ \citep{Ricci2017}. 

Figure \ref{NGC4500_chandra} shows that the X-ray emission appears to be preferentially distributed along the N-S direction in both the soft and hard bands, especially towards the north. The soft X-ray emission extends to 4.5$\arcsec$ ($\sim$ 1 kpc) in radius. The extended component in the hard band is not detected in the 1.5$\arcsec$-8$\arcsec$ annular region. The radial profiles (Figure \ref{NGC4500_radial_profiles}) do show some excess emission in the hard band out to 2$\arcsec$ ($\sim$426 pc), with a total of 21.5 excess counts over the PSF and a total excess fraction of 14.6\%, although not statistically significant.

\subsection{ESO 005-G004}

ESO 005-G004 is an Sb edge-on galaxy hosting a Seyfert 2 nucleus at $z = 0.00623$ ($D$ $\sim$ 26.8 Mpc; 1" $\sim$ 128 pc) with a CT column density of log $N_{\rm H}$ = 24.2-24.3 cm$^{-2}$ (\citealt{Ricci2015,Ricci2017,Balokovic2017}). 

As shown in Figure \ref{ESO005-G004_chandra}, the extended soft X-ray emission is not detected in ESO 005-G004, without any concentrated emission in the center. This extremely low level of soft X-ray emission is likely due to high column density absorption in the disk of the host galaxy in an edge-on view. A high surface brightness core appears in the hard band, and the radial profile (Figure \ref{ESO005-G004_radial_profiles}) shows excess emission out to at least 3$\arcsec$ (384 pc) in radius with a total of 29.5 $\pm$ 9.3 excess counts above the PSF, taking up 17.1\% of the total counts.

\subsection{2MASXJ00253292+6821442}

J0025+6821 is a Seyfert 2 galaxy (morphology not classified) at $z = 0.0120$ ($D$ $\sim$ 51.9 Mpc; 1$\arcsec$ $\sim$ 246 pc) with a column density of log $N_{\rm H}$ = 23.9-24.3 cm$^{-2}$ (\citealt{Ricci2017,Balokovic2017}). 

This source also has a low level of soft X-ray emission (Figure \ref{J0025_chandra}), and the extended component is not detected above 3$\sigma$. The surface brightness of the hard X-ray emission is concentrated in the center and then quickly drops to the PSF level (Figure \ref{J0025_radial_profiles}). No extended emission is detected in the hard band either.

\section{Discussion}
\label{sec:discussion}

\subsection{Comparison with extended hard X-ray detected sources in the literature}

So far, only a handful of sources in the literature have detections of extended hard X-ray emission (e.g., Circinus, \citealt{Arevalo2014}; NGC 1068 \citealt{Bauer2015}), and ESO 428-G014 \citep{Fabbiano2017,Fabbiano2018a,Fabbiano2018b} and NGC 7212 \citep{Jones2020} exhibit prominent, kpc-scale hard X-ray emission. For ESO 428-G014 and NGC 7212, both the 3-6 keV hard X-ray continuum and Fe K$\alpha$ show extended emission out to 15$\arcsec$ and 8$\arcsec$, corresponding to $\sim$1.8 kpc and $\sim$4.3 kpc in radius, respectively. Unlike the 7 CT AGN in this relatively shallow pilot survey, ESO 428-G014 and NGC 7212 have comparable, deep cumulative {\it Chandra} exposures of 154 ks and 150 ks and a similar total number of 1554 and 1529 counts in the 3-7 keV band (hence similar count rates), although the distances or physical scales (per arcsec) differ by a factor of 4.6. We extracted the excess counts and measured the extended fractions and total excess fractions in the 3-7 keV and 6-7 keV bands for ESO 428-G014 and NGC 7212 in the same manner as our 7 CT AGN (Table \ref{table5}). For ESO 428-G014, the excess emission at 3-7 keV in the extended 1.5$\arcsec$-8$\arcsec$ region is well detected above the PSF at 16$\sigma$, contributing 20.8\% of the total emission in this band. Adding the excess counts in the inner 0.5$\arcsec$-1.5$\arcsec$ region, we measured a total of 462.9 $\pm$ 31.3 excess counts above the PSF and a total excess fraction of 30.2\%. The 6-7 keV band (Fe K$\alpha$) also shows significant extended emission with a total of 72.8 $\pm$16.2 excess counts above the PSF, accounting for 16.4\% of the total emission in this band. For NGC 7212 (the most distant source among the 9 CT AGN), the excess counts in the extended 1.5$\arcsec$-8$\arcsec$ region are also well detected above the PSF with an extended fraction of 14.7\%. The total excess counts, 549.5 $\pm$ 23.4, make up 35.7\% of the total emission at 3-7 keV, the highest fraction reported so far. The extended emission is also prominent in the 6-7 keV band with a total of 95.7 $\pm$ 9.8 excess counts above the PSF and a total excess fraction of 25.5\%. 

ESO 428-G014 and NGC 7212 bookend the redshift/distance range of our 7 CT AGN, most of which are at $z \sim 0.01$. Five out of seven CT AGN in our sample show total excess emission in the 3-7 keV band ranging from $\sim$12\% to 22\% that are $>$3$\sigma$ detections. Among the 7 CT AGN, ESO 137-G034 and NGC 3281 are the only two that display biconical ionization structures, and their extended 3-7 keV emission is well detected above $3 \sigma$ along the ionization cones. ESO 137-G034 has a very similar 3-7 keV count rate (0.01 cts/s) to ESO 428-G014 and NGC 7212, although the exposure time is only $\sim$1/3 with a total of 441 counts in the 3-7 keV band. The physical extent of the 3-7 keV hard X-ray emission measured along the ionization cone direction is $\sim$1.0 kpc in radius, comparable to that of ESO 428-G014, with a hint of further extended emission in the N-S direction. For NGC 3281, the 3-7 keV count rate (0.04 cts/s) is a factor of $\sim$4 higher than those of ESO 428-G014, NGC 7212, and ESO 137-G034, and is the highest among the entire sample, but the hard extended fraction is the lowest among the four, i.e., more concentrated in the nuclear region. Among the 7 CT AGN, only ESO 137-G014 shows extended 6-7 keV (Fe K$\alpha$) emission extending $>$0.65 kpc in radius). Deeper {\it Chandra} data are required to consolidate this extended component and reveal the hinted full structure. 

Our CT AGN cover a 3-7 keV source count rate range of 0.004-0.04 cts/s, and we do not observe any correlation between the count rate (or source flux) and the excess counts or extended fraction, although the extended fraction has a strong linear correlation with the excess counts as expected. There appears to be a count rate threshold at $\sim$0.01 cts/s combined with a total excess fraction threshold at $\sim$20\% (or an extended fraction threshold at $\sim$7\%) above which a prominent extended hard X-ray morphology is revealed and the excess emission is well detected above the PSF. Of course, a large, complete sample of CT AGN is needed to verify this trend.

\begin{figure} 
\centering
\includegraphics[width=9cm]{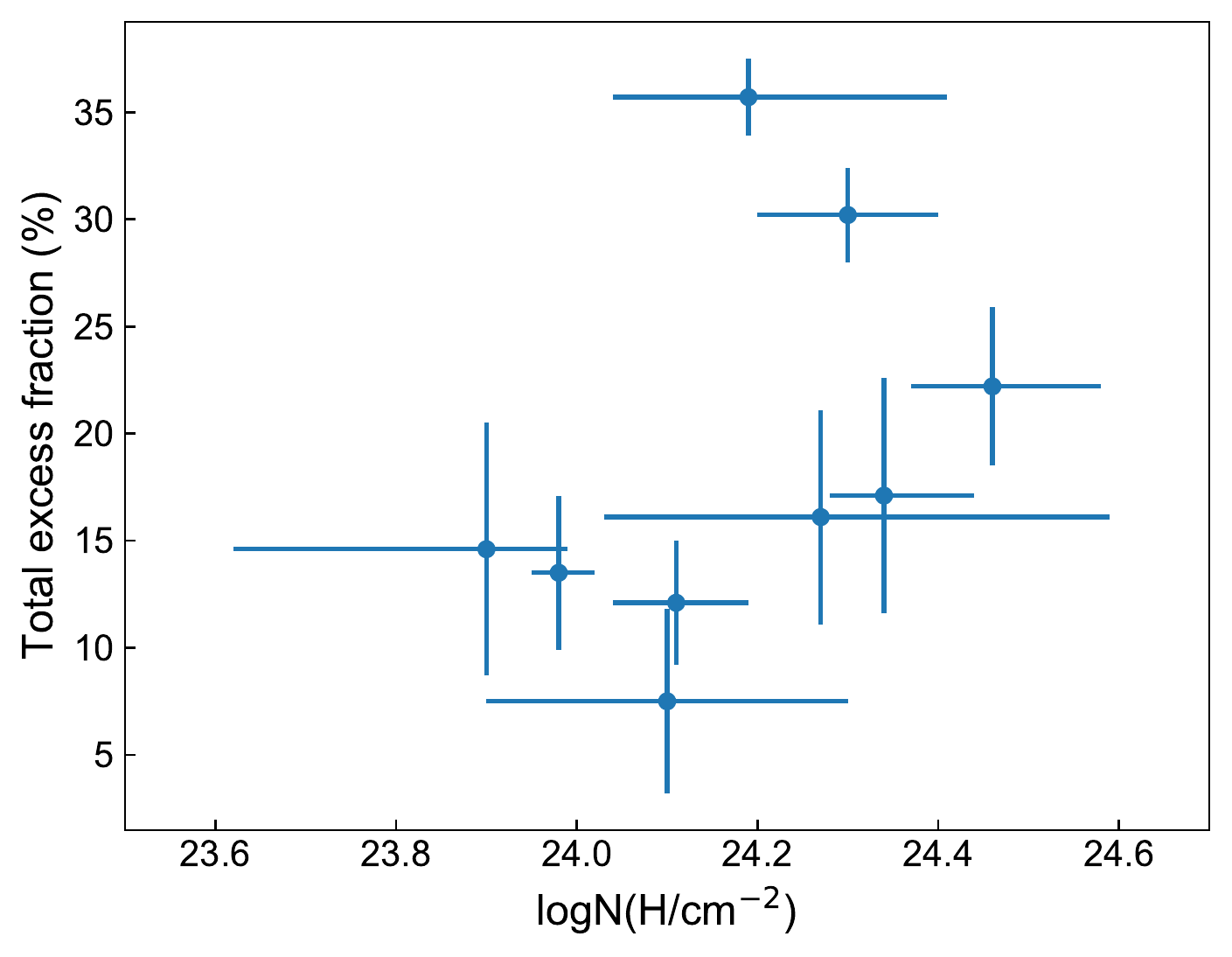} 
\caption{Total excess fraction at 3-7 keV versus log$N_{\rm H}$ of the 9 CT AGN. There appears to be a general trend of increasing total excess fraction with log$N_{\rm H}$.}
\label{excfrac_Nh}
\end{figure}

Given a total of 9 CT AGN with systematically measured excess counts and extended excess fractions, we further investigated whether the measured quantities correlate with any physical parameters (e.g., AGN bolometric luminosity, $N_{\rm H}$ etc.), which may give clues to the origin of the extended hard X-ray emission. Figure \ref{excfrac_Nh} shows the total excess fraction as a function of log$N_{\rm H}$. We obtained the $N_{\rm H}$ values from the literature that were mostly constrained though broad-band X-ray spectral fitting with a torus model \citep{Ricci2015,Ricci2017,Balokovic2017,Georgantopoulos2019,Marchesi2019}. These $N_{\rm H}$ values are consistent with our ongoing, preliminary analysis based on {\it NuSTAR} spectra. There appears to be a general trend of increasing total excess fraction with log$N_{\rm H}$. The Pearson correlation coefficient of this sample is 0.43, which suggests a moderate correlation. The correlation seems to be strengthened by a larger sample (Jones et al. in prep). This may imply a connection between the amounts of nuclear obscuration and dense molecular clouds in the host galaxy disk, which supply the materials needed to scatter the X-ray photons and produce the extended component as suggested in ESO 428-G014 \citep{Fabbiano2017,Fabbiano2018a,Fabbiano2018b}. This trend also suggests that some of the obscuration may be coming from farther out in the galaxy (not just the torus). Given the current small sample size and the difficulty in constraining $N_{\rm H}$ accurately in this regime, we would need a larger sample with well constrained $N_{\rm H}$ to verify this trend. No trend is found in other parameter pairs yet.

\subsection{Implications on torus modeling and AGN feedback}

As to the question of whether the extended hard X-ray emission is ubiquitous in CT AGN, although not all the sources in our sample have extended hard X-ray emission reaching kpc-scales ($\geq$ 1 kpc in radius), most of them (5 out of 7) do show extended hard X-ray emission above the PSF out to at least $\sim$360 pc in radius, which is beyond the dusty torus in the AGN unified model (e.g., \citealt{Urry1995,Ramos2017}). We consider the outer edge of the torus to be the radius at which the central BH gravity ceases to dominate over the host galaxy potential, usually corresponding to dust temperature at about 150 K. Using a simple formula (Eq.5) in \cite{Barvainis1987}, we estimated that the outer torus radii for the 7 CT AGN are in the range of $\sim$33 pc to 113 pc. Detecting hard X-ray emission beyond the traditional dusty torus in the standard AGN unified model implies that we need to test and improve torus modeling and update our knowledge of SMBH-host galaxy interactions.

The 7 CT AGN in our sample and ESO 428-G014 all have high quality, broad band {\it NuSTAR} spectra covering 3-79 keV, which we will present in a future publication. In principle, aspects of the obscuring torus geometry, e.g., its scale height, opening angle, inclination angle, and torus covering factor in clumpy torus models \citep{Nenkova2008,Elitzur2008}, can all be extracted from observed broadband X-ray spectra of AGN (e.g., with {\it NuSTAR}) with the help of parameterized spectral models (e.g., \citealt{Murphy2009,Balokovic2018,Tanimoto2019}). However, the existence of reprocessed emission on $>$100 pc scales (i.e., outside of the compact torus) is currently not accounted for by any published models. This newly discovered component, if ignored, can lead to biased estimates of the torus covering factor and/or its average column density. For example, our preliminary analysis of spatially resolved data for ESO 428-G014 suggests that its torus covering factor may be significantly lower (by almost a factor of two depending on the modeling details) compared to a spatially unresolved spectral analysis (Balokovi{\'c} et al. in prep).

It is possible that the opening angle and the internal covering factor within the torus can be derived from the extended X-ray and other measures. This may reduce the degrees of freedom in torus models as they usually contain many free parameters (e.g., \citealt{Murphy2009,Balokovic2018,Tanimoto2019}), and therefore reduce the uncertainties in the remaining torus model parameters. The {\it Chandra} detections of the extended hard X-ray emission in ESO 428-G014, ESO 137-G034, and NGC 3281 can serve as testbeds for this purpose.

The extended hard X-ray emission does not only appear in the direction of ionization cones but also in the cross-cone regions, as clearly shown in ESO 428-G014 \citep{Fabbiano2017,Fabbiano2018a} and NGC 7212 \citep{Jones2020}. Rather than a completely obscuring torus in the standard AGN unified model, the appearance of extended hard X-ray emission supports the increasingly popular scenario of a clumpy structure of the torus (e.g., \citealt{Nenkova2002,Nenkova2008,Elitzur2012}), which allows for the transmission of radiation on kpc scales. The interactions of the photons escaping the nuclear region with the ISM clouds in the host galaxy would give rise to the extended diffuse emission in both the ionization cone and cross-cone directions \citep{Fabbiano2018a,Fabbiano2018b}. The X-ray emission of these sources is also accompanied by the presence of radio jets \citep{Fabbiano2018b,Jones2020}. The jet-cold disk ISM interactions may form a very hot cocoon enclosing the interaction region, which could contribute to the X-ray emission seen in the cross-cone region. This hot cocoon has been predicted by relativistic hydrodynamical simulations (e.g., \citealt{Mukherjee2018}) and provides an alternative explanation to the X-ray emission perpendicular to the ionization cone direction.

The extended 3-6 keV continuum and Fe K$\alpha$ emission, as in ESO 428-G014, are explained as due to electron scattering off dense ISM clouds in the host galaxy \citep{Fabbiano2017,Fabbiano2018a}. ALMA CO(2-1) and SINFONI H$_2$ 2.12 $\mu$m observations reveal that the hard X-ray emission overlaps with warm H$_2$, where CO(2-1) shows a cavity \citep{Feruglio2020}. A similar case was reported in NGC 2110 by \cite{Fabbiano2019} and \cite{Rosario2019}. This suggests that the scattering material giving rise to the extended hard X-ray emission is warm molecular gas, and CO may be excited to higher-J levels by AGN hard X-ray irradiation and/or shocks. The spatial anti-correlation of hard X-ray emission and dense molecular clouds is also reported in the Circinus galaxy \citep{Kawamuro2019}. It is likely that the molecular gas is efficiently dissociated by AGN X-ray irradiation, leaving an X-ray dominated region. The X-ray irradiation has the potential to consequently suppress star formation. Such observations of molecular gas in the circumnuclear regions of ESO 137-G034 and NGC 3281 will help uncover the origin and mechanism behind the extended hard X-ray emission detected by {\it Chandra}.

\section{Summary and conclusions}
\label{sec:conclusions}

We performed a {\it Chandra} spatial analysis of a sample of 7 uniformly selected nearby CT AGN to investigate the extended hard X-ray component, and measure the excess counts, extended fractions, and physical scales. Five of them show extended emission in the 3-7 keV band detected at $>$ 3$\sigma$ with total excess fractions ranging from $\sim$12\% to 22\%. Among the 7 CT AGN, ESO 137-G034 and NGC 3281 are the two sources that display biconical ionization structures with the extended hard X-ray emission reaching kpc-scales. ESO 137-G034 exhibits the most prominent extended hard X-ray emission with 22\% of the 3-7 keV emission detected in the extended component. The spatial extent reaches $\sim$ 1.9 kpc in diameter with a tendency to extend farther. The extended hard X-ray emission is also well detected in NGC 3281 in the ionization biconical region with a total excess fraction of $\sim$14\% and a spatial extent of $\sim$ 3.5 kpc in diameter. Three additional sources show extended hard X-ray emission above the PSF out to $\sim$360 pc in radius at least. The extended emission in the 0.3-3.0 keV soft band is well detected in most of the sources except ESO 005-G004 and J0025+6821.

We compared the properties of the 7 CT AGN with two other CT AGN with well-detected kpc-scale extended hard X-ray emission, ESO 428-G014 and NGC 7212. Based on these 9 CT AGN, we find a trend that the detection of a prominent extended hard component requires a (3-7 keV) count rate threshold at $\sim$0.01 cts/s combined with a total extended excess fraction of $>$ $\sim$20\%. Since this is a relatively shallow pilot survey, a complete sample with deep {\it Chandra} data will improve the statistics and reveal full X-ray ionization structures, e.g., in ESO 137-G034 where a Z-shaped morphology has been revealed in the optical emission line maps.

We discussed the need to incorporate the extended hard X-ray component in torus modeling in order both to remove emission not associated with the torus that may bias the results, and to provide input parameters to the torus model, reducing the number of degrees of freedom. More torus modeling work is under way to test the effects induced by including this new component (Balokovi{\'c} et al. in prep). The Chandra detections of the extended hard X-ray emission in e.g., ESO 428-G014 and ESO 137-G034, can serve as testbeds for improving torus modeling.

The extended hard X-ray component provides a new window to investigating how the central SMBH interacts with and impacts the host galaxy at large scales. Future combined, high-resolution observations and analyses of the optical emission lines, radio jets, molecular gas, and X-ray emission will unveil the interaction/feedback mechanisms and improve our understanding of AGN-host galaxy connection. 

\section*{acknowledgments}

We thank Raffaella Morganti for providing the ATCA radio data for our analysis. This work makes use of data from the {\it Chandra} data archive, and the NASA-IPAC Extragalactic Database (NED). The analysis makes use of CIAO and Sherpa, developed by the {\it Chandra} X-ray Center; SAOImage ds9; XSPEC, developed by HEASARC at NASA-GSFC; and the Astrophysics Data System (ADS). This work also uses observations made with the NASA/ESA Hubble Space Telescope, obtained from the data archive at the Space Telescope Science Institute. This work was supported by the Chandra Guest Observer program, grant no. GO9-20088X (PI: Elvis). M.B. acknowledges support from the Black Hole Initiative at Harvard University, which is funded in part by the Gordon and Betty Moore Foundation (grant GBMF8273) and in part by the John Templeton Foundation.

\begin{appendix}
\section{Additional images and radial profiles}

\begin{figure*} 
\centering
\includegraphics[width=13.7cm]{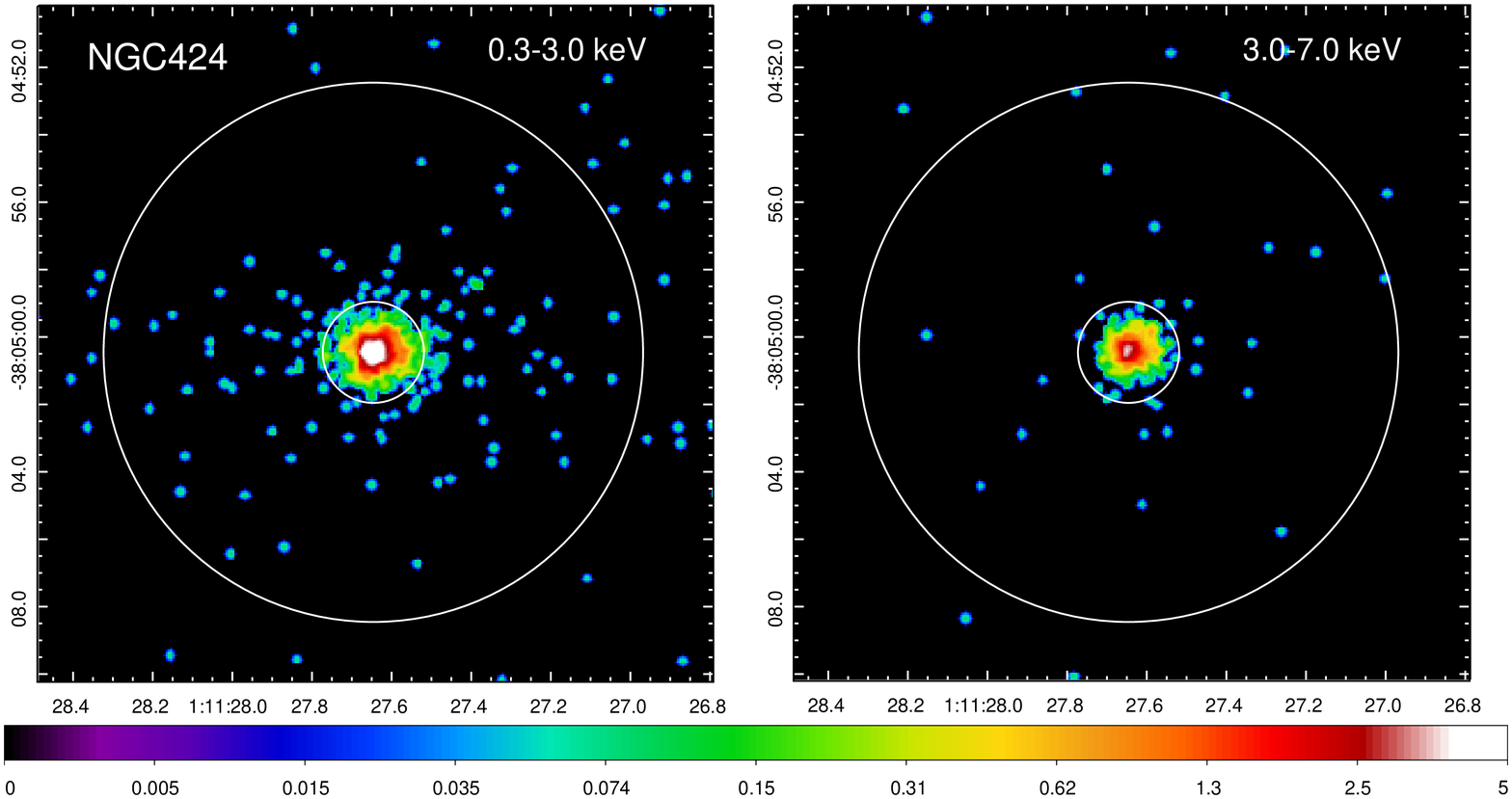}
\includegraphics[width=13.7cm]{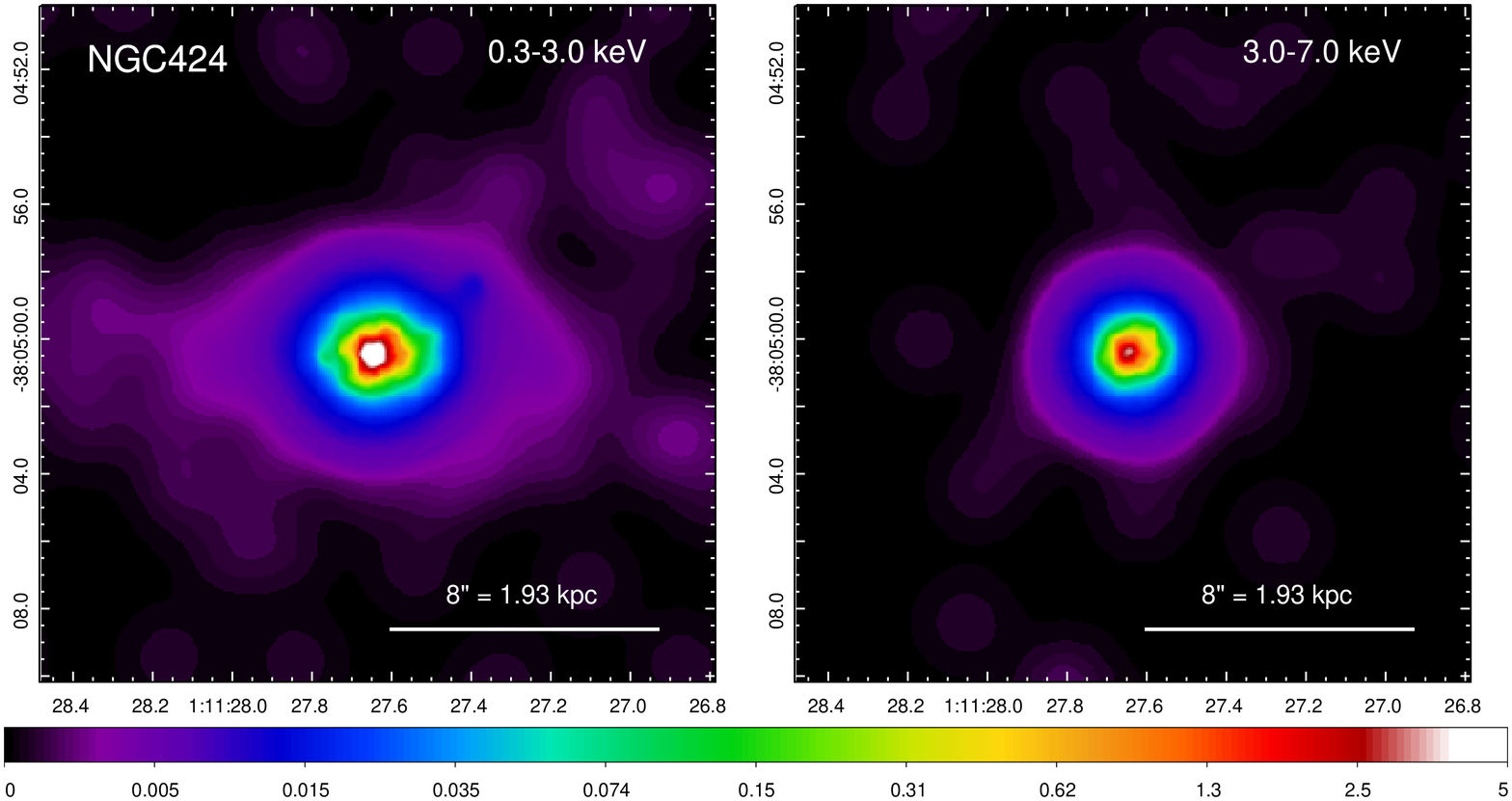}
\caption{{\bf Top:} 20$\arcsec$ $\times$ 20$\arcsec$ merged {\it Chandra} ACIS-S 0.3-3.0 keV (left) and 3.0-7.0 keV (right) band images of NGC 424 at 1/8 subpixel binning. The inner 1.5$\arcsec$ radius circle and the outer 8$\arcsec$ circle define the region in between for extracting excess counts in the extended emission. {\bf Bottom:} Adaptively smoothed images ({\it dmimgadapt}; 0.5-15 pixel scales, 5 counts under kernel, 30 iterations) on 1/8 binned pixel. All the images are displayed in logarithmic scale with colors corresponding to the counts per image pixel.  }
\label{NGC424_chandra}
\end{figure*}

\begin{figure*} 
\centering
\includegraphics[width=5.952cm]{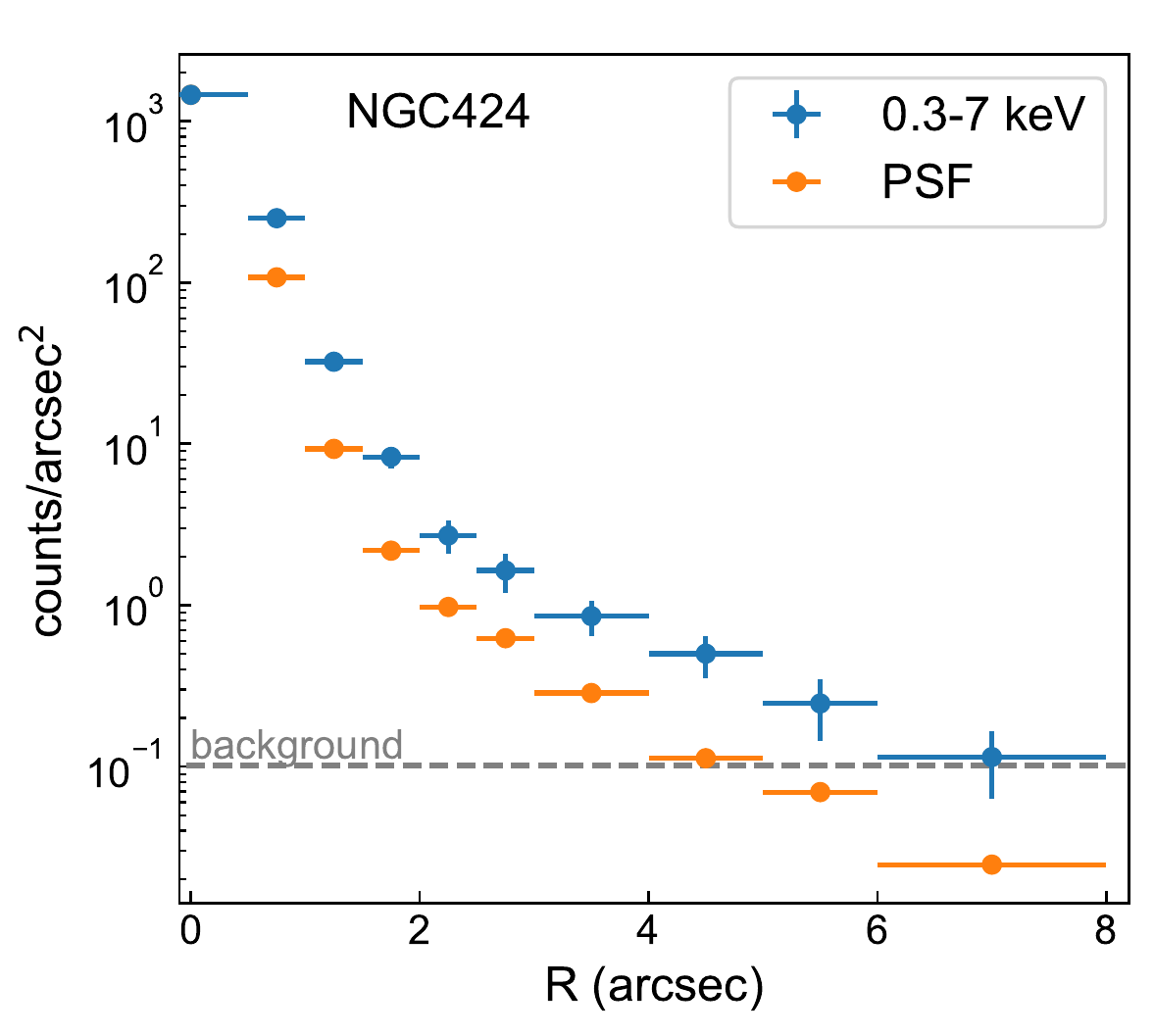}  
\includegraphics[width=5.952cm]{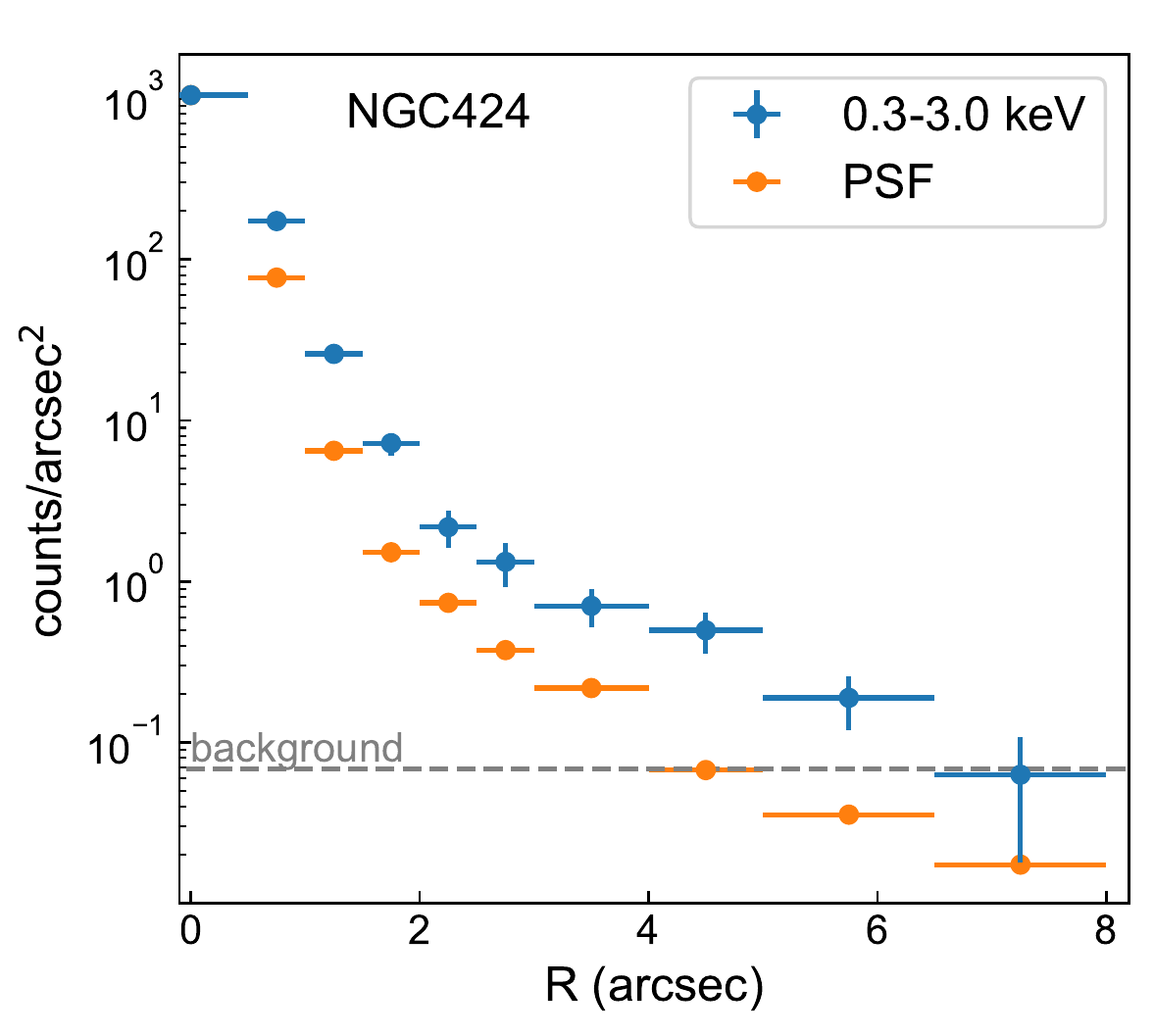} 
\includegraphics[width=5.952cm]{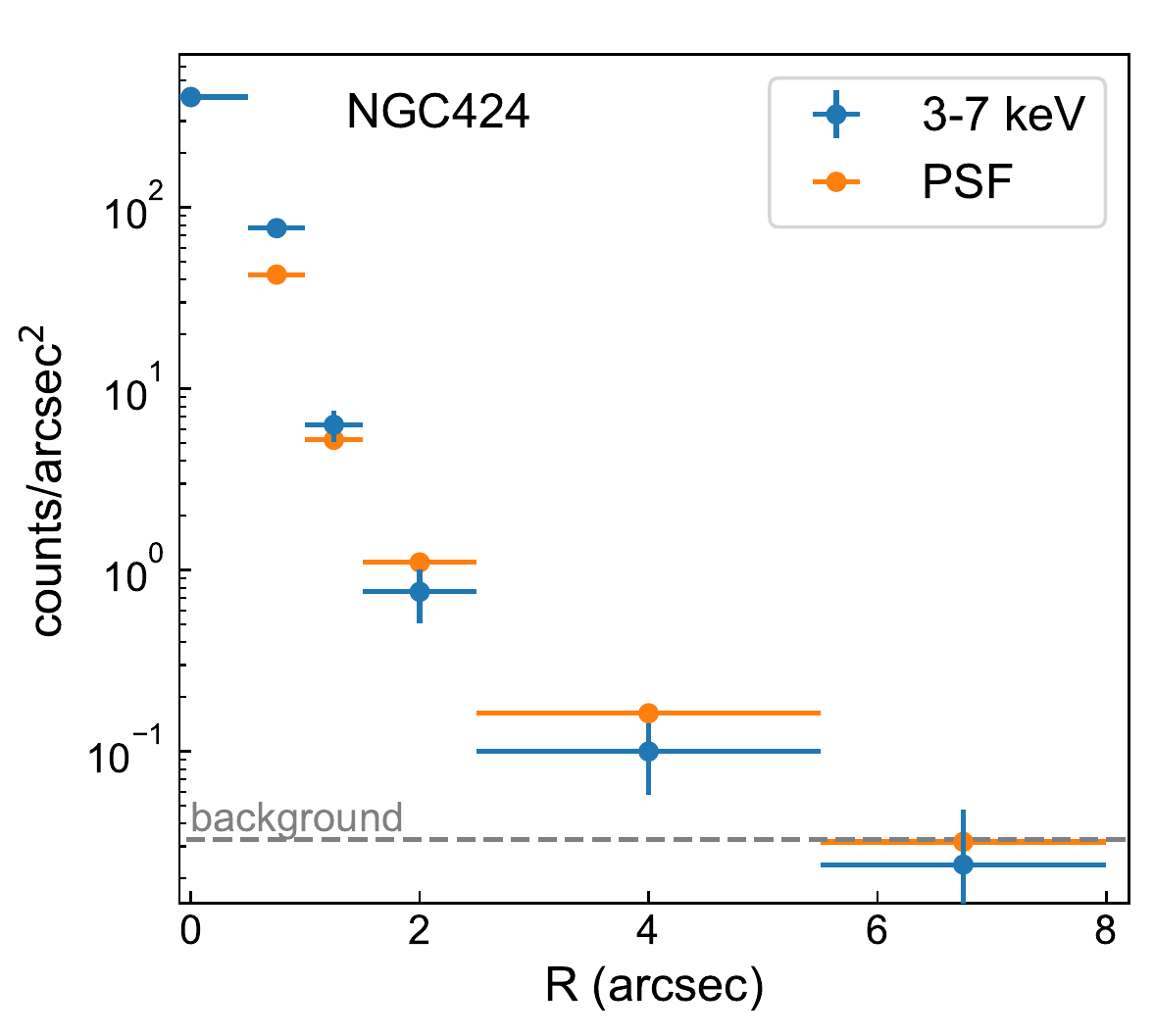} 
\caption{Radial profiles of NGC 424 for the full band, soft band, and hard band. The background has been subtracted off from the radial profiles, and the level of which is indicated as the grey dashed horizontal line. The PSF is normalized to the counts in the central 0.5$\arcsec$ radius bin.}
\label{NGC424_radial_profiles}
\end{figure*}

\begin{figure*} 
\centering
\includegraphics[width=13.7cm]{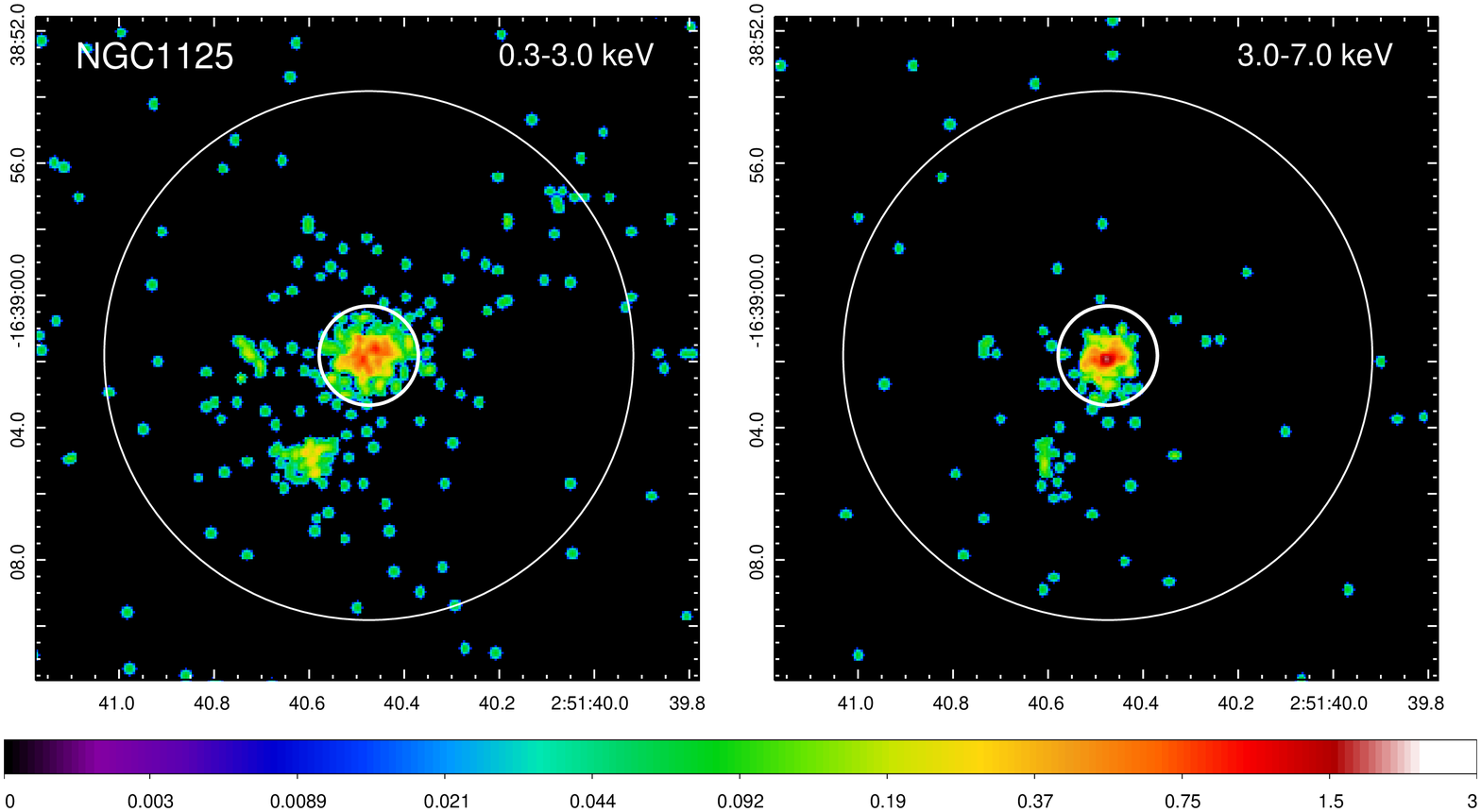}
\includegraphics[width=13.7cm]{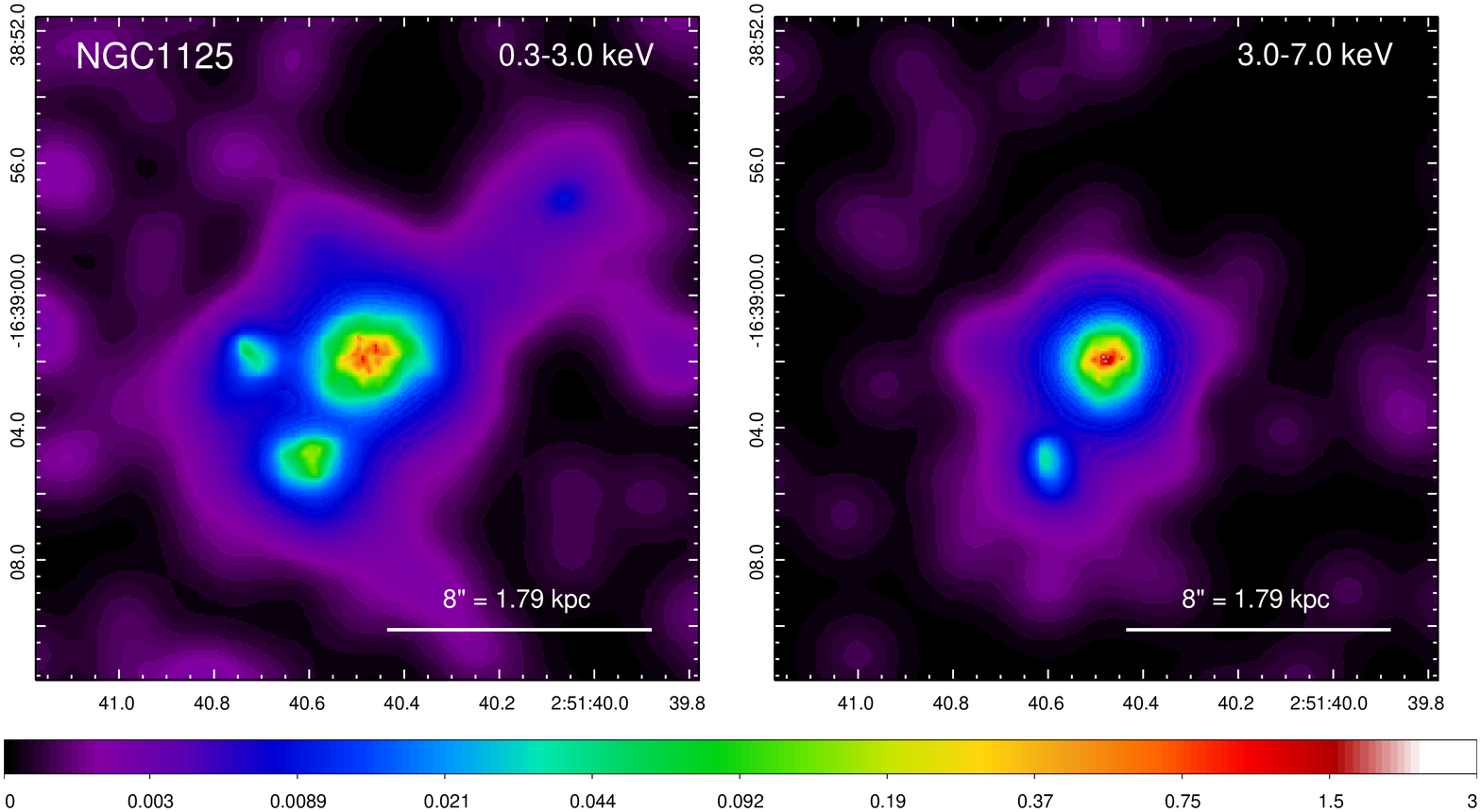}
\caption{{\bf Top:} 20$\arcsec$ $\times$ 20$\arcsec$ {\it Chandra} ACIS-S 0.3-3.0 keV (left) and 3.0-7.0 keV (right) band images of NGC 1125 at 1/8 subpixel binning. The inner 1.5$\arcsec$ radius circle and the outer 8$\arcsec$ circle define the region in between for extracting excess counts in the extended emission. {\bf Bottom:} Adaptively smoothed images ({\it dmimgadapt}; 0.5-15 pixel scales, 5 counts under kernel, 30 iterations) on 1/8 binned pixel. All the images are displayed in logarithmic scale with colors corresponding to the counts per image pixel. }
\label{NGC1125_chandra}
\end{figure*}

\begin{figure*} 
\centering
\includegraphics[width=5.952cm]{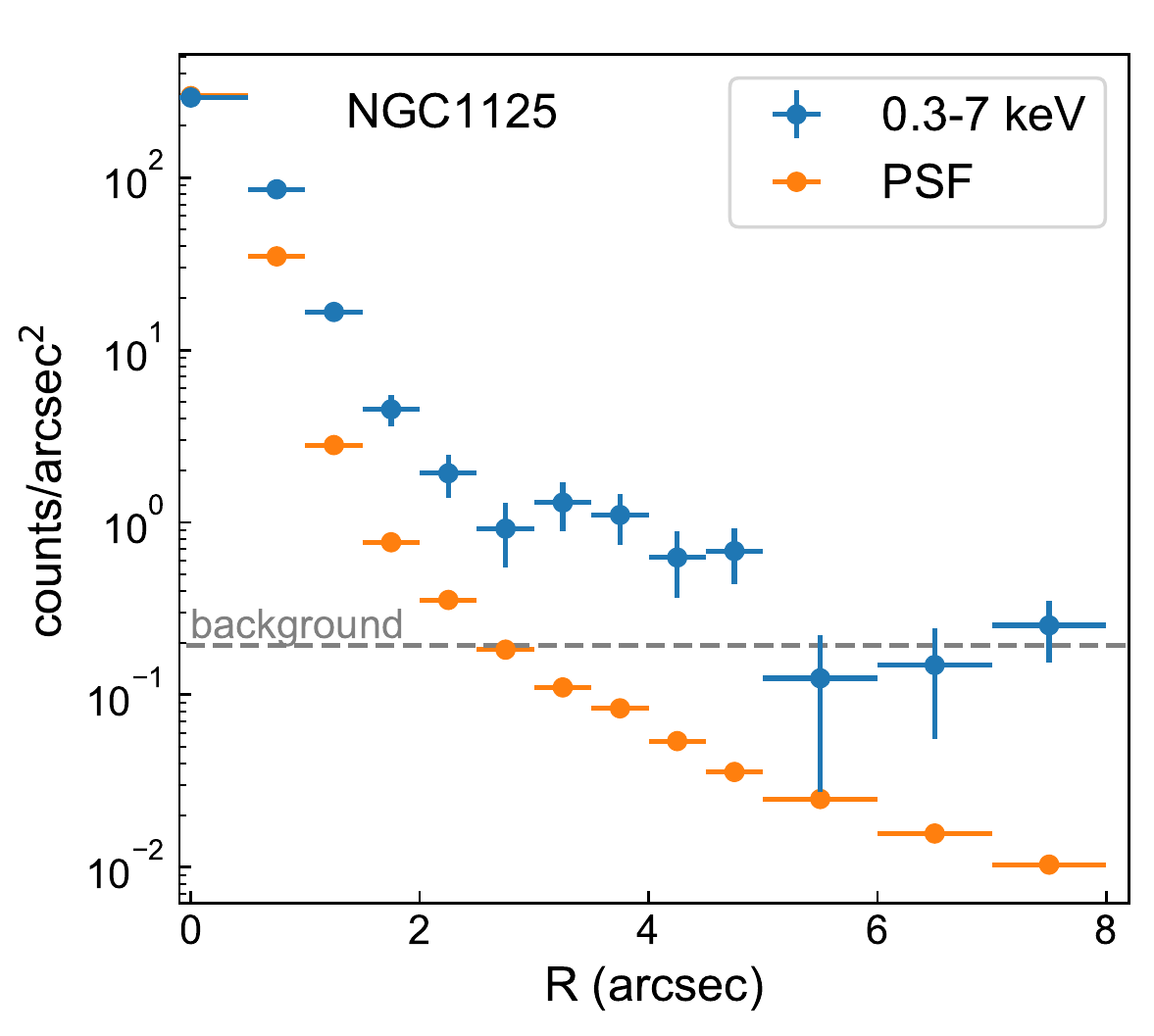}
\includegraphics[width=5.952cm]{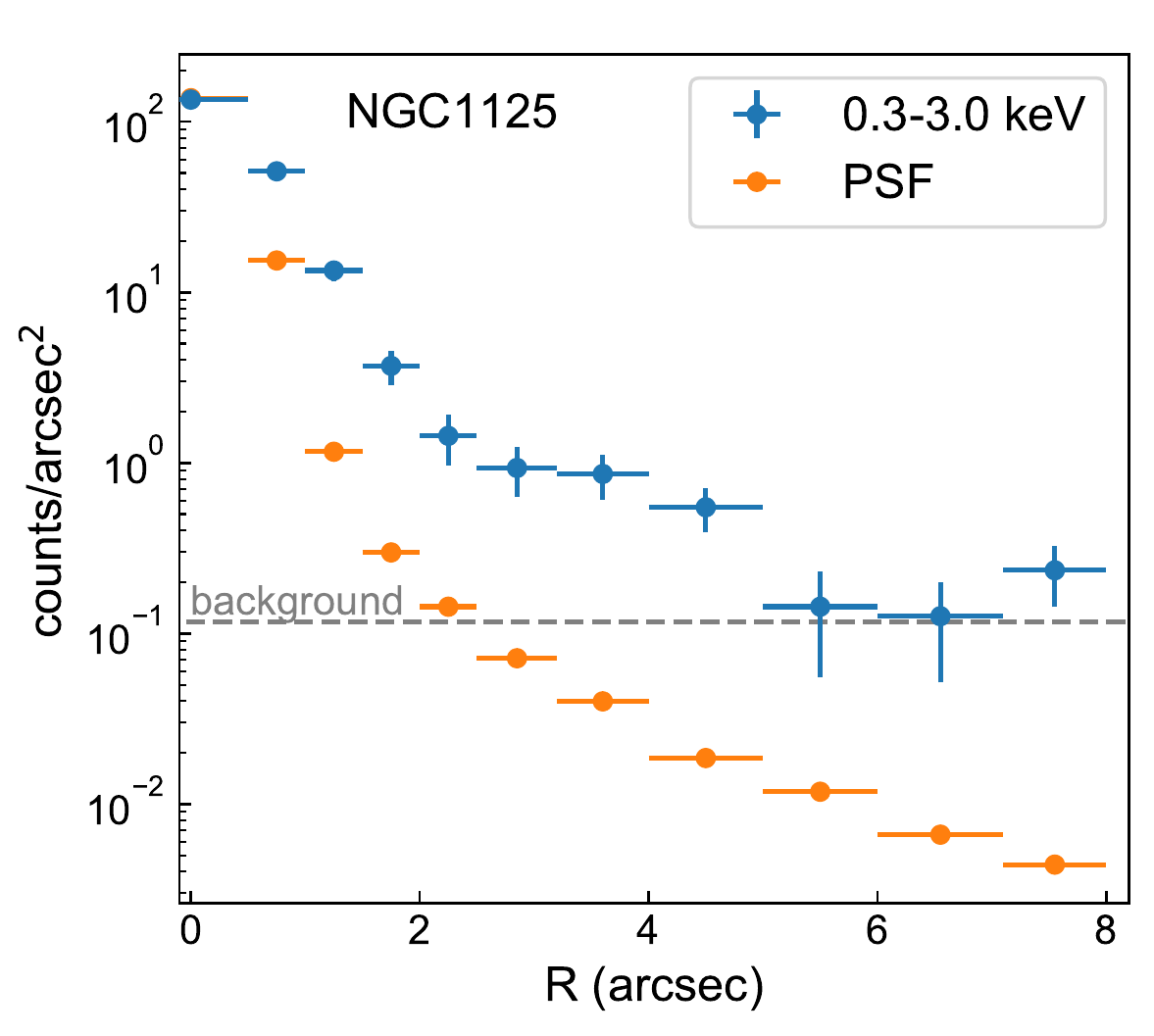}
\includegraphics[width=5.952cm]{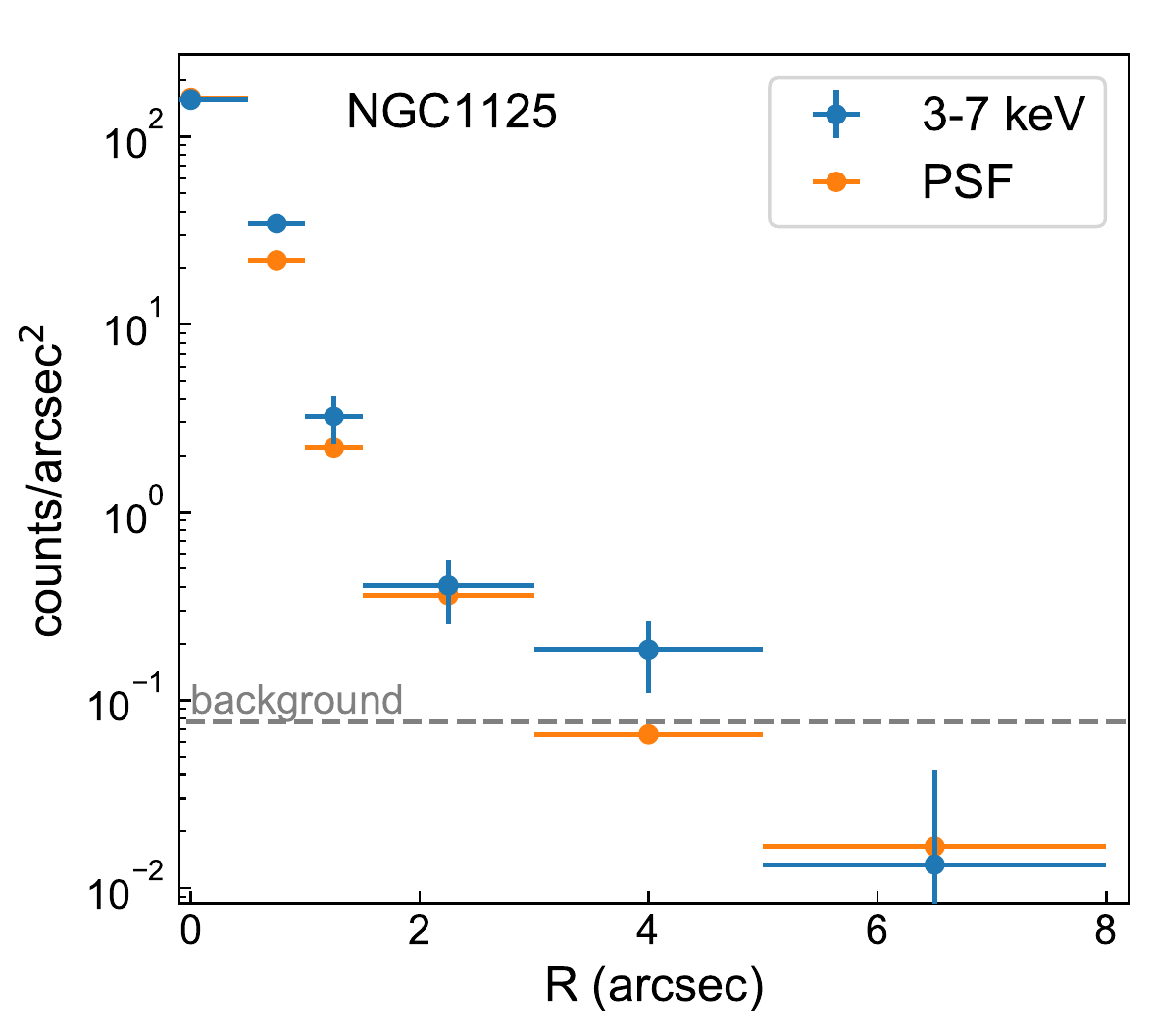}
\caption{Radial profiles of NGC 1125 for the full band, soft band, and hard band. The background has been subtracted off from the radial profiles, and the level of which is indicated as the grey dashed horizontal line. The PSF is normalized to the counts in the central 0.5$\arcsec$ radius bin. The two X-ray sources detected off-center were removed before generating the radial profiles.}
\label{NGC1125_radial_profiles}
\end{figure*}

\begin{figure*} 
\centering
\includegraphics[width=13.7cm]{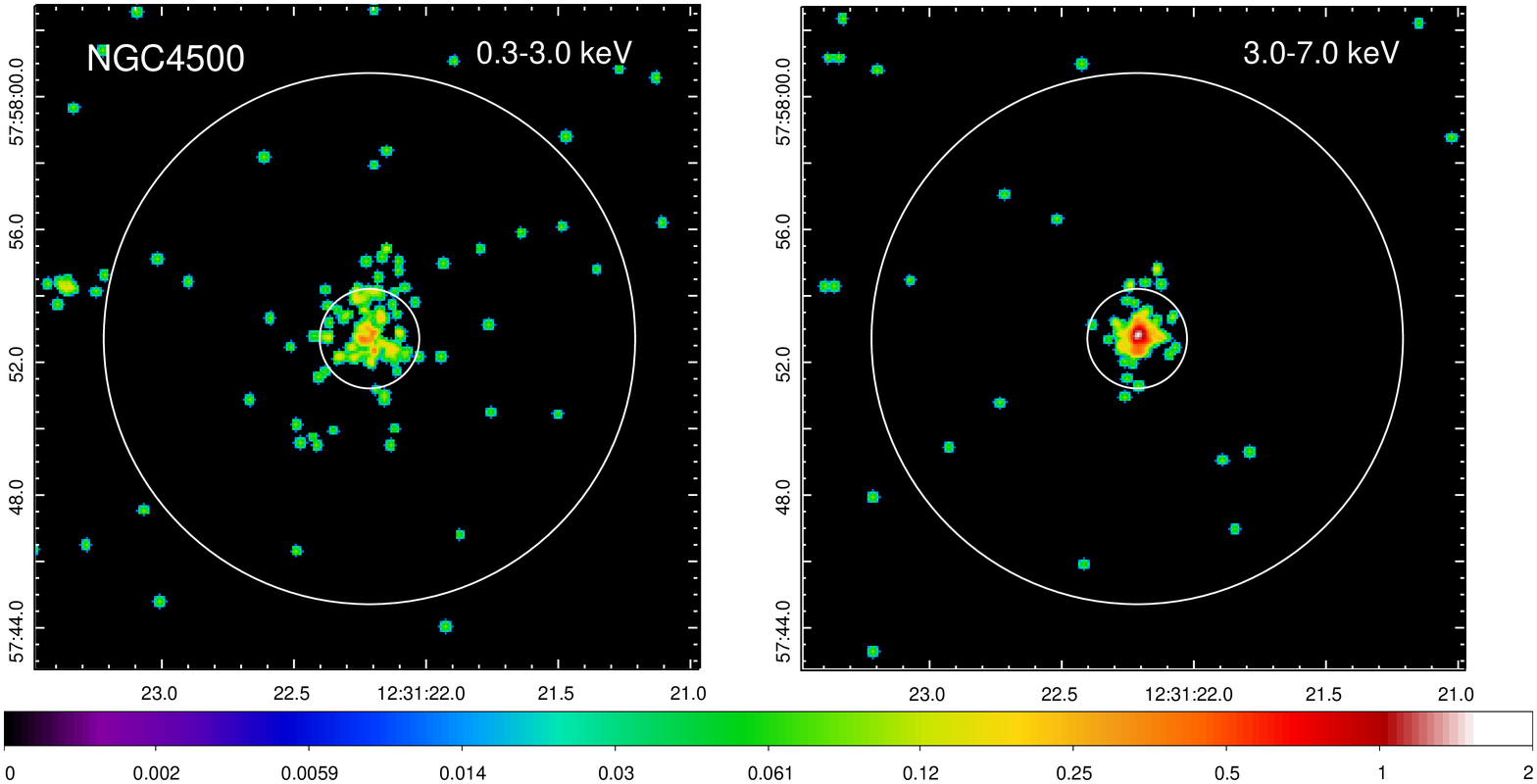}
\includegraphics[width=13.7cm]{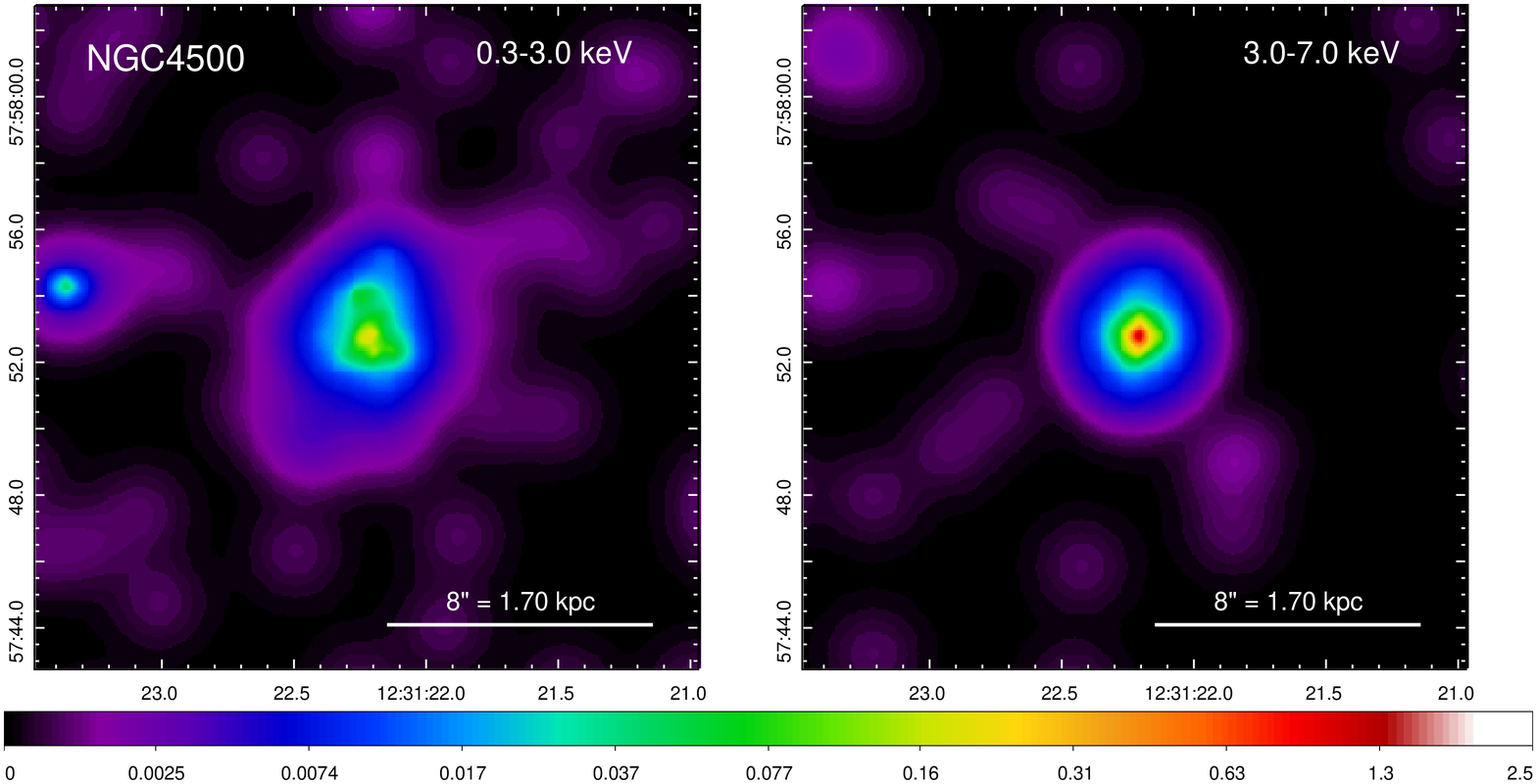}
\caption{{\bf Top:} 20$\arcsec$ $\times$ 20$\arcsec$ {\it Chandra} ACIS-S 0.3-3.0 keV (left) and 3.0-7.0 keV (right) band images of NGC 4500 at 1/8 subpixel binning. The inner 1.5$\arcsec$ radius circle and the outer 8$\arcsec$ circle define the region in between for extracting excess counts in the extended emission. {\bf Bottom:} Adaptively smoothed images ({\it dmimgadapt}; 0.5-15 pixel scales, 5 counts under kernel, 30 iterations) on 1/8 binned pixel. All the images are displayed in logarithmic scale with colors corresponding to the counts per image pixel.}
\label{NGC4500_chandra}
\end{figure*}

\begin{figure*} 
\centering
\includegraphics[width=5.952cm]{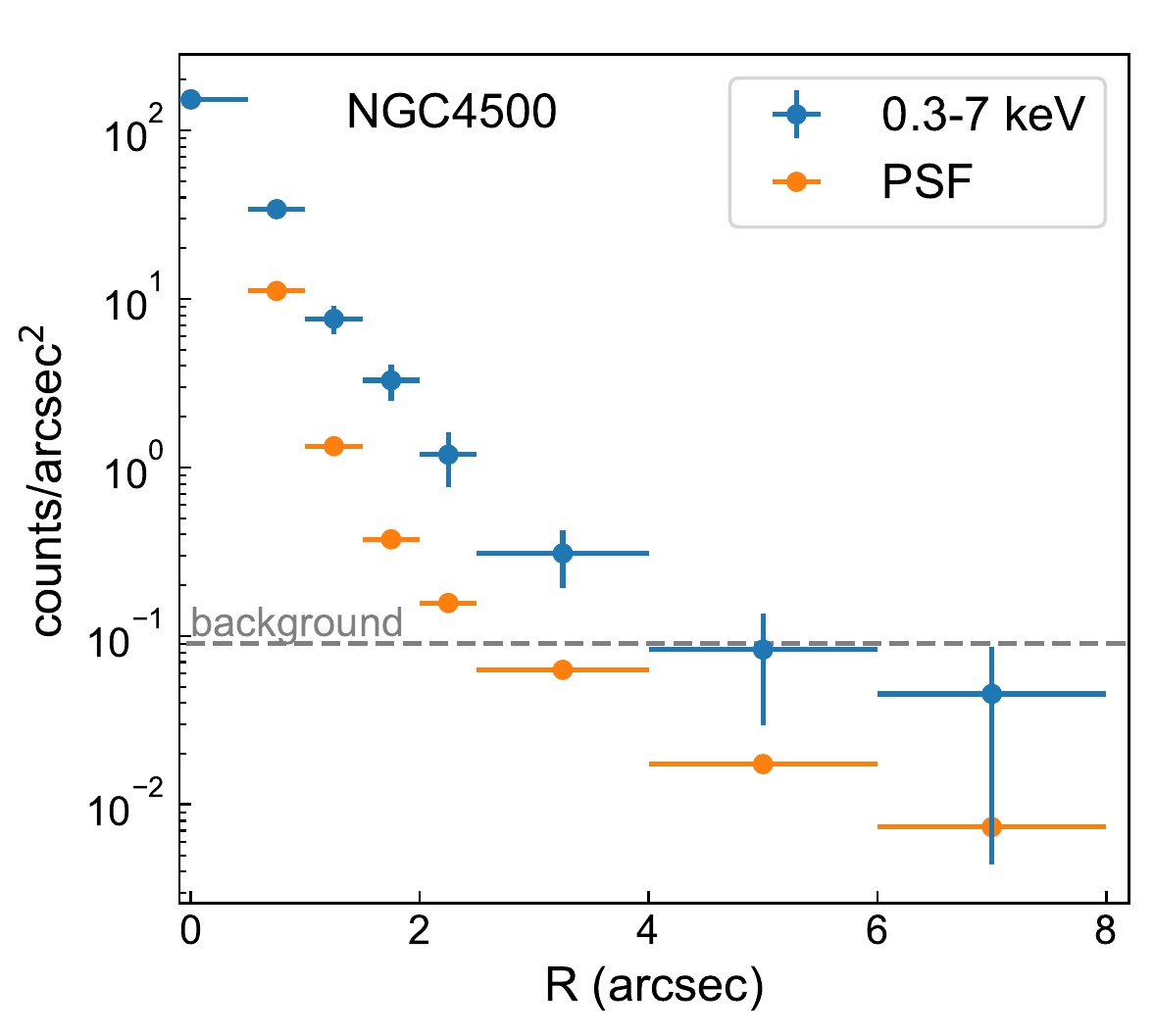}  
\includegraphics[width=5.952cm]{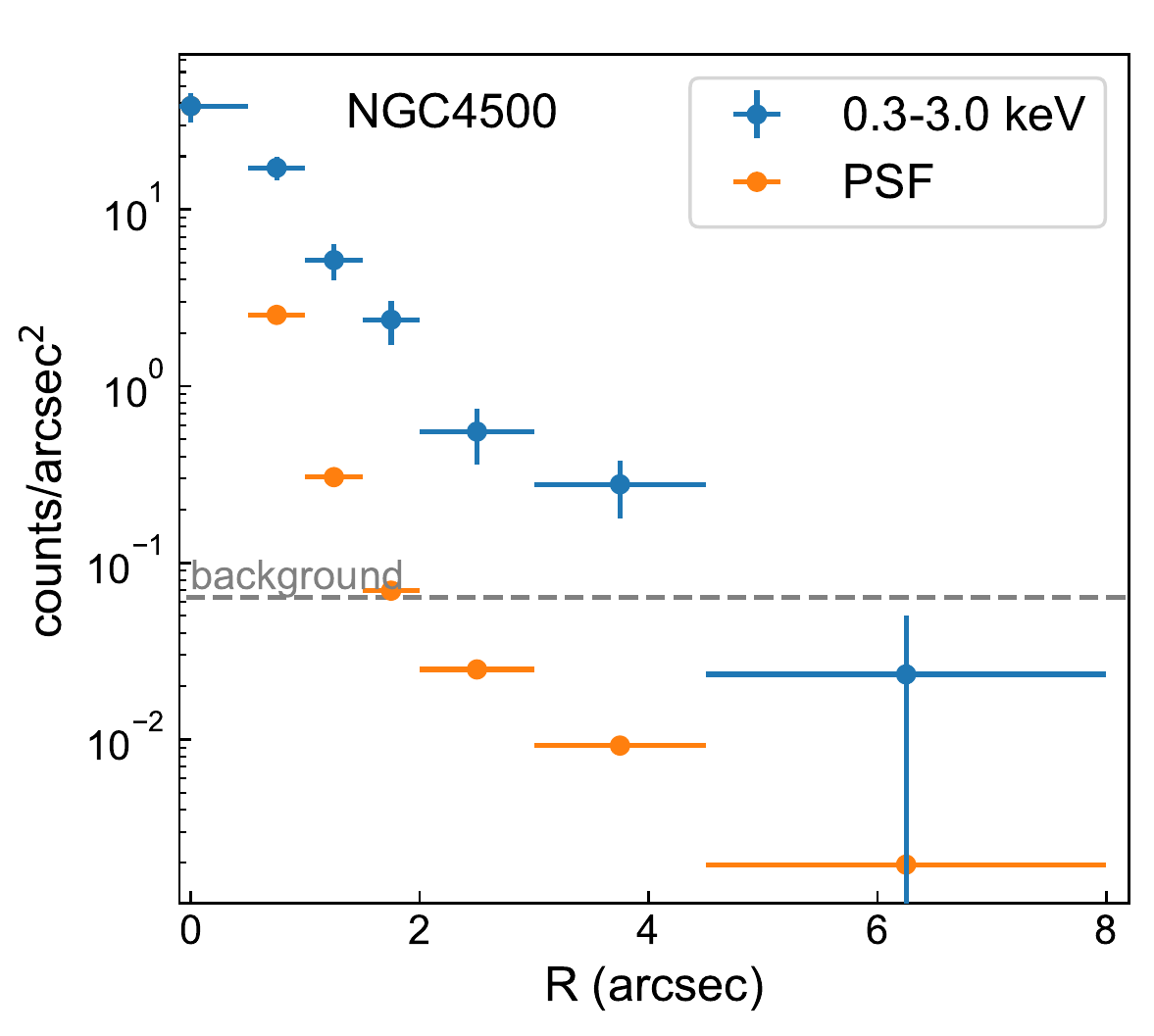}
\includegraphics[width=5.952cm]{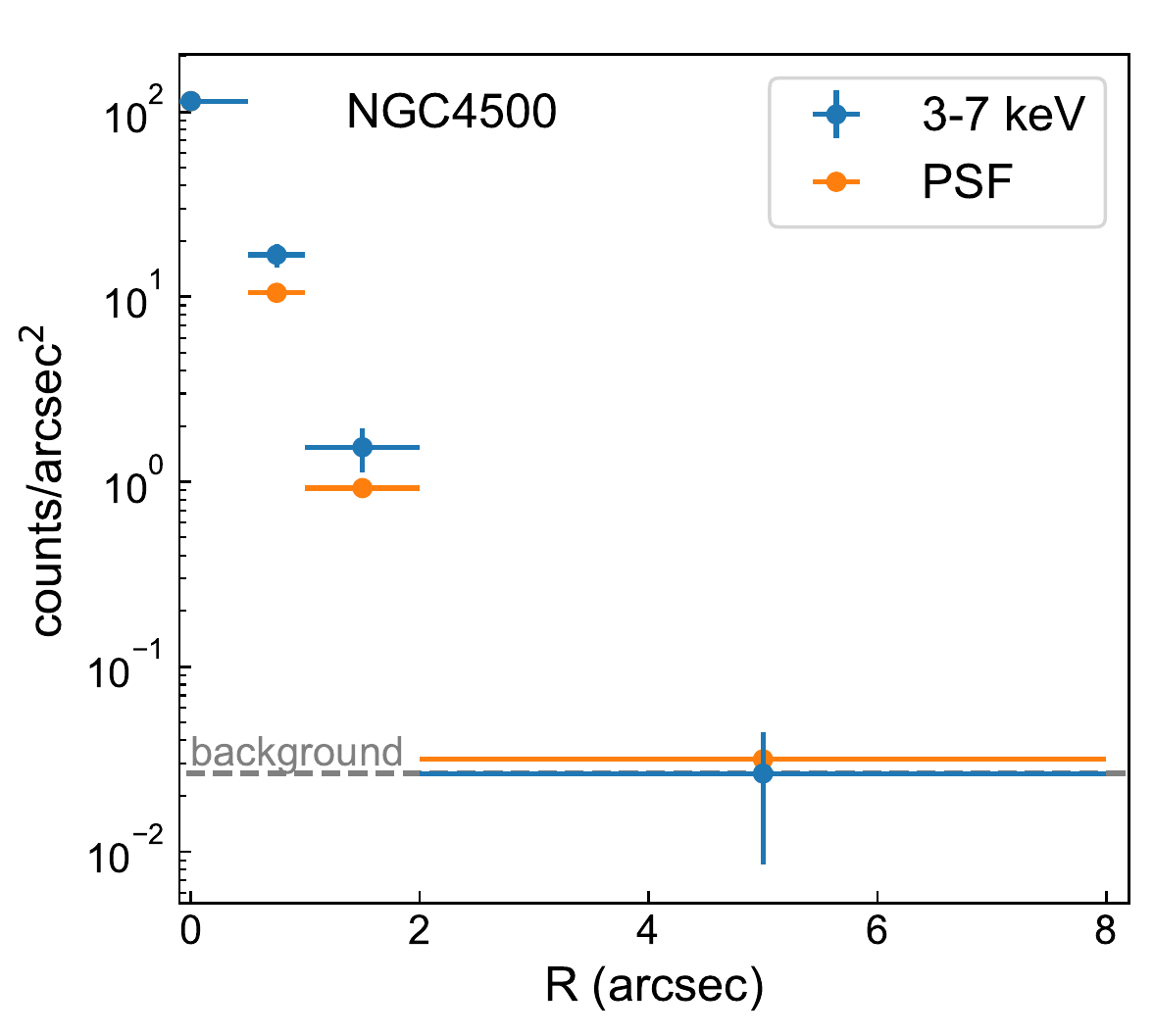}
\caption{Radial profiles of NGC 4500 for the full band, soft band, and hard band. The background has been subtracted off from the radial profiles, and the level of which is indicated as the grey dashed horizontal line. The PSF is normalized to the counts in the central 0.5$\arcsec$ radius bin. }
\label{NGC4500_radial_profiles}
\end{figure*}

\begin{figure*} 
\centering
\includegraphics[width=13.7cm]{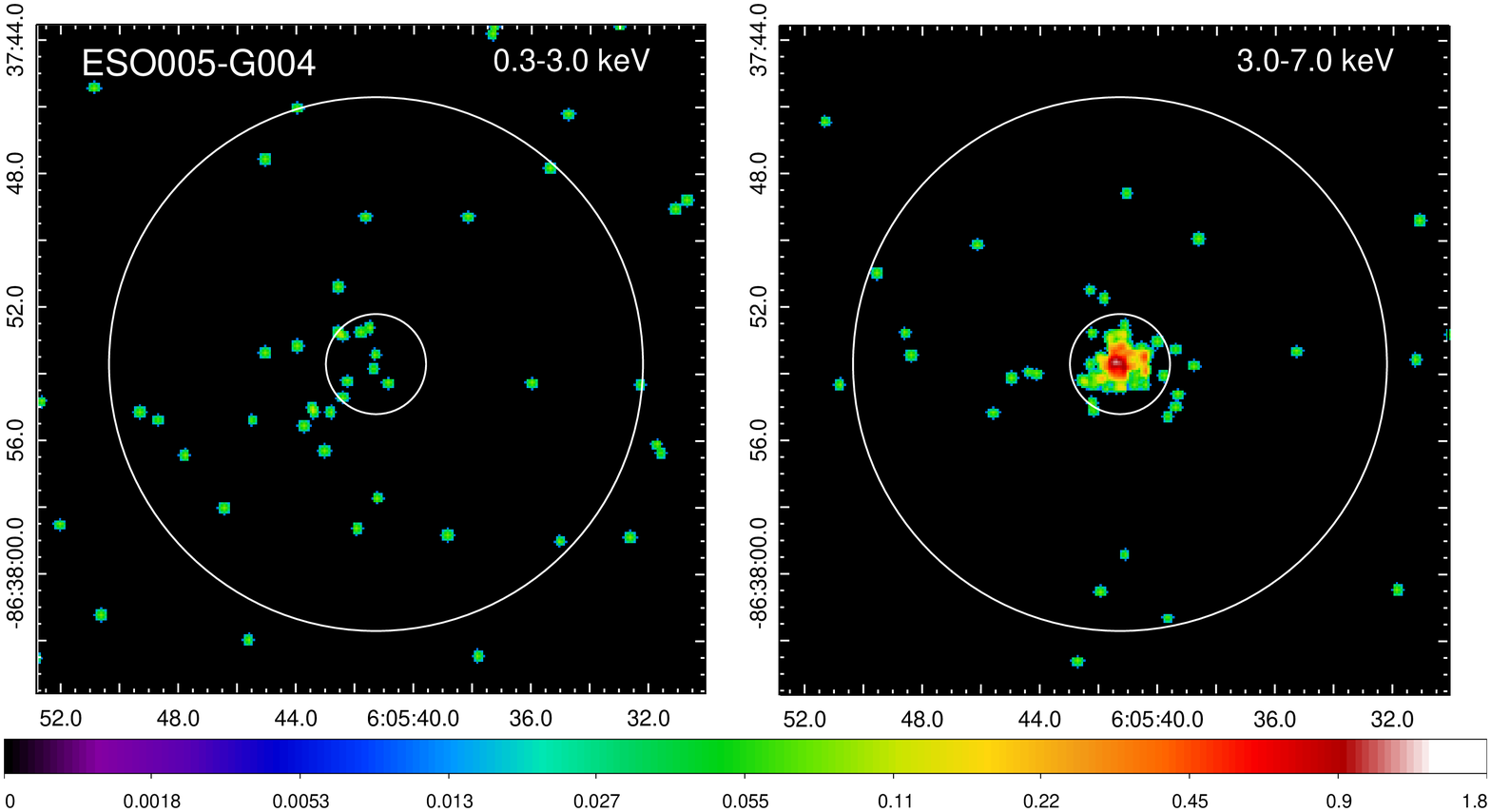}
\includegraphics[width=13.7cm]{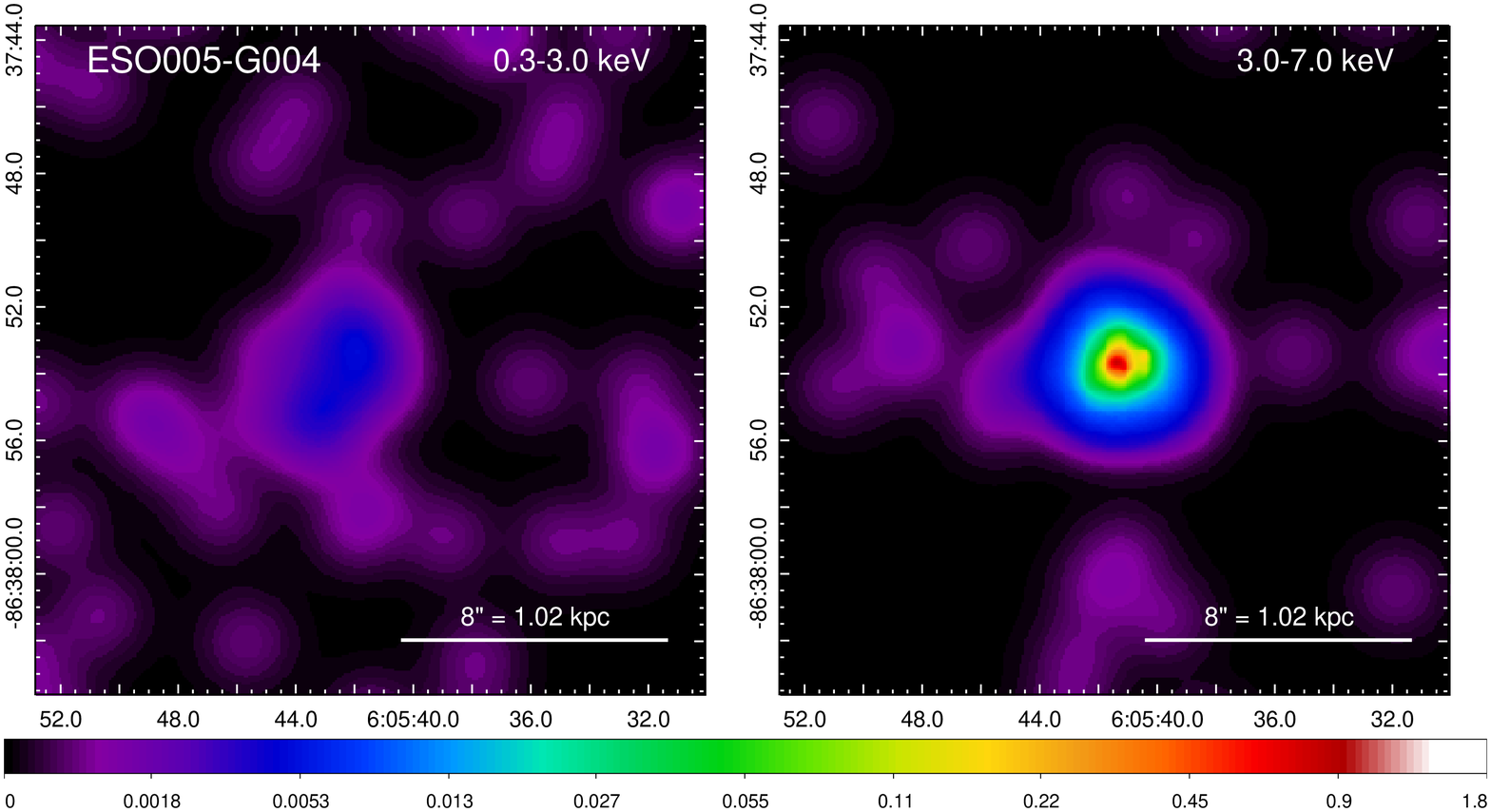}
\caption{{\bf Top:} 20$\arcsec$ $\times$ 20$\arcsec$ {\it Chandra} ACIS-S 0.3-3.0 keV (left) and 3.0-7.0 keV (right) band images of ESO 005-G004 at 1/8 subpixel binning. The inner 1.5$\arcsec$ radius circle and the outer 8$\arcsec$ circle define the region in between for extracting excess counts in the extended emission. {\bf Bottom:} Adaptively smoothed images ({\it dmimgadapt}; 0.5-15 pixel scales, 5 counts under kernel, 30 iterations) on 1/8 binned pixel. All the images are displayed in logarithmic scale with colors corresponding to the counts per image pixel. }
\label{ESO005-G004_chandra}
\end{figure*}

\begin{figure*} 
\centering
\includegraphics[width=5.952cm]{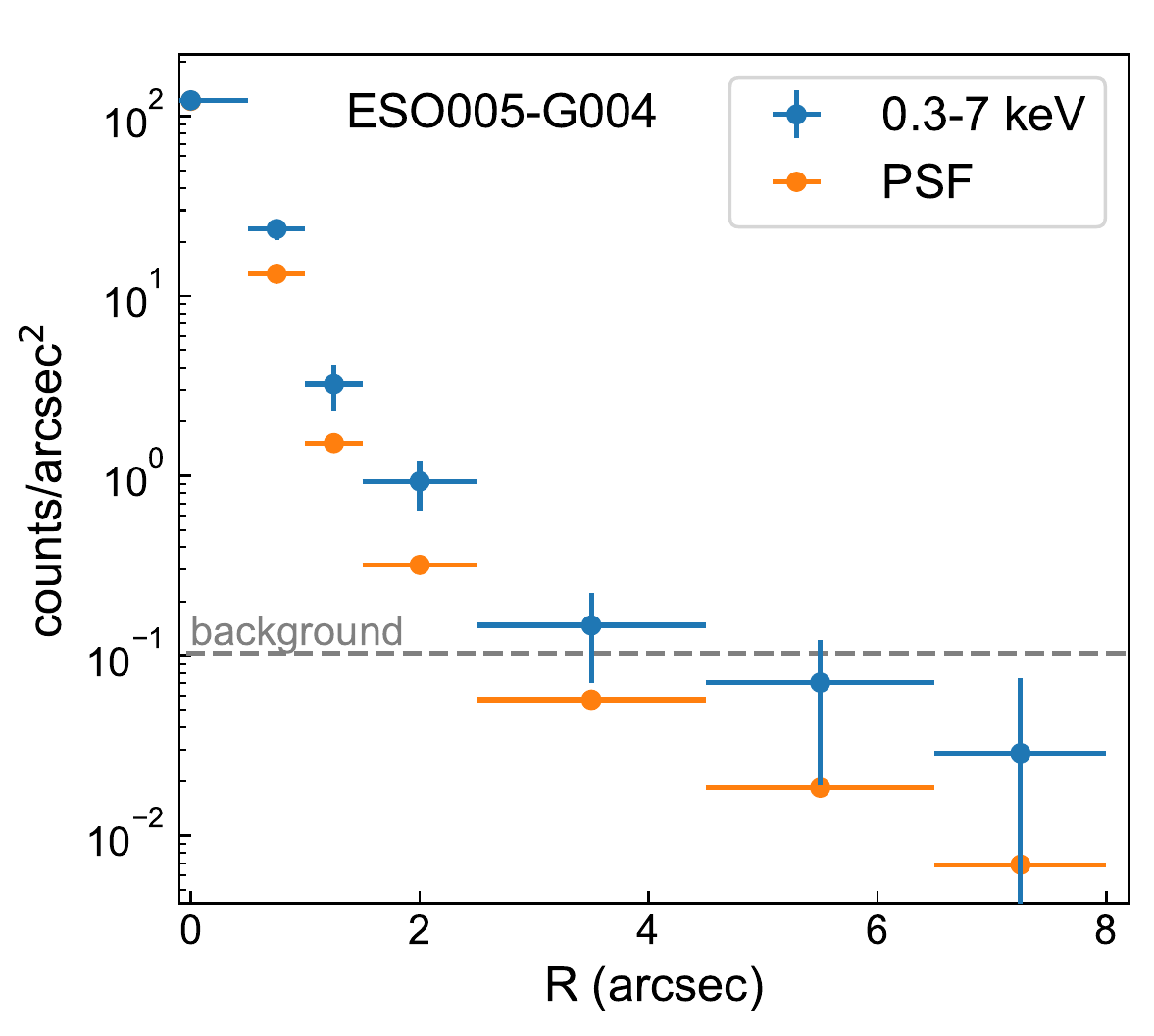}
\includegraphics[width=5.952cm]{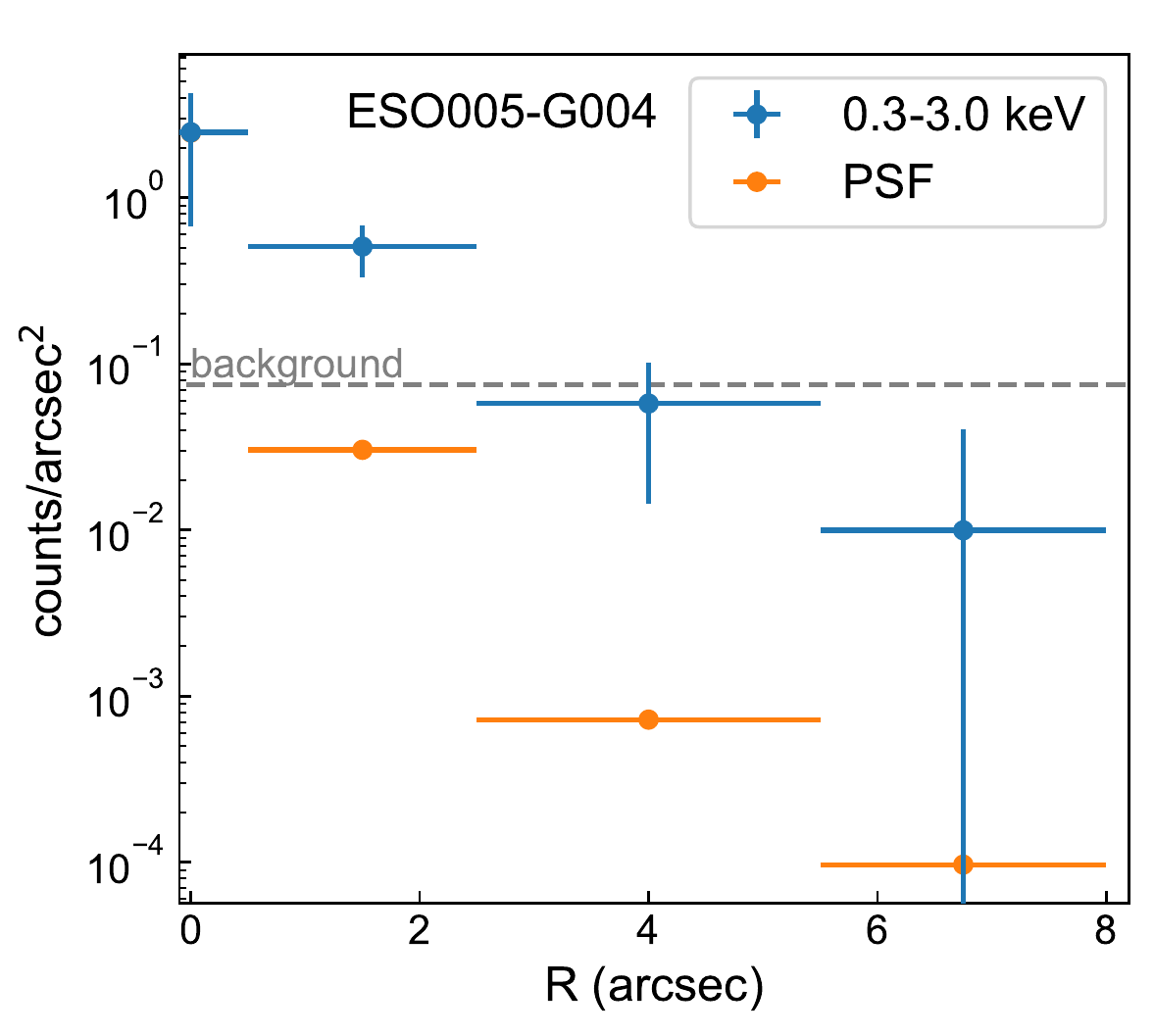}
\includegraphics[width=5.952cm]{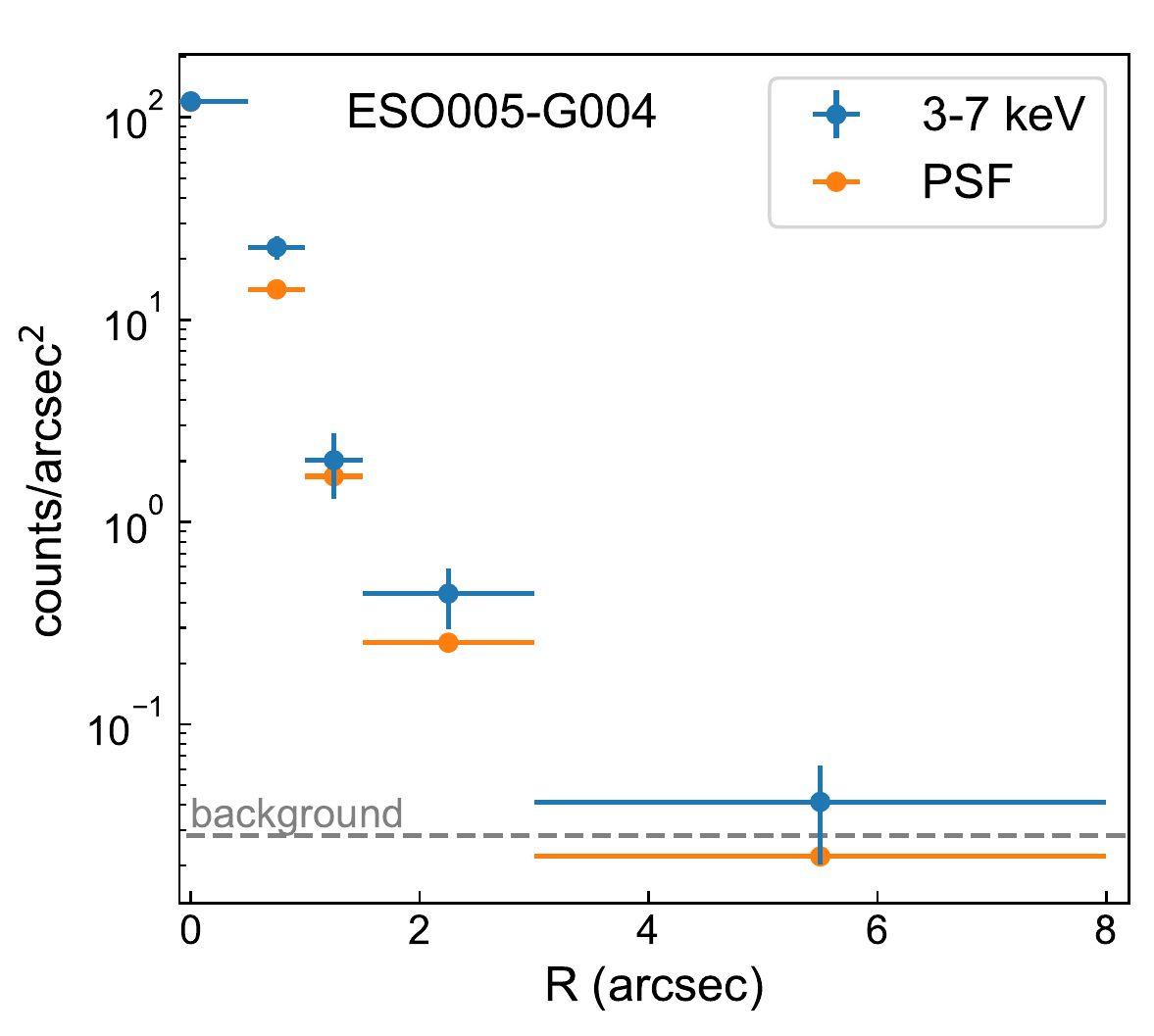}
\caption{Radial profiles of ESO 005-G004 for the full band, soft band, and hard band. The background has been subtracted off from the radial profiles, and the level of which is indicated as the grey dashed horizontal line. The PSF is normalized to the counts in the central 0.5$\arcsec$ radius bin. }
\label{ESO005-G004_radial_profiles}
\end{figure*}

\begin{figure*} 
\centering
\includegraphics[width=13.7cm]{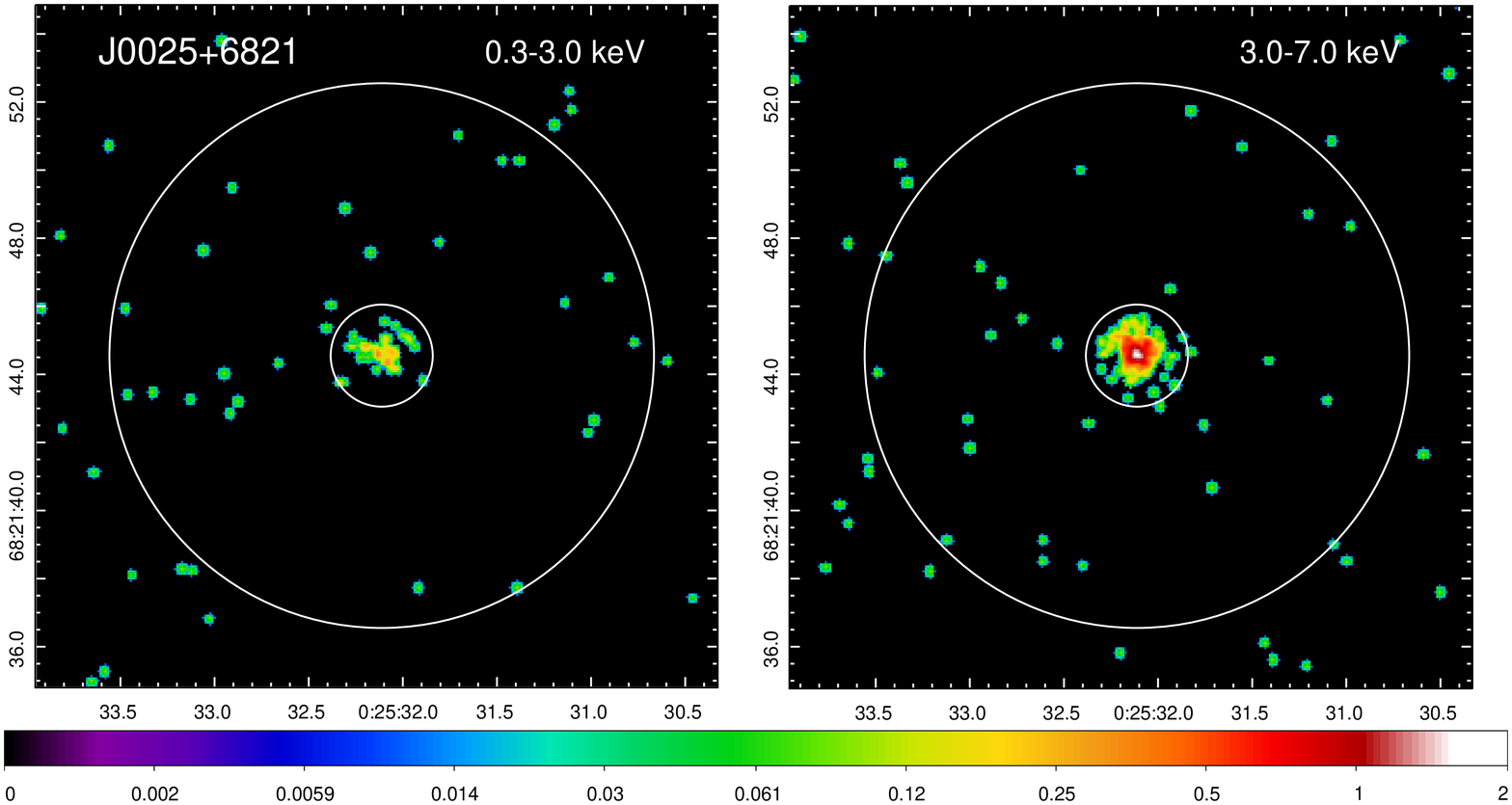}
\includegraphics[width=13.7cm]{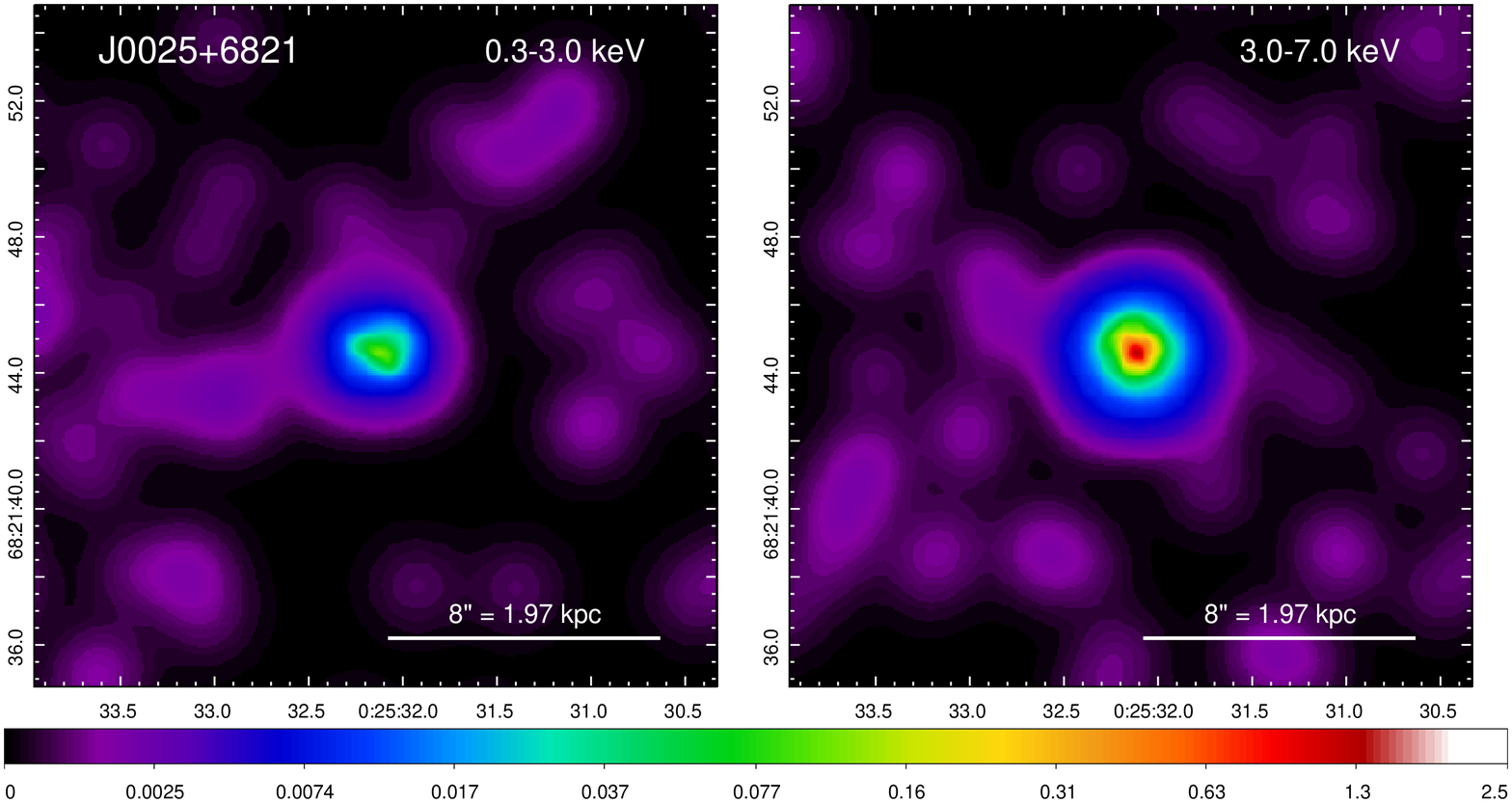}
\caption{{\bf Top:} 20$\arcsec$ $\times$ 20$\arcsec$ merged {\it Chandra} ACIS-S 0.3-3.0 keV (left) and 3.0-7.0 keV (right) band images of J0025+6821 at 1/8 subpixel binning. The inner 1.5$\arcsec$ radius circle and the outer 8$\arcsec$ circle define the region in between for extracting excess counts in the extended emission. {\bf Bottom:} Adaptively smoothed images ({\it dmimgadapt}; 0.5-15 pixel scales, 5 counts under kernel, 30 iterations) on 1/8 binned pixel. All the images are displayed in logarithmic scale with colors corresponding to the counts per image pixel.}
\label{J0025_chandra}
\end{figure*}

\begin{figure*} 
\centering
\includegraphics[width=5.952cm]{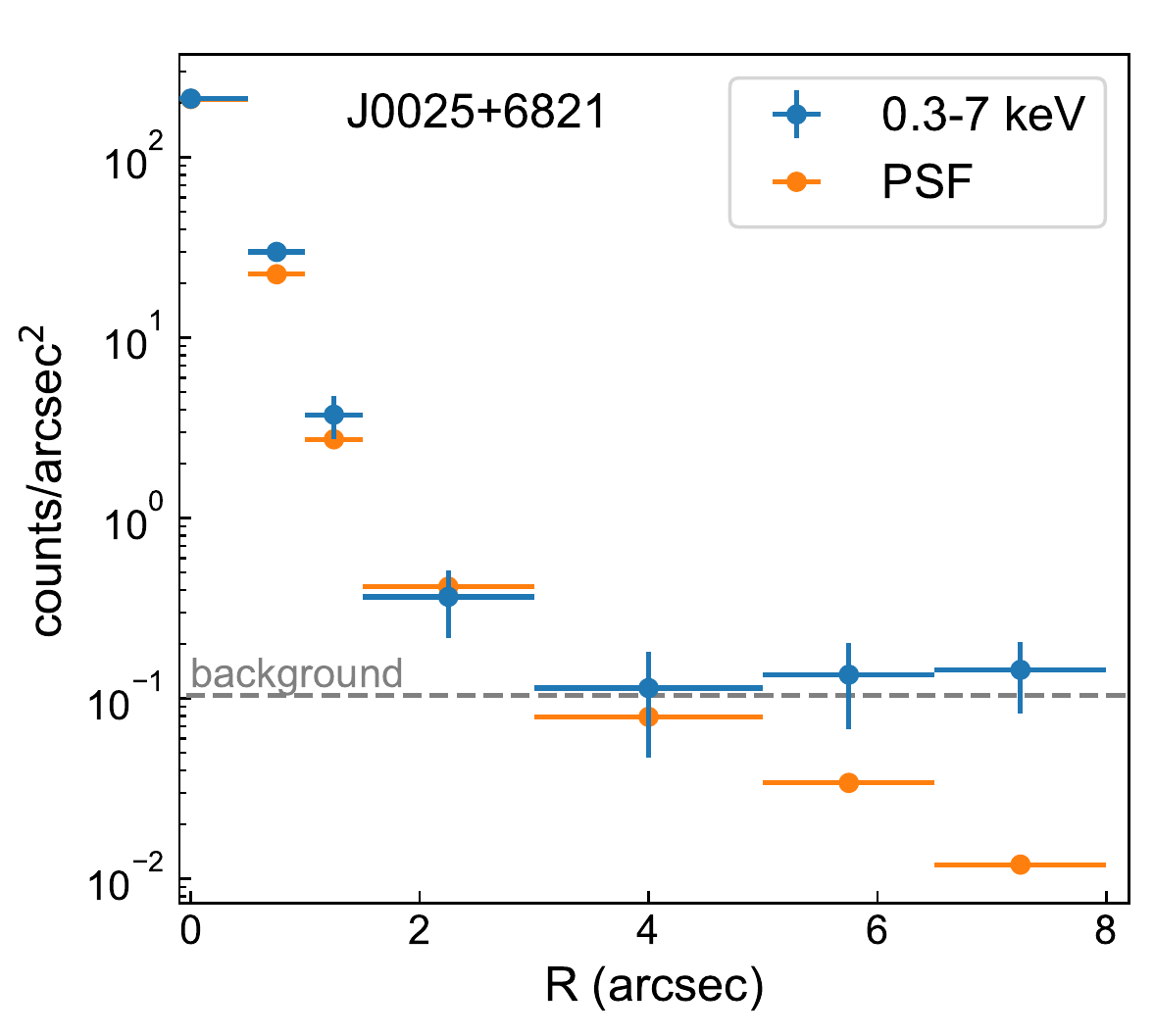}
\includegraphics[width=5.952cm]{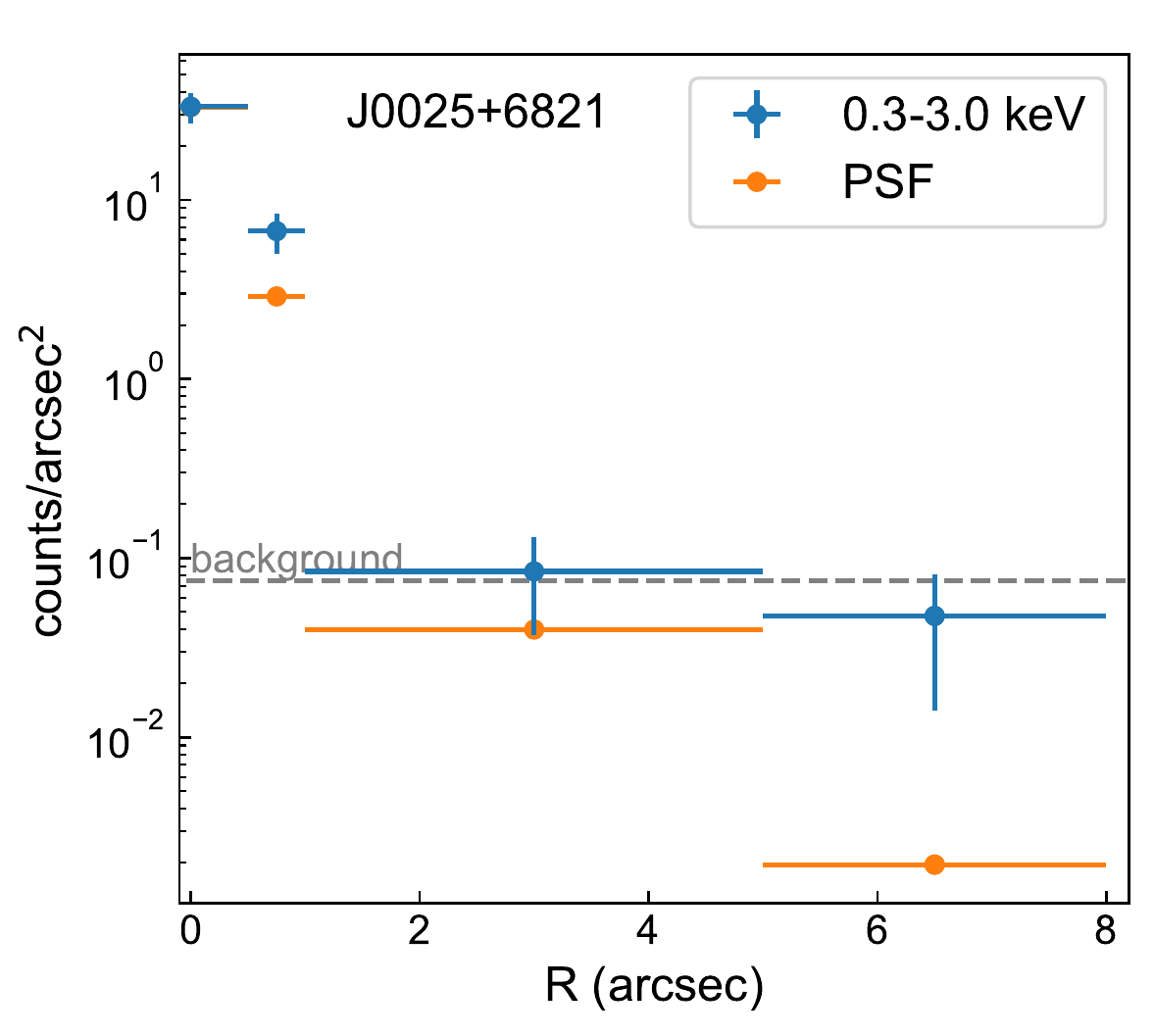}
\includegraphics[width=5.952cm]{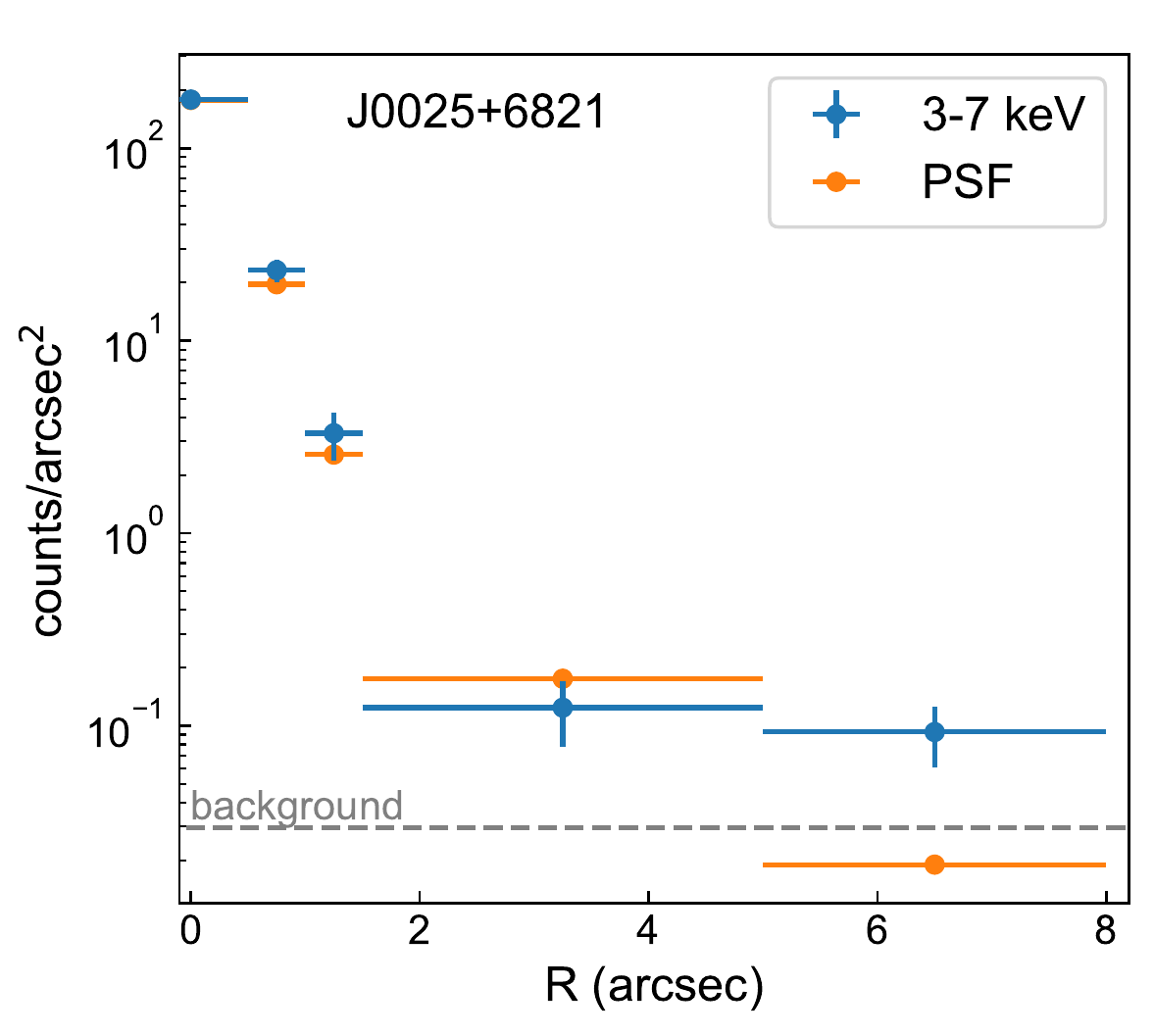}
\caption{Radial profiles of J0025+6821 for the full band, soft band, and hard band. The background has been subtracted off from the radial profiles, and the level of which is indicated as the grey dashed horizontal line. The PSF is normalized to the counts in the central 0.5$\arcsec$ radius bin. }
\label{J0025_radial_profiles}
\end{figure*}

For NGC 424, NGC 1125, NGC 4500, ESO 005-G004, and J0025+6821, the {\it Chandra} images do not show obvious ionization cones and cross-cones, so we generated the radial profiles over all azimuthal angles. The {\it Chandra} soft band (0.3-3 keV) and hard band (3-7 keV) images and their adaptively smoothed versions are shown in Figures \ref{NGC424_chandra}, \ref{NGC1125_chandra}, \ref{NGC4500_chandra}, \ref{ESO005-G004_chandra}, and \ref{J0025_chandra}. The corresponding radial profiles in the 0.3-3 keV, 3-7 keV, and 0.3-7 keV bands are shown in Figures \ref{NGC424_radial_profiles}, \ref{NGC1125_radial_profiles}, \ref{NGC4500_radial_profiles}, \ref{ESO005-G004_radial_profiles}, and \ref{J0025_radial_profiles}, respectively.

\end{appendix}

\end{document}